\documentclass[a4paper,12pt]{article}
\usepackage{aas_macros}
\pdfoutput=1 

\usepackage{jcappub} 

\usepackage[utf8x]{inputenc}
\usepackage[T2A]{fontenc}
\usepackage[english]{babel}
\usepackage{caption}
\usepackage{subcaption}
\usepackage{amsmath}
\usepackage{amssymb}
\usepackage{bm}
\usepackage[dvipsnames,usenames]{xcolor} 
\usepackage[colorlinks=true,citecolor=blue]{hyperref}
\usepackage{todonotes}
\usepackage[numbers,sort&compress]{natbib}

\graphicspath{{./plots/}}

\newcommand{\be}{\begin{equation}}
\newcommand{\ee}{\end{equation}}
\newcommand{\bea}{\begin{eqnarray}}
\newcommand{\eea}{\end{eqnarray}}

\title{Towards an improved  model of self-interacting dark matter haloes}

\author[1]{Anastasia Sokolenko,}
\emailAdd{anastasia.sokolenko@fys.uio.no}
\author[2]{Kyrylo Bondarenko,}
\emailAdd{bondarenko@lorentz.leidenuniv.nl}
\author[3]{Thejs Brinckmann,}
\emailAdd{brinckmann@physik.rwth-aachen.de}
\author[4]{Jes\'us Zavala,}
\emailAdd{jzavala@hi.is}
\author[5]{Mark Vogelsberger,}
\emailAdd{mvogelsb@mit.edu}
\author[1]{Torsten Bringmann}
\emailAdd{torsten.bringmann@fys.uio.no}
\author[2]{and Alexey Boyarsky}
\emailAdd{boyarsky@lorentz.leidenuniv.nl}

\affiliation[1]{Department of Physics, University of Oslo,Box 1048, NO-0371 Oslo, Norway}
\affiliation[2]{Intituut-Lorentz, Leiden University, Niels Bohrweg 2, 2333 CA Leiden, The Netherlands}
\affiliation[3]{Institute for Theoretical Particle Physics and Cosmology (TTK), RWTH Aachen University, Otto-Blumenthal-Strasse, 52057, Aachen, Germany}
\affiliation[4]{Center for Astrophysics and Cosmology, Science Institute,University of Iceland, Dunhagi 5, 107 Reykjavik, Iceland}
\affiliation[5]{Department of Physics, Kavli Institute for Astrophysics and Space Research, Massachusetts Institute of Technology, Cambridge, MA 02139, USA}

\abstract{
We discuss the relation between the strength of the self-interaction of dark matter particles and the predicted properties of the inner density distributions of dark matter haloes. We present the results of $N$-body simulations for 28 galaxy cluster sized haloes performed with the same initial conditions for cold dark matter and for self-interacting dark matter with cross-sections ranging from 0.1 to 10 cm$^2$/g. 
We provide a simple phenomenological description of these results and compare them to the semi-analytical 
model typically used in the literature. We find that some of the assumptions made in this model are not satisfied in the simulations. We identify the reasons for this disagreement and improve the semi-analytical 
model correspondingly. We discuss how simulation results can be properly compared with 
observations and in particular how quantities like the core radius and the inner dark matter surface density 
depend on the self-interaction cross-section.}

\begin{document}
\maketitle

\section{Introduction}

The Cold Dark Matter (CDM) paradigm has been proven to be very successful in describing the large-scale distribution of galaxies and serves as a cornerstone of our current understanding of galaxy formation and evolution (e.g.~\cite{2014Natur.509..177V,2014MNRAS.444.1518V,2015MNRAS.446..521S,2018MNRAS.475..676S}).
Self-interacting dark matter (SIDM)~\cite{Spergel:1999mh} is an interesting and well-motivated hypothesis, both from the astrophysics and particle physics perspectives of dark matter (e.g. \cite{Yoshida:2000uw,Gnedin:2000ea,Firmani:2000qe,Dave:2000ar,Colin:2002nk,ArkaniHamed:2008qn,Ackerman:mha,Feng:2009mn,Buckley:2009in,Feng:2009hw,Loeb:2010gj,Aarssen:2012fx,Tulin:2012re,Vogelsberger:2012ku,Rocha:2012jg,Elbert:2014bma,Robertson:2015faa,Cyr-Racine:2015ihg,Vogelsberger:2015gpr,Harvey:2016bqd,Kamada:2016euw,Kim:2016ujt} or see~\cite{Tulin:2017ara} for a review).
It currently stands as a viable alternative to the CDM paradigm, and as such, the task of constraining the strength of self-interactions from astrophysical observations remains of paramount importance. 

Often quoted {\it upper} bounds on the SIDM transfer cross-section per unit mass are at around $1-2$~ cm$^2/$~g, derived from e.g. the structural properties of elliptical galaxies or galaxy clusters, or from the Bullet cluster and other merging systems (e.g.~\cite{Randall:2007ph,Peter:2012jh,Harvey:2015hha,Robertson:2016xjh,Wittman:2017gxn}). However, we note that the robustness of current constraints on the SIDM cross-section is still debated, in particular due to difficulties in relating observables to quantities that constrain the cross-section (e.g. due to the impact of gas and stars on structure formation or due to projection effects) or in properly measuring such observables directly (e.g. the offsets between the dark matter distribution and luminous matter in merging clusters). A {\it lower} bound of around $0.1$~ cm$^2/$~g can be derived from the requirement that the self-interaction is strong enough for SIDM to be distinct from CDM on small scales (see e.g.~\cite{Tulin:2017ara,Bullock:2017xww,Buckley:2017ijx} for a review), in particular, in order to change the inner structure of dark matter (DM) haloes distinctly from CDM and explain the sizes of DM density cores (e.g.~\cite{Zavala:2012us,Brinckmann:2017uve,Vogelsberger:2018bok}), if the latter are robustly confirmed by observations~\cite{Kaplinghat:2015aga}. Apart from systematic errors in observational data and the uncertainties in modeling baryonic effects~\cite{Kaplinghat:2015aga,Elbert:2016dbb,Robles2017}, the properties of the haloes and, in particular, the sizes of the cores (if they exist) are expected to have significant scatter, due to individual merger history and specific initial conditions, see e.g.~\cite{Strigari:2018bcn} and references therein. 

In order to use observational data to determine (or constrain) an intrinsic quantity of DM particles such as its 
self-interaction cross-section, it is perhaps more efficient to fit the data to the whole ensemble of haloes at the 
same time. Such a procedure was discussed in~\cite{Bondarenko:2017rfu}, where the inner DM surface density, 
a quantity obeying a well-known scaling law~\cite{Boyarsky:2009rb,Boyarsky:2009af} for a halo mass range 
spanning 6 orders of magnitude, was used to compare SIDM predictions for different cross-sections with 
observations. The main theoretical uncertainty of this type of analysis is the relation between the observable 
quantity, the core radius $r_{\text{core}}$, where the DM density is close to constant $\rho(0)$, and the 
radius $r_{\text{SIDM}}$, where self-interactions become important and the velocity dispersion of DM 
particles is expected to be close to constant. While the former radius, $r_{\text{core}}$, is more directly 
connected to observations, the latter, $r_{\text{SIDM}}$, is more directly predicted by 
theory~\cite{Kaplinghat:2015aga,Kamada:2016euw,Valli:2017ktb}. The explicit relation between these two 
radii for every cross-section has not been discussed in detail in the literature. In order to take into account 
this theoretical uncertainty, a free phenomenological parameter $\kappa$,
\begin{equation}
 \kappa \equiv \rho(0)/\langle \rho(r_{\text{SIDM}})\rangle\,,
 \label{eq:kappa_def}
\end{equation}
was introduced in Ref.~\cite{Bondarenko:2017rfu}. Here $\rho(0)$ is the central halo density 
and $\langle \rho(r_{\text{SIDM}})\rangle$ is the average density of the halo within the radius $r_{\text{SIDM}}$, 
i.e.~$\langle \rho(r_{\text{SIDM}})\rangle = {M(r_{\text{SIDM}})}/\left({\frac{4}{3}\pi r_{\text{SIDM}}^3}\right)$.
In this article, we use numerical simulations to remove this uncertainty as far as possible.

The main goal of this work is to develop and test an (spherical) analytical phenomenological model that predicts (potentially) observable properties of pure SIDM haloes for arbitrary values of the self-interaction transfer cross-section per unit mass $\sigma/m$ and to compare this with the semi-analytical models typically used in the literature.
In particular, by knowing the density profile of a given halo at large radii we would like to predict the inner structure of the
same halo for a given value of the cross-section.
As the properties of SIDM haloes are believed to be the same as for CDM haloes at large radii, we can calibrate the model by taking as an input the properties of a simulated CDM halo and predict the DM density and velocity dispersion profiles of the SIDM halo forming from the same initial conditions for a given value of $\sigma/m$. We do not discuss here the effects of baryons on SIDM haloes, as we would like to check the phenomenological description for the DM only case first. For discussion in this direction, see instead~\cite{Kaplinghat:2013xca,Vogelsberger:2014pda,Kamada:2016euw,Robertson:2017mgj,Sameie:2018chj}, as well as a recent SIDM review~\cite{Tulin:2017ara}. We would like to remark that the model presented in this paper is an improvement over previous models discussed in the literature \cite{Kaplinghat:2013xca,Kaplinghat:2015aga}.

This article is organized as follows. We start, in Section \ref{sec:sim}, by describing the numerical simulations
we performed and provide a brief overview of the results, i.e.~how SIDM halos differ from their CDM counterparts. 
In Section
\ref{sec:anal_model}, we develop an analytic model to describe SIDM halos, which we further refine in
Section \ref{sec:rm}. We then compare predictions of our model to that commonly adopted in the literature, in 
Section \ref{sec:tulin}, before presenting our conclusions in Section \ref{sec:conc}. In three Appendices we provide
further technical details about the simulation results that support the discussion in the main part of the article.

\section{Simulations}
\label{sec:sim}

\subsection{Setup}
The initial simulation suite used in this work was performed using the \texttt{AREPO} code~\cite{Springel:2009aa}, with an added module for dark matter self-interactions~\cite{Vogelsberger:2012ku,Vogelsberger:2015gpr}. This simulation suite is described in detail in \cite{Brinckmann:2017uve}; in the following we briefly summarize the main aspects relevant for this work. The suite consists of a sample of zoom-in simulations of massive cluster-sized haloes with initial conditions generated with the \texttt{MUSIC} code\footnote{https://www-n.oca.eu/ohahn/MUSIC/}~\cite{Hahn:2011uy} at a redshift of $z=50$, with an effective resolution of $512^{3}$ particles, a softening length of $\epsilon = 5.4$~kpc~$h^{-1}$ and particle mass m$_p = 1.271 \times 10^9$~M$_{\odot}$~$h^{-1}$. In addition, one halo was also simulated with a factor of 2 better resolution. 
The suite presented in \cite{Brinckmann:2017uve} consists of $3\times28$ haloes 
(and additional $3\times1$ halo at the higher resolution level) in a CDM and SIDM cosmology, with cross-sections of $\sigma/m=0.1$ and $1~\text{cm}^2/\text{g}$, starting from matching initial conditions. For this work, we expand on that suite by re-simulating all 28 haloes with a cross-section of $\sigma/m=0.5~\text{cm}^2/\text{g}$, as well as 10 of those haloes with cross-sections of $\sigma/m=5~\text{cm}^2/\text{g}$ and $\sigma/m=10~\text{cm}^2/\text{g}$. Finally we also re-simulated 3 haloes in the sample at the higher resolution level for the CDM and SIDM cosmologies with cross-sections $\sigma/m=0.1$, $0.5$ and $1~\text{cm}^2/\text{g}$.

All simulations were computed with a cosmology consistent with Planck~\cite{Ade:2015xua}: with contributions to the energy density of the universe from matter $\Omega_{\rm m}=0.315$ and cosmological constant $\Omega_\Lambda=0.685$, dimensionless Hubble parameter $h=0.673$, root-mean-square amplitude of perturbations in $8$~Mpc~$h^{-1}$ spheres today $\sigma_8=0.83$, and tilt of the primordial power spectrum $n_s=0.96$.

The haloes we study were identified with the \texttt{SUBFIND} algorithm~\cite{Springel:2000qu} and are very massive dynamically relaxed\footnote{We used the relaxation criteria of~\cite{Ludlow:2013bd}, see Section 3.1 of~\cite{Brinckmann:2017uve} for details.} cluster-sized haloes 
in the mass range M$_{200} \approx 0.5-1.9 \times 10^{15}$~M$_{\odot}$~$h^{-1}$ and radius R$_{200} \approx 1300-2000$~kpc~$h^{-1}$, with a peak in the distribution at around M$_{200} \sim 0.9 \times 10^{15}$~M$_{\odot}$~$h^{-1}$ and R$_{200} \sim 1550$~kpc~$h^{-1}$ (see Fig. A1 of ~\cite{Brinckmann:2017uve}), where R$_{200}$ denotes the radius within which the average density is 200 times the critical density of the universe today, $\rho_c=3H^2_0/(8\pi G)$, and M$_{200}$ the enclosed mass within this radius.

\subsection{Properties of SIDM haloes}

In a $\Lambda$CDM cosmology, DM haloes are expected to have a cuspy density profile that
is well described by the universal form suggested by Navarro, Frenk and White (NFW) 
\cite{Navarro:1995iw,Navarro:1996gj}
\be
    \rho_{\text{NFW}}(r) = 
    \frac{\rho_s}{(r/r_s) (1 + (r/r_s))^2}\,,
    \label{eq:NFW}
\ee
where $\rho_s$ and $r_s$ are referred to as the scale density and radius, respectively.
An alternative pair of parameters  to describe such profiles is given by the virial mass $M_{200}$ 
and the halo concentration
\begin{equation}
  c\equiv\frac{R_{200}}{r_{-2}} \ .
   \label{eq:concentration}
\end{equation}
where $r_{-2}$ is the radius where the logarithmic slope of the density profiles equals $-2$ (i.e.~$r_{-2}=r_s$ for 
the NFW profile). As it turns out, these two parameters (concentration and mass) are not independent of each other but strongly correlated
(which, for example, explains the observed scaling of surface densities \cite{Bondarenko:2017rfu}). 
Our simulated CDM haloes follow exactly these general expectations.

\begin{figure}[t!]
  \centering
  \includegraphics[width=0.48\textwidth]{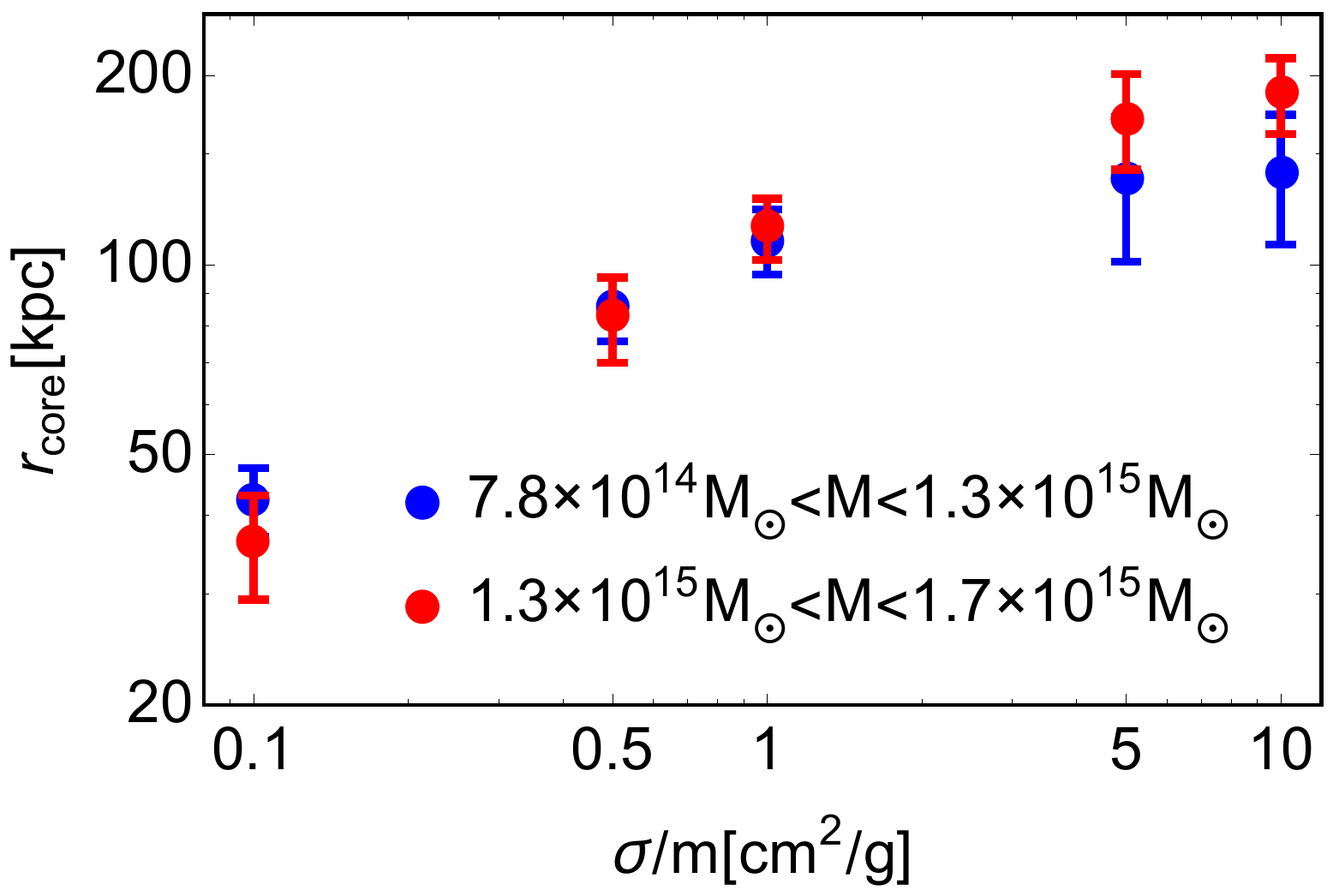}~ \includegraphics[width=0.48\textwidth]{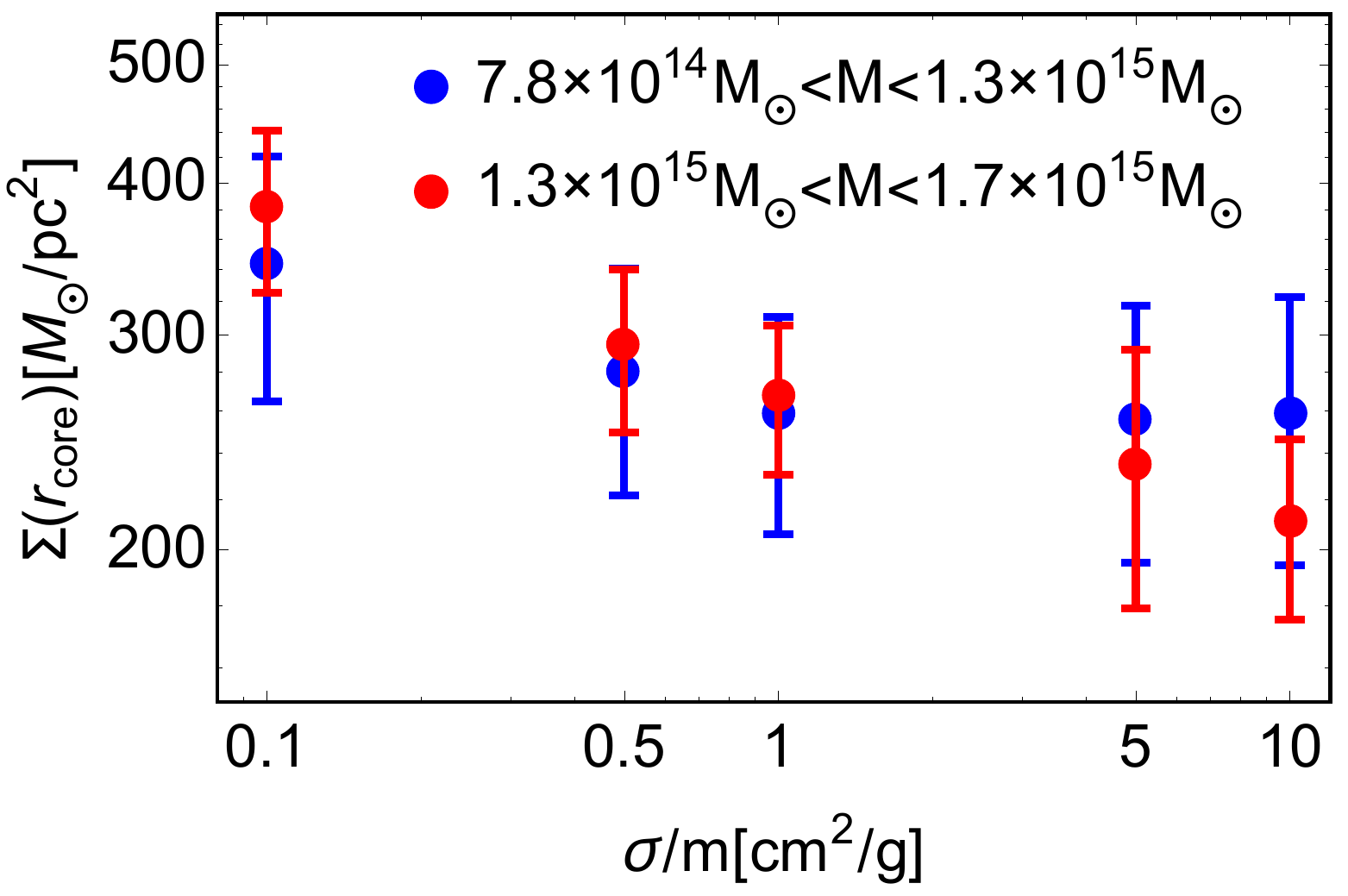}
  \caption{The dependence of $r_{\textrm{core}}$ (left) and the surface density (right) in SIDM haloes on the self-interaction strength. The data is presented for 2 mass bins with approximately equal scatters of masses. Error bars represent the statistical 
  spread in our suite of simulated haloes and correspond to one standard  deviation.
 }
  \label{fig:rcoreofsigma}
\end{figure}

If DM is self-interacting, on the other hand, the inner part of a DM halo should develop a core of constant density instead of a cuspy profile, while the outer part of the halo should be unaffected and hence follow the standard NFW profile \cite{Spergel:1999mh}. We confirm this general expectation in all our simulated SIDM haloes.
As shown in the left panel of Fig.~\ref{fig:rcoreofsigma} we also confirm the general expectation that the core radius should
grow with the interaction strength. Here and in the following we use the following definition of the core radius  
$r_{\text{core}}$:
\begin{equation}
    \rho_{\text{SIDM}}(0) = \rho_{\text{CDM}} (r_{\text{core}})\,.
    \label{eq:coredef}
\end{equation}
The core radius is thus defined as the radius at which the central DM density (from our SIDM simulations or observations)
equals the DM density in the CDM case, parameterized here as NFW profile. 
The advantage of this definition is that it is in principle observable and independent of the functional
form that is used to parameterize the cored profile. A more direct observable quantity that can serve to distinguish
between CDM and SIDM haloes is the surface density $\Sigma$ at this radius \cite{Bondarenko:2017rfu}. In the right panel of Fig.~\ref{fig:rcoreofsigma} we show this quantity as a function of the interaction strength. In Fig.~\ref{fig:rcoreofsigma} we have grouped both $r_{\text{core}}(\sigma/m)$ and $\Sigma(\sigma/m)$ into two mass bins. We see that both quantities saturate around $\sigma/m\sim5\text{ cm}^2/\text{g}$, the maximal $r_{\text{core}}$ being larger for haloes with larger masses.

In the remainder of this article, we will critically assess, and improve, analytical models commonly used in the 
literature to predict the behaviour shown in Fig.~\ref{fig:rcoreofsigma} in order to constrain DM self-interactions
from observations.
While we defer a detailed discussion to later, we can already at this point draw two important 
conclusions directly from  inspection of this figure:

\begin{itemize}
\item The core radius shows a relatively weak dependence on the interaction strength. This implies that a small
error in estimating the former results in a significant error when deriving constraints on the latter.
\item It is fundamentally impossible to constrain cross sections larger than a given limiting value from observations of the core size. For the cluster-sized objects that we have simulated here, this applies to
interaction strengths of $\sigma/m\gtrsim 3\,\mathrm{cm}^2/\mathrm{g}$. This is because the core has a maximum size set roughly by the radius where the velocity dispersion peaks. Once the core size reaches this value, it becomes insensitive to larger cross sections.
\end{itemize}

Let us stress that these conclusions are directly based on simulation data, and hence
independent of the analytical model that is used to describe self-interactions.
In particular, the first point motivates the main goal of this article, which is to obtain a detailed modelling of the effect of DM self-interactions on halo profiles.

\section{Analytical model of SIDM haloes}
\label{sec:anal_model}

If DM particles interact with each other, the scattering probability 
decreases with the density and hence the distance from the halo centre, becoming negligible
at large enough distances. At these distances, SIDM particles essentially behave as collisionless particles, the same as in the CDM model~\cite{Moore:2000fp,Yoshida:2000uw,Yoshida:2000bx,Rocha:2012jg,Peter:2012jh,Vogelsberger:2012ku,Zavala:2012us,Vogelsberger:2014pda}. At the radii where collisions occur frequently we expect that {\it thermal} equilibrium has been established. To model these two regimes, we can define a characteristic radius, $r_\text{SIDM}$, which separates the two regions of interest (see~\cite{Tulin:2017ara} for a review): outside $r_\text{SIDM}$ collisions are insignificant and the collision integral in the Boltzmann equation can be neglected; inside $r_\text{SIDM}$ we assume that interactions are efficient enough to establish {\it thermal} equilibrium and the collision integral also vanishes. Therefore, in both regions, one can use the Jeans equation relating the 3-dimensional DM velocity dispersion $\sigma_{\bm{v}}(r)$ and the density profile $\rho(r)$:
\begin{equation}
    \frac{d}{d r}\left(
    \frac{\sigma_{\bm{v}}^2}{3}
    \frac{r^2}{\rho}\frac{d \rho}{d r}
    \right) = - 4 \pi G r^2\rho \, .
    \label{eq:Jeans}
\end{equation}

In SIDM haloes, the mean-free path $\lambda$ between collisions is expected to be quite large, much larger than the radius $r_\text{SIDM}$:
\begin{equation}
    \lambda \equiv \frac{1}{(\sigma/m) \rho}
\simeq 4.8 \text{ kpc } \left( \frac{1 \text{ cm}^2/\text{g}}{\sigma/m} \right)
\left( \frac{1 M_{\odot}/\text{pc}^3}{\rho} \right) \, .
\label{eq:meanfreepath}
\end{equation}
This implies that if a kinetic equilibrium can be established within $r_\text{SIDM}$, this can only be a global equilibrium, with the same velocity dispersion for all $r< r_\text{SIDM}$.
Nevertheless, as we see from the simulations, a few collisions per particle in a Hubble time are sufficient to redistribute energy resulting in an isothermal core (constant velocity dispersion) within $r_\text{SIDM}$. With this condition, the 
Jeans equation becomes
\begin{equation}
     \frac{\bar{\sigma}_{\bm{v}}^2}{3} \frac{d}{d r}\left(
    \frac{r^2}{\rho}\frac{d \rho}{d r}
    \right) = - 4 \pi G r^2\rho \, ,
\label{eq:Jeans_2}
\end{equation}
where the constant $\bar{\sigma}_{\bm{v}}$ describes the average value of the velocity dispersion $\sigma_{\bm{v}}$ inside $r_\text{SIDM}$.
This equation has solutions with different asymptotic behaviour at the centre. We anticipate that we do not consider the unphysical solutions where the density goes to zero. We note that the thermalization of the inner core in SIDM haloes is only a quasi-stable configuration. Given enough time, collisions eventually trigger a runaway instability of the core, analogous to the well-known gravothermal catastrophe in globular clusters \cite{LyndenBell1968}. The collapse of the core results in a central density profile that is even cuspier than in CDM haloes \cite{Koda2011,Pollack2015}. For this process to be relevant within a Hubble time, however, large cross sections $\gtrsim10~\text{cm}^2/\text{g}$ are required. Our model does not cover this regime since it is not relevant for the purposes of this work. We will be looking for solutions to the Jeans equation that have a constant density at the center, which is the quasi-stable configuration for relevant cross sections as shown by SIDM simulations in the past.
In this model, as we mentioned before, the collision integral is equal to zero on both sides of $r_\text{SIDM}$, for different reasons in each regime. In reality, however, there is an intermediate region where the collisions cannot be neglected, but they are still not frequent enough to establish thermodynamical equilibrium. In other words, the model implicitly assumes that the thickness of the intermediate region is much smaller than $r_{\text{SIDM}}$ and that it can be approximated by a thin spherical shell at the radius $r_{\text{SIDM}}$.  
In this simple, but often adopted model the central region $r<r_{\text{SIDM}}$ is then in thermodynamical equilibrium and the outer region $r>r_{\text{SIDM}}$, where DM particles are effectively collisionless, is connected to the inner region by some boundary conditions at $r_{\text{SIDM}}$. It is clear that in this approximation some quantities will be continuous at $r_{\text{SIDM}}$, but not necessarily all.

The solution to the Jeans equation~\eqref{eq:Jeans_2} with a core ($\rho'(0)=0$) depends on 2 parameters: $\bar{\sigma}_{\bm{v}}$ and $\rho_0\equiv\rho(0)$. Therefore, we need two boundary conditions to fix the SIDM profile within $r_{\text{SIDM}}$ that we choose by considering the following approximations, which will be verified below by direct comparison with simulation data. Let us assume that despite self-interactions, DM particles will not leave the radius $r_\text{SIDM}$, but will only be redistributed within it. This means that we can choose, as the first boundary condition, the requirement that the mass at $r_\text{SIDM}$ is the same in the SIDM halo as in the CDM halo:
\begin{equation}
    M_{\text{SIDM}}(r_\text{SIDM}) = M_{\text{CDM}}(r_\text{SIDM}) \, .
    \label{eq:boundarycondition1}
\end{equation}
As for the second boundary condition, we will assume that the kinetic energy defined as
\begin{equation}
    E_{\text{kin}}(r) = 2\pi \int\limits_0^{r} \rho_{\text{CDM}}(r) \sigma_{\bm{v}}^2(r) r^2 dr
    \label{eq:boundarycondition2}
\end{equation}
is equal inside $r_\text{SIDM}$ for CDM and SIDM
\begin{equation}
    E_{\text{kin}}^{\text{SIDM}}(r_\text{SIDM}) = E_{\text{kin}}^{\text{CDM}}(r_\text{SIDM}) \, .
\end{equation}
These two boundary conditions for the Jeans equation, together with the requirement of a constant density at the centre, allow one to fix the constant velocity dispersion $\bar{\sigma}_{\bm{v}}$ and find a unique solution for the DM density profile. We would like to emphasize that the ansatz where the isothermal profile~\eqref{eq:Jeans_2} inside the radius $r_{\text{SIDM}}$ is connected to a CDM profile at larger radii was already used previously~\cite{Kaplinghat:2013xca,Kaplinghat:2015aga,Tulin:2017ara}. However, as motivated by a direct comparison with our simluation data, we use different boundary conditions compared to earlier works (see also Section~\ref{sec:tulin}).

\subsection{Verifying the model assumptions with numerical simulations}
\label{sec:verification}
To verify the validity of the simple model formulated above, we need to explicitly check whether there exists a radius 
for the simulated haloes
inside which, to a certain precision, (i) the masses of CDM and SIDM haloes are equal to each other; (ii) the total kinetic energies of CDM and SIDM haloes are equal; (iii) the velocity dispersions for the SIDM haloes become flat. In this subsection, we will check these assumptions with simulated data for different cross-sections and demonstrate that such a radius exists. In the next subsection, we will check if the Jeans equation~\eqref{eq:Jeans_2} with our boundary conditions at 
this radius 
indeed describes the inner density profile correctly. 

\begin{figure}[t!]
  \centering
    \includegraphics[width=0.75\textwidth]{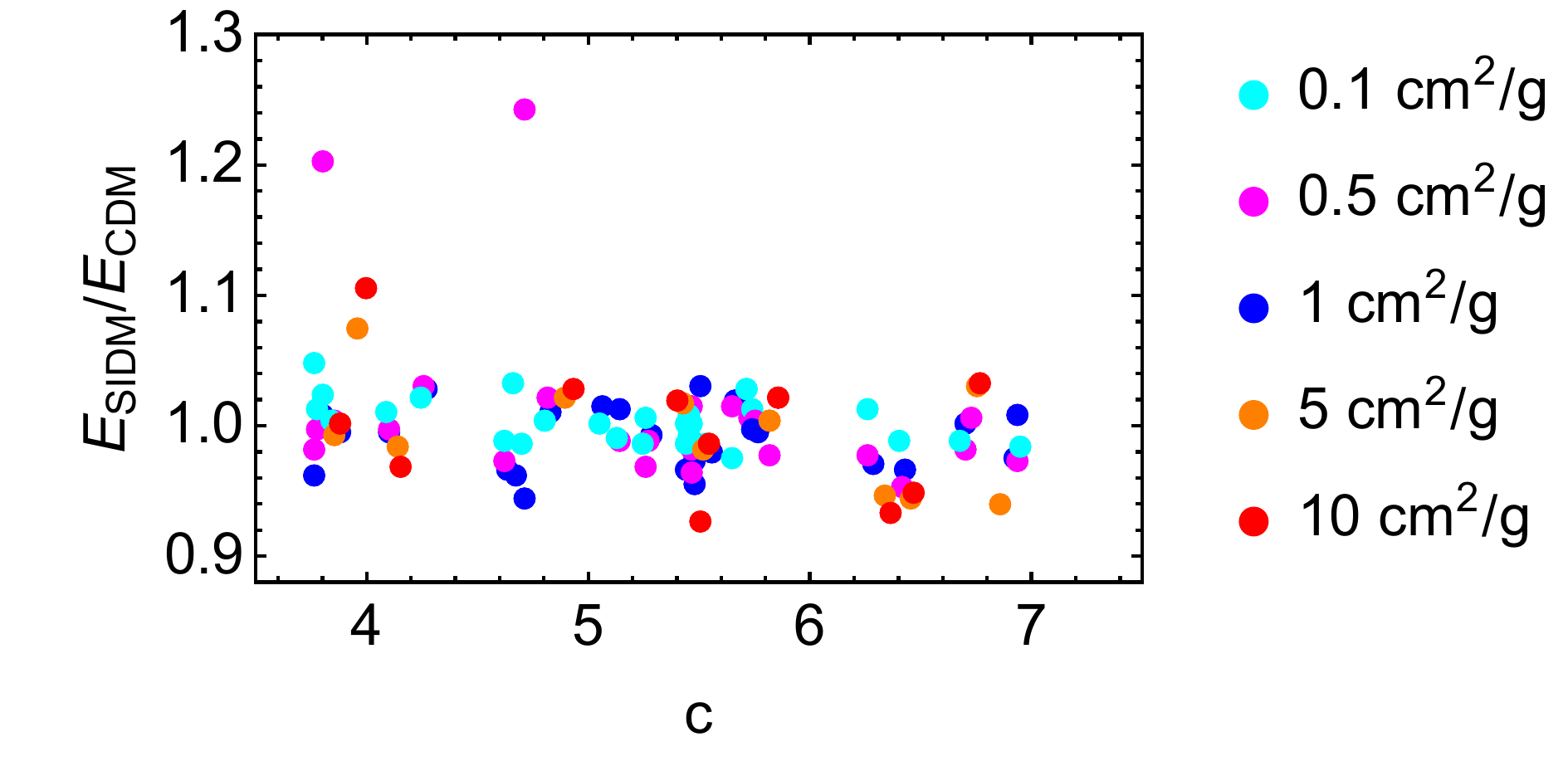}
    \caption{Ratio of kinetic energies of SIDM haloes to those of the corresponding CDM haloes at the radius of equal masses $r_M$ for different values of the self-interaction cross-section $\sigma/m$ as a function of halo concentration.}
    \label{fig:energy_conservation}
\end{figure}
 
We start by defining $r_M$ as the radius where, for a given halo, the masses in SIDM and CDM are equal (see Appendix~\ref{sec:rM} for examples of $r_M$ for different cross-sections),
and check the hypothesis of equal kinetic energies at this radius. The ratio between kinetic energies in SIDM and CDM simulations is shown in Fig.~\ref{fig:energy_conservation}, as a function of the halo concentration as defined in Eq.~(\ref{eq:concentration}).
One can see that the kinetic energies of SIDM and CDM profiles inside radius $r_M$ agree with an accuracy of $\lesssim 5\%$ for most of the haloes.

\begin{figure}[t!]
  \centering
    \includegraphics[width=0.48\textwidth]{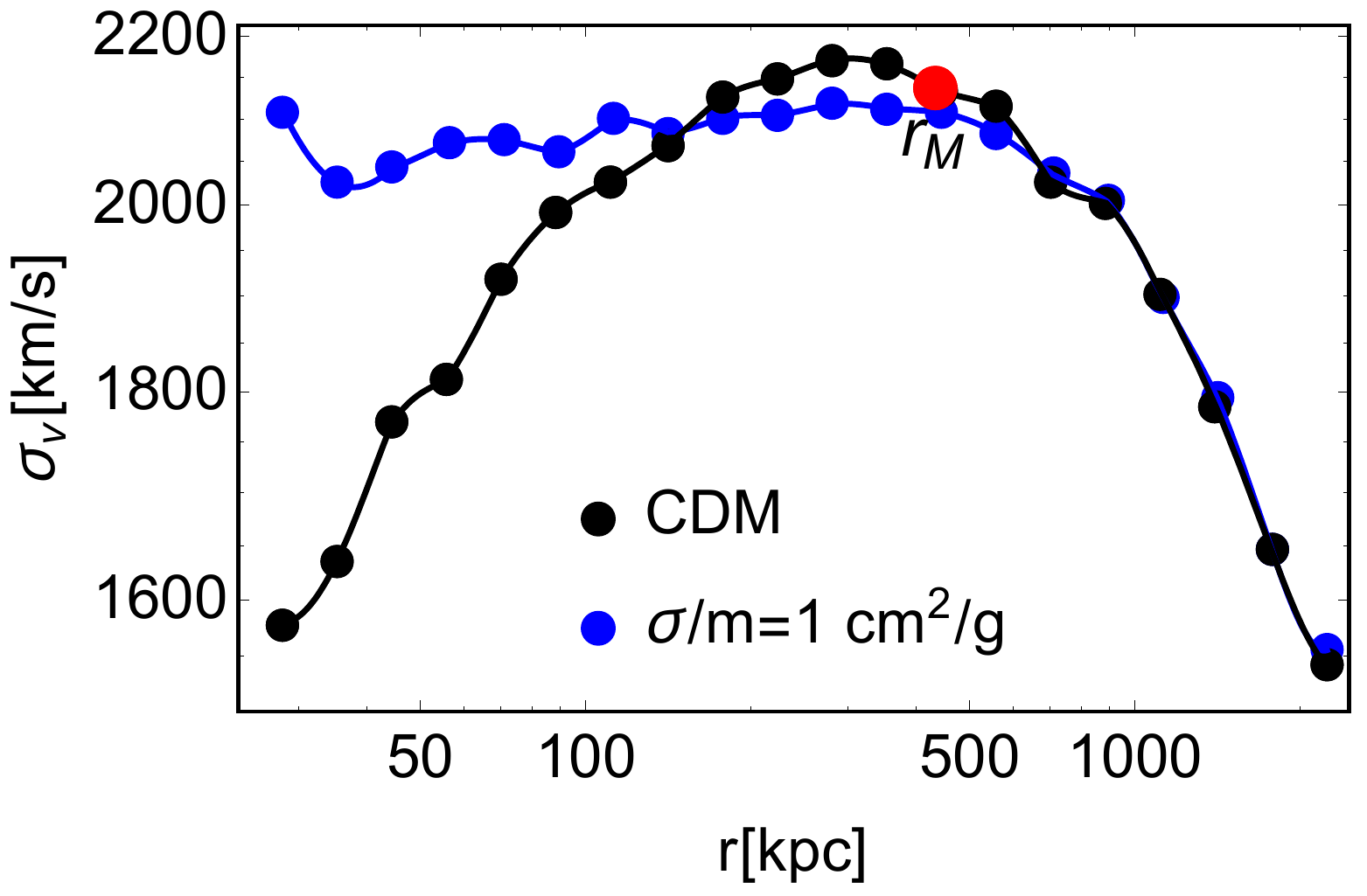}~\includegraphics[width=0.49\textwidth]{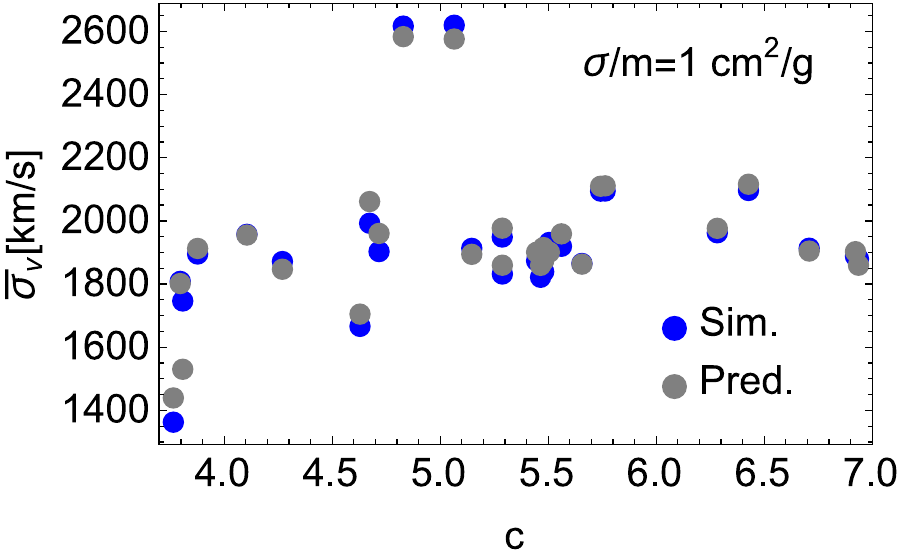}
    \caption{\textit{Left panel:} Total velocity dispersion profile for an example of a halo in our simulation suite: CDM (black) and SIDM (blue) with $\sigma/m = 1$ cm$^2/$g. The red point represents the radius where the enclosed mass is equal in both cases. \textit{Right panel:} 
    The constant central velocity dispersion $\bar{\sigma}_{\bm{v}}$ refers to the average value of the velocity dispersion inside $r_M$ (blue) and the predicted value (gray) obtained from the assumption of equal mass and kinetic energy at $r_M$, see Eq.~\eqref{eq:sigma2kin}. The SIDM case with $\sigma/m = 1$ cm$^2/$g is shown as a function of halo concentration.}
    \label{fig:sigmatot}
\end{figure}

The assumed boundary conditions on equal kinetic energies and masses result in the following average velocity dispersion $\bar{\sigma}_{\bm{v}}$ in SIDM haloes:
\begin{equation}
    (\bar{\sigma}_{\bm{v}}^{\text{pred}})^2 = \frac{2 E_{\text{kin}}^{\text{CDM}}(r_M)}{M_{\text{CDM}}(r_M)} \ .
   \label{eq:sigma2kin}
\end{equation}
The predicted value of the velocity dispersion is compared with the simulation data for $\sigma/m = 1$ cm$^2/$g in the right panel of Fig.~\ref{fig:sigmatot} where we can see that the agreement is quite good for most haloes.
Another assumption we would like to check is the flatness of the velocity dispersion inside $r_M$ (see Fig.~\ref{fig:sigmatot}, left panel).
\begin{figure}[t!]
  \centering
    \includegraphics[width=0.75\textwidth]{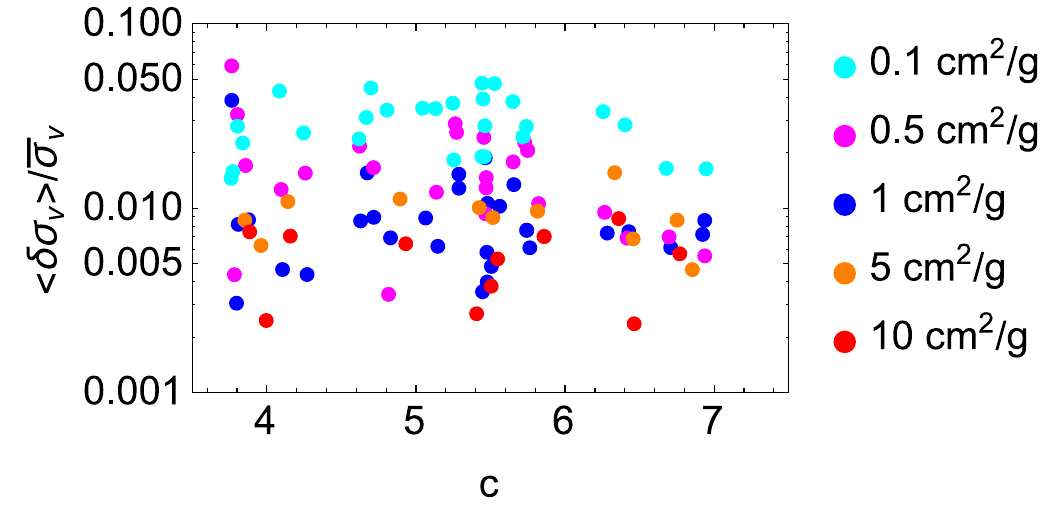}
    \caption{The average deviation of the total velocity dispersion in  SIDM simulations from the best fit constant value of $\bar{\sigma}_{\bm{v}}$ inside $r_M$ for different values of the self-interaction cross-section $\sigma/m$ as a function of halo concentration.}
    \label{fig:deltasigma}
\end{figure}
Fig.~\ref{fig:deltasigma} shows the deviation from the best fit constant value of $\bar{\sigma}_{\bm{v}}$, averaged over all radii within $r_M$, for a given halo as a function of its concentration.
We can see that for most of the haloes this deviation is 
$\lesssim1 \%$ for $\sigma/m\ge 1$ cm$^2/$g and $\lesssim5 \%$ for smaller cross-sections.

We can thus conclude that $r_M$  is indeed a good proxy for 
$r_{\text{SIDM}}$ and our boundary condition (kinetic energy conservation) is also a reasonable assumption that provides a good prediction for the constant velocity dispersion in SIDM haloes, using CDM data only as input.
This means that we have all elements required to predict the DM density profile as a solution of the Jeans equation with constant velocity dispersion $\bar{\sigma}_{\bm{v}}$. This prediction will provide a validity check of our model.

\subsection{Verifying the model prediction for the density profile}

\begin{figure}[t!]
  \centering
    \includegraphics[width=0.545\textwidth]{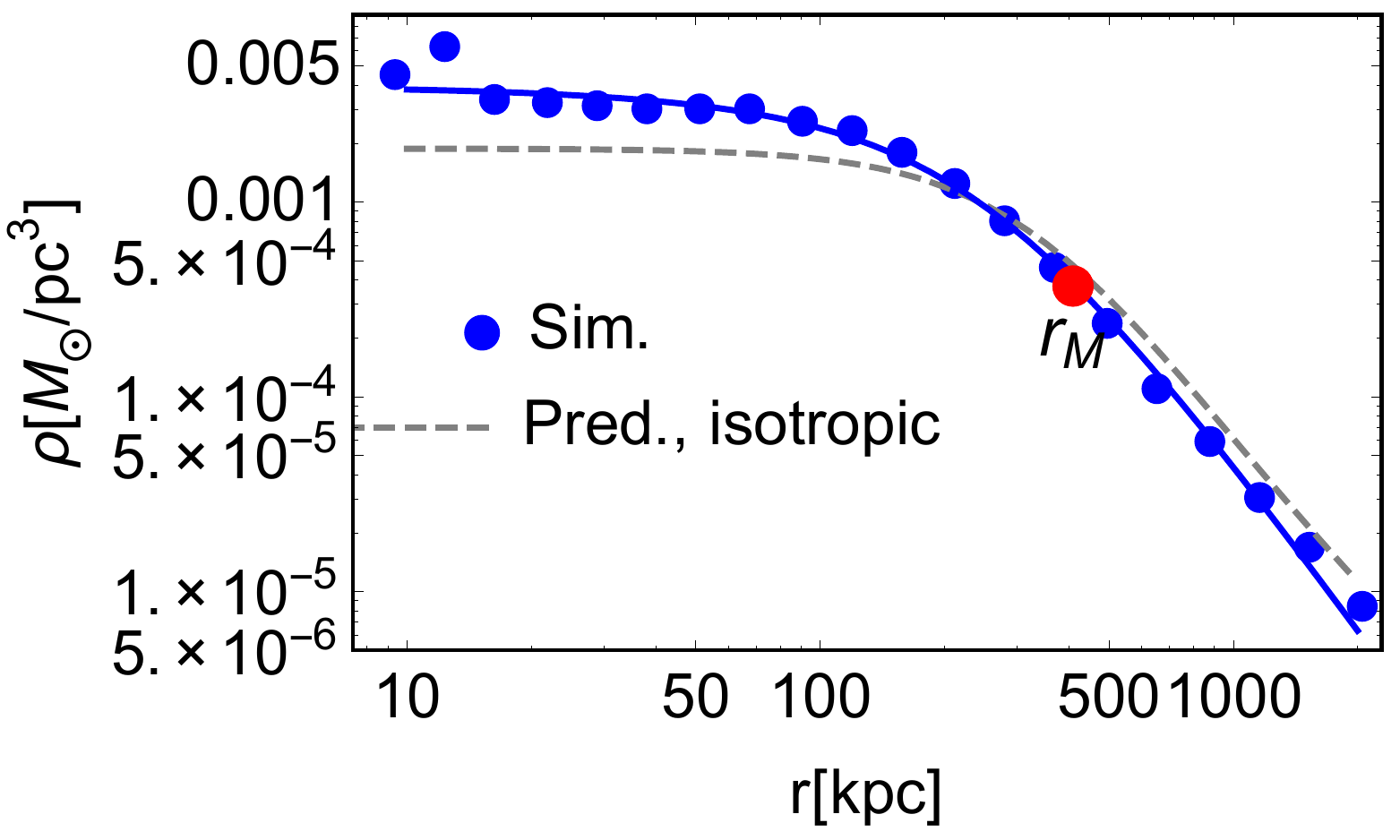}~\includegraphics[width=0.48\textwidth]{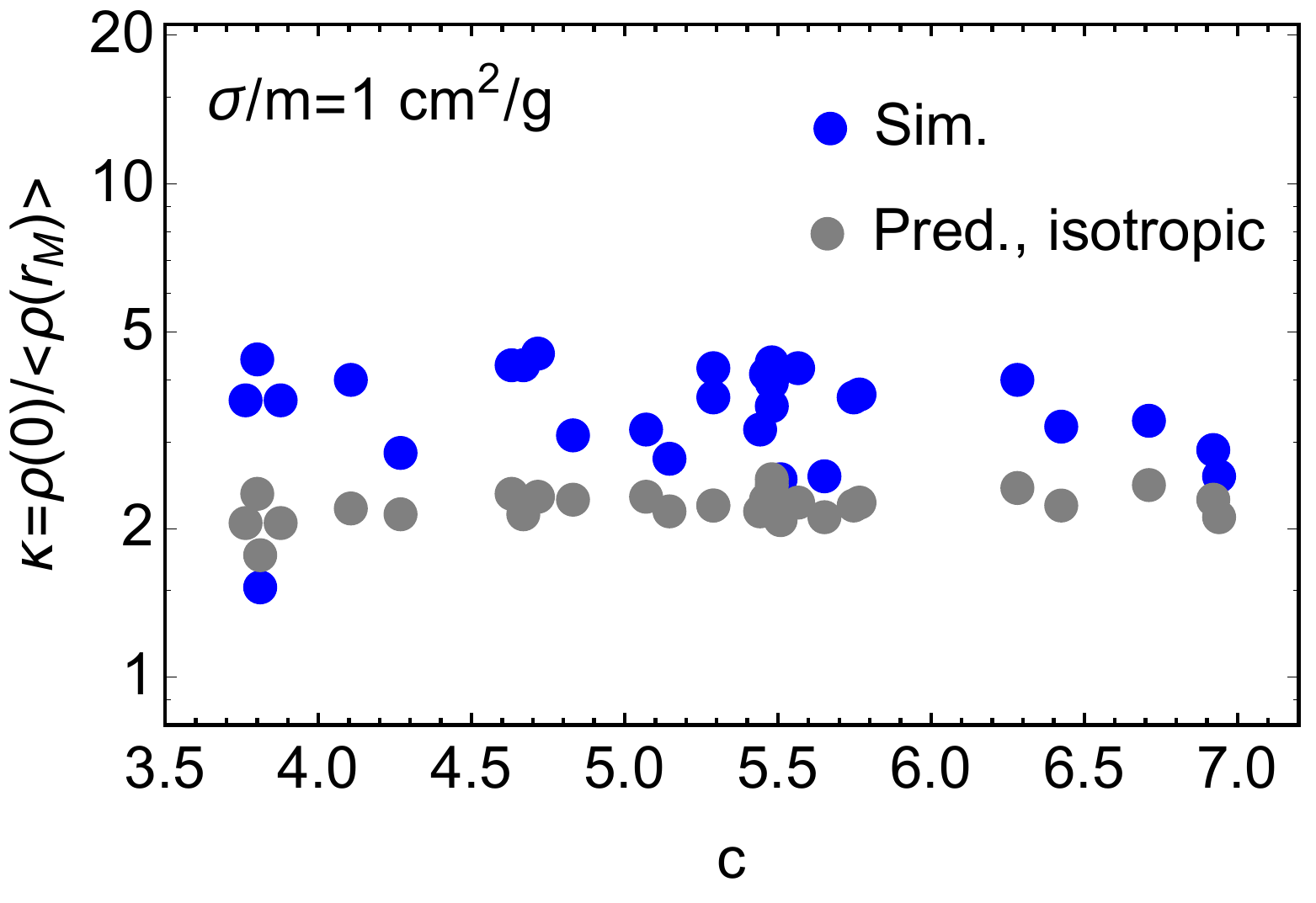}
    \caption{\textit{Left panel:} Density profiles for a SIDM halo with $\sigma/m = 1$ cm$^2/$g from simulations (blue) and the prediction from our isotropic (dashed gray line).  
    \textit{Right panel:} Ratio between central density and enclosed density at $r_M$ for the ensemble of SIDM haloes with $\sigma/m = 1$ cm$^2/$g as a function of halo concentration.}
    \label{fig:jeans_solution_iso}
\end{figure}

\begin{figure}[t!]
  \centering
    \includegraphics[width=0.75\textwidth]{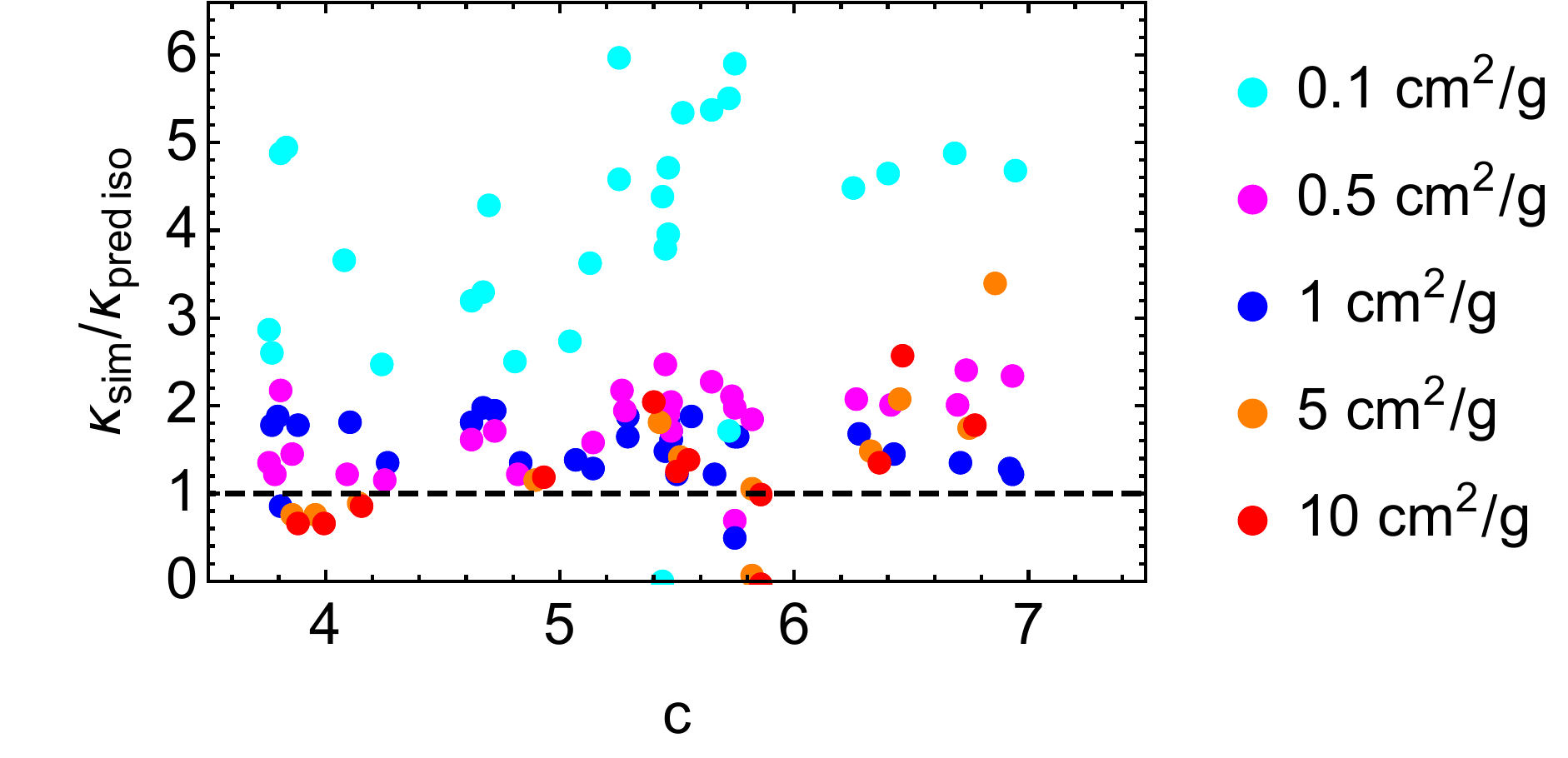}
    \caption{{The ratio of $\kappa=\rho(0)/\langle\rho(r_M)\rangle$ between simulated and predicted values for our isotropic model versus halo concentration in SIDM haloes for different values of the self-interaction cross-section $\sigma/m$.}}
    \label{fig:kappaalliso}
\end{figure}

An example of a predicted SIDM density profile for $\sigma/m = 1$ cm$^2/$g is given in the left panel of Fig.~\ref{fig:jeans_solution_iso}. It is clear that
the predicted central density is lower than the simulation data. The right panel of the same figure shows that 
the predicted density for $\sigma/m=1$ cm$^2/$g is systematically too low for all haloes (see Eq.~\eqref{eq:kappa_def} for the definition of $\kappa$). On average, $\langle\kappa\rangle_{\text{sim}}/\langle\kappa\rangle_{\text{pred}} = 1.6$. A similar behaviour is observed for other cross-sections as well (see Fig.~\ref{fig:kappaalliso}).
The reason for this systematic discrepancy,
as detailed in the next section, 
is the fact that the velocity dispersion still remains anisotropic in SIDM haloes.
\begin{figure}[t!]
  \centering
    \includegraphics[width=0.48\textwidth]{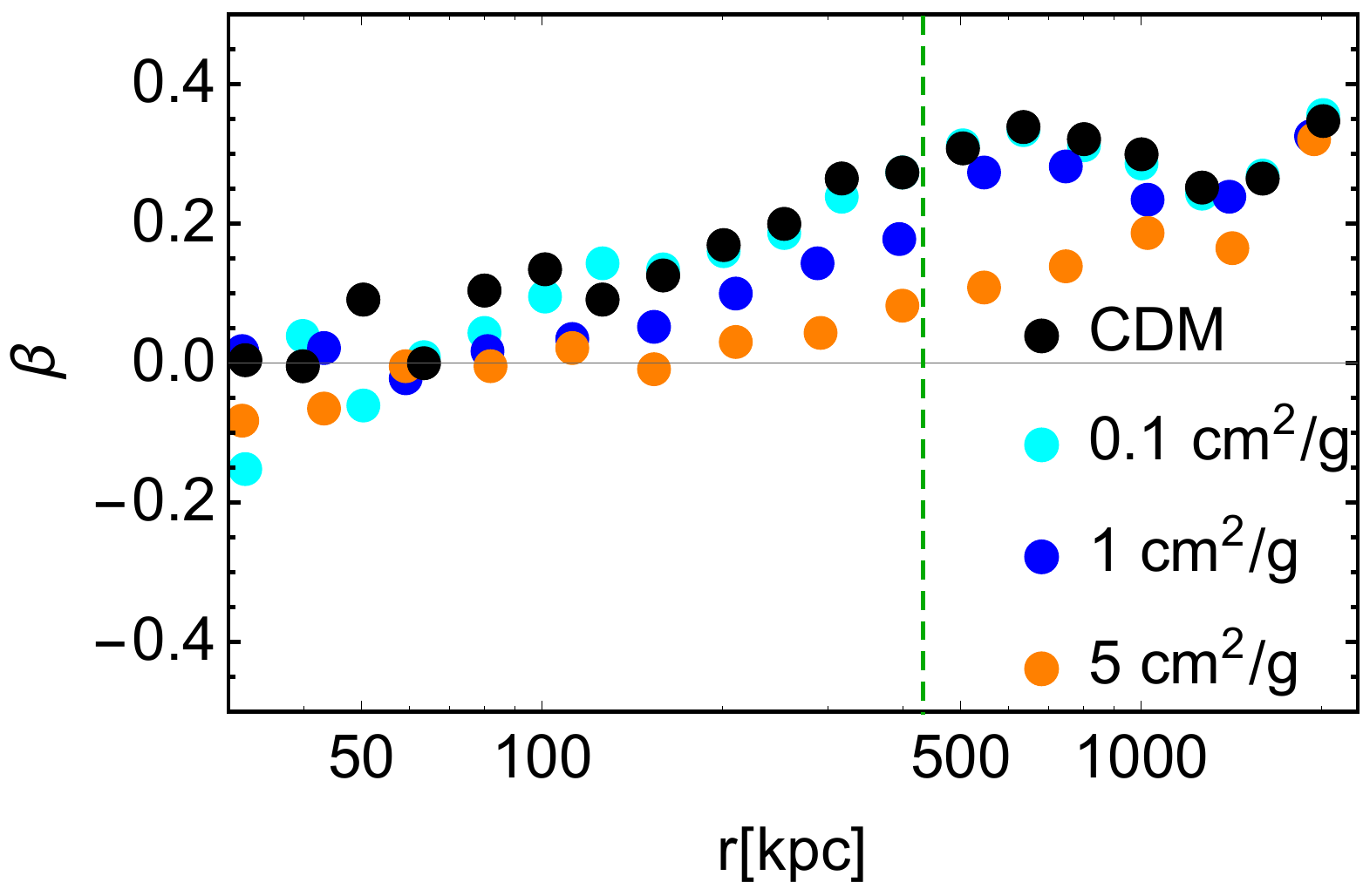}~\includegraphics[width=0.48\textwidth]{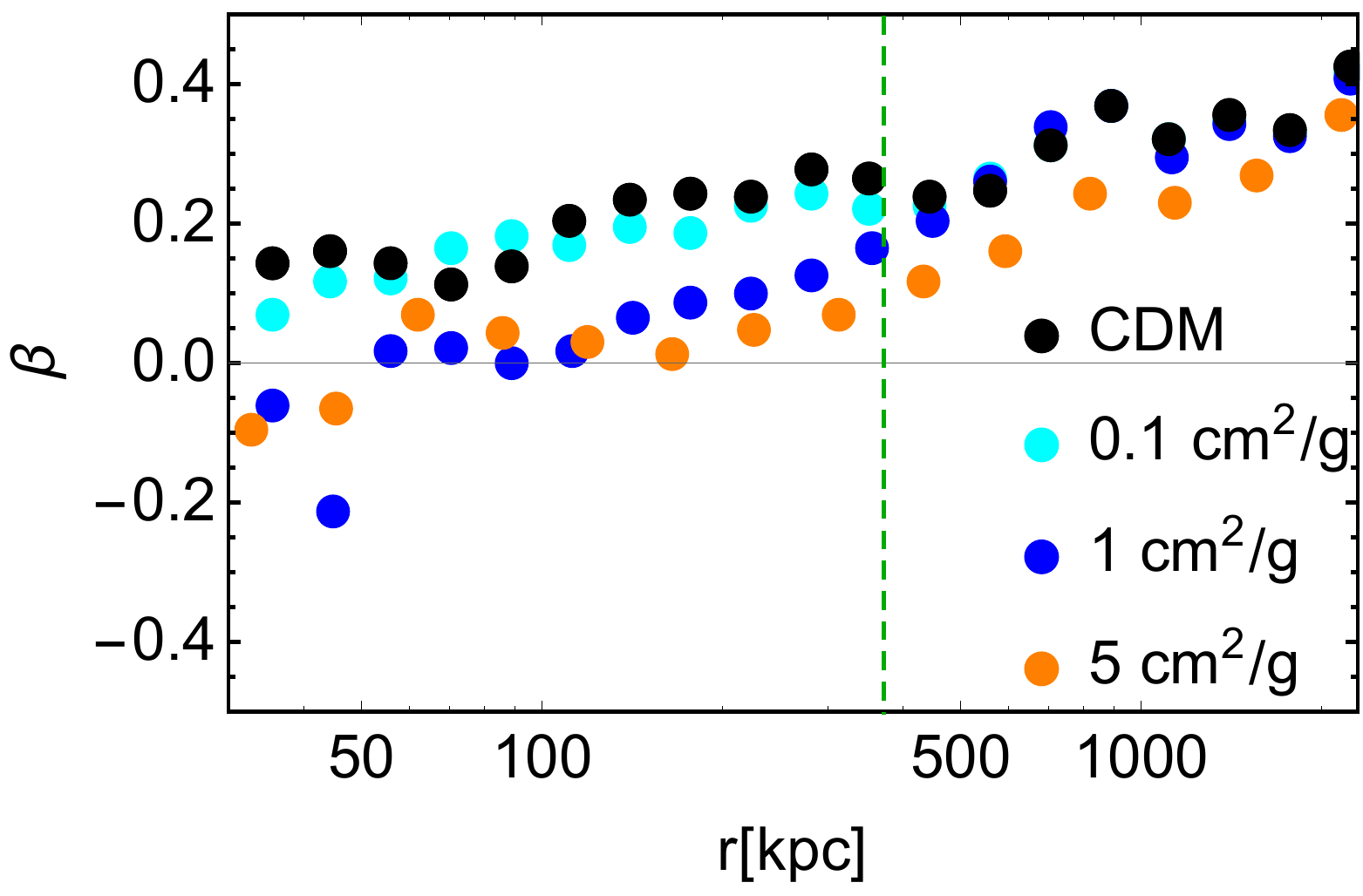}
    \caption{Velocity anisotropy profiles $\beta(r) = 1 - (\sigma_{\theta}^2 + \sigma_{\phi}^2)/(2\sigma_{r}^2)$ from simulations for CDM (black) and SIDM (cyan, blue and orange for $\sigma/m=0.1$, 1 and 5 cm$^2/$g, respectively) for two different example haloes. The vertical green line marks the $r_M$ radius.}
    \label{fig:beta}
\end{figure}

\subsection{Anisotropy of the velocity dispersion}

A perfect equilibrium would imply that all components of the velocity dispersion are isotropic.  Therefore the anisotropy of the velocity dispersion
\begin{equation}
    \beta(r) = 1 - \dfrac{\sigma_{\theta}^2 + \sigma_{\phi}^2}{2\sigma_{r}^2} \
\end{equation}
should be equal to zero, where $\sigma_{\theta,\phi}$ are the velocity dispersions in the tangential directions, while $\sigma_{r}$ is the radial velocity dispersion. However, in the simulations, we observe that the anisotropy $\beta$ for SIDM haloes does not vanish inside the radius $r_M$, where full equilibrium has been assumed. This is shown in Fig.~\ref{fig:beta}.
In fact, the anisotropy remains comparable to that of CDM haloes 
and does not drop to zero fast enough inside $r_M$ to be neglected, which means that thermal equilibrium is not fully established; at least not for $\sigma/m\lesssim1$~cm$^2/$g.

\begin{figure}[t!]
  \centering
    \includegraphics[width=0.58\textwidth]{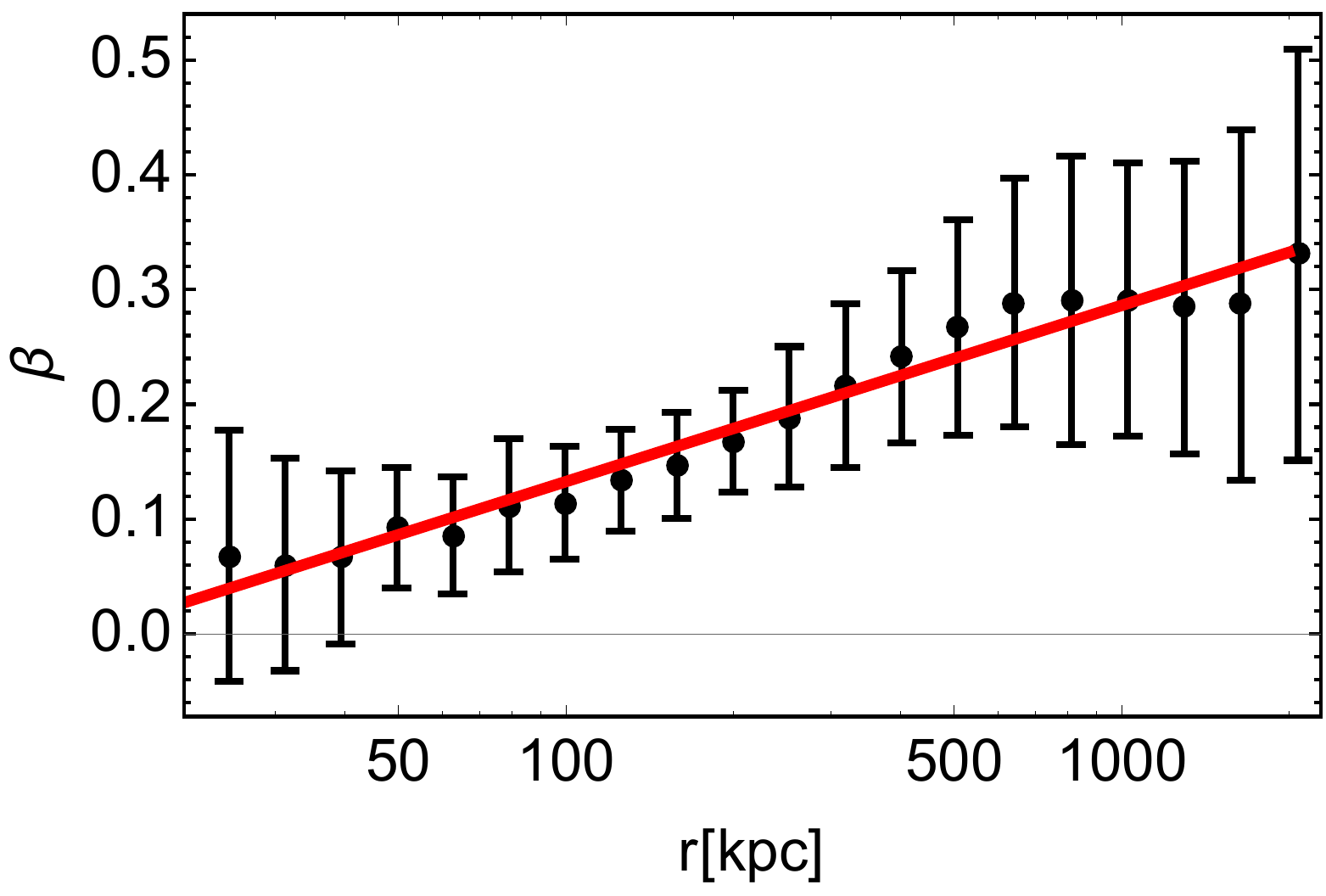}
    \caption{Velocity anisotropy versus radius for the  CDM haloes. Black dots are mean values for the given radius bin and error bars represent the standard deviation. The red line is the best fit with ansatz~\eqref{eq:betaansatz} (see Table~\ref{tab:beta} for values of the fit parameters).}
    \label{fig:betaCDM}
\end{figure}

The velocity anisotropy of DM particles is of course not directly observable. To make our model less dependent on simulation input,
we would like to come up with a prescription which can be applied not only to simulated data, where we know all quantities but to observational data in the future. We have found that a simple two-parametric ansatz
\begin{equation}
    \beta(r) = \left\{ \begin{matrix}
    A \ln (r/r_{\beta}),&\text{ for }r\ge r_{\beta} \\
    0,&\text{ for }r< r_{\beta}
    \end{matrix}
    \right.
    \label{eq:betaansatz}
\end{equation}
describes $\beta(r)$ for a given cross-section reasonably well, for both the SIDM and CDM cases. The best fit values of the parameters $A$ and $r_{\beta}$ are presented in Table~\ref{tab:beta}, see also Appendix~\ref{sec:betafit} for more details. The quality of the fit for the ensemble of CDM haloes is shown in Fig.~\ref{fig:betaCDM}.

\begin{table}[t!]
\centering
\begin{tabular}{ |c|c|c|  }
 \hline
 \multicolumn{3}{|c|}{Best fit values of $\beta=A \ln (r/r_{\beta})$} \\
 \hline
 $\sigma/m$ $[\text{cm}^2/\text{g}]$ & $A$ & $r_{\beta}$[kpc]\\
 \hline
 CDM   & 0.067 & 13.7 \\
 0.1   & 0.074 & 21.8 \\
 0.5   & 0.089 & 50.1 \\
 1     & 0.102 & 76.4 \\
 5     & 0.124 & 194 \\
 10    & 0.166 & 387 \\
 \hline
\end{tabular}
\caption{The best fit values for the anisotropy profile $\beta(r)$ (Eq.~\ref{eq:betaansatz}) from simulated haloes in CDM and SIDM for different cross sections.}
\label{tab:beta}
\end{table}

We will use the above ansatz for $\beta(r)$ to improve our model and predict the density for SIDM using the anisotropic Jeans equation in the following.

\subsection{Prediction for the density profile with anisotropic Jeans equation}

\begin{figure}[t!]
  \centering
    \includegraphics[width=0.51\textwidth]{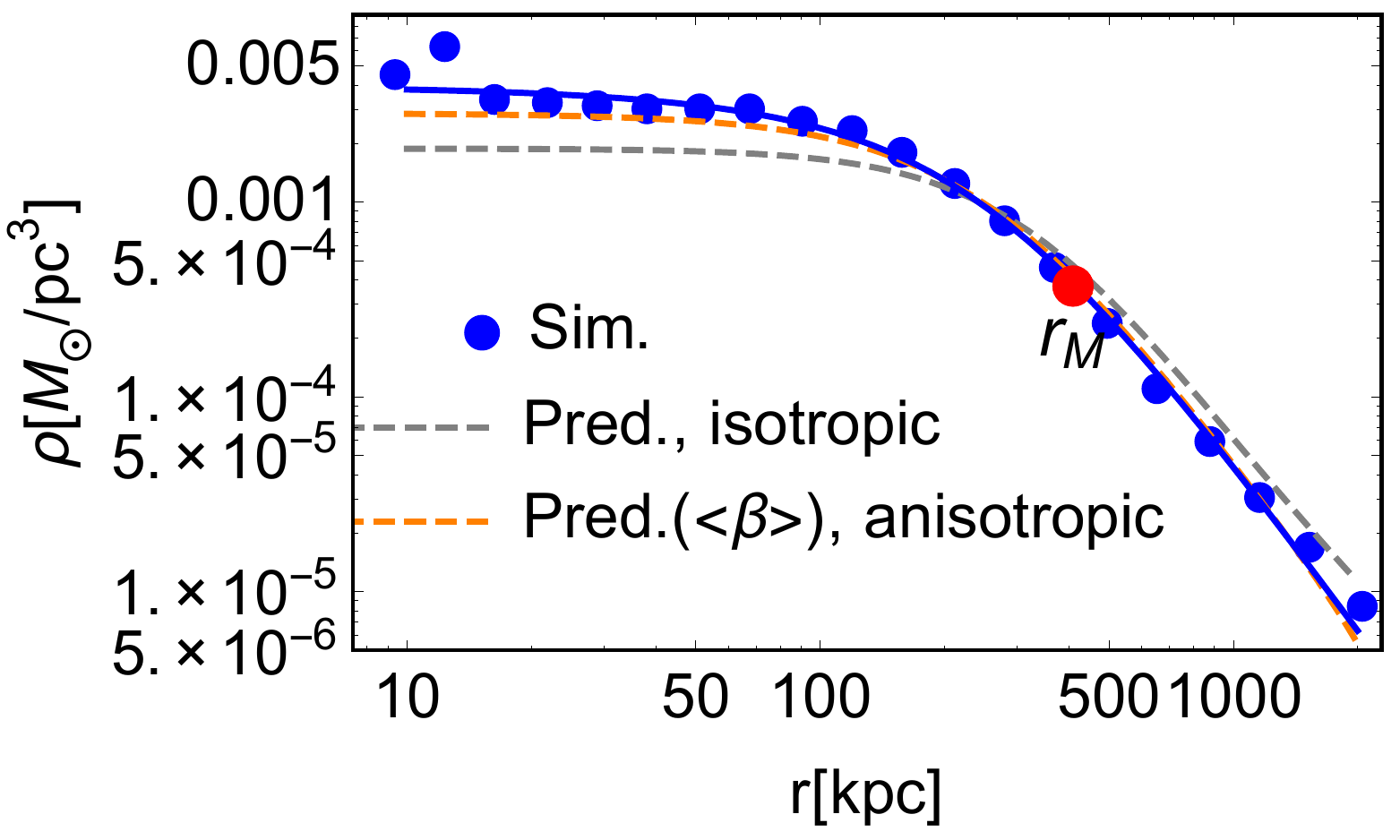}~\includegraphics[width=0.45\textwidth]{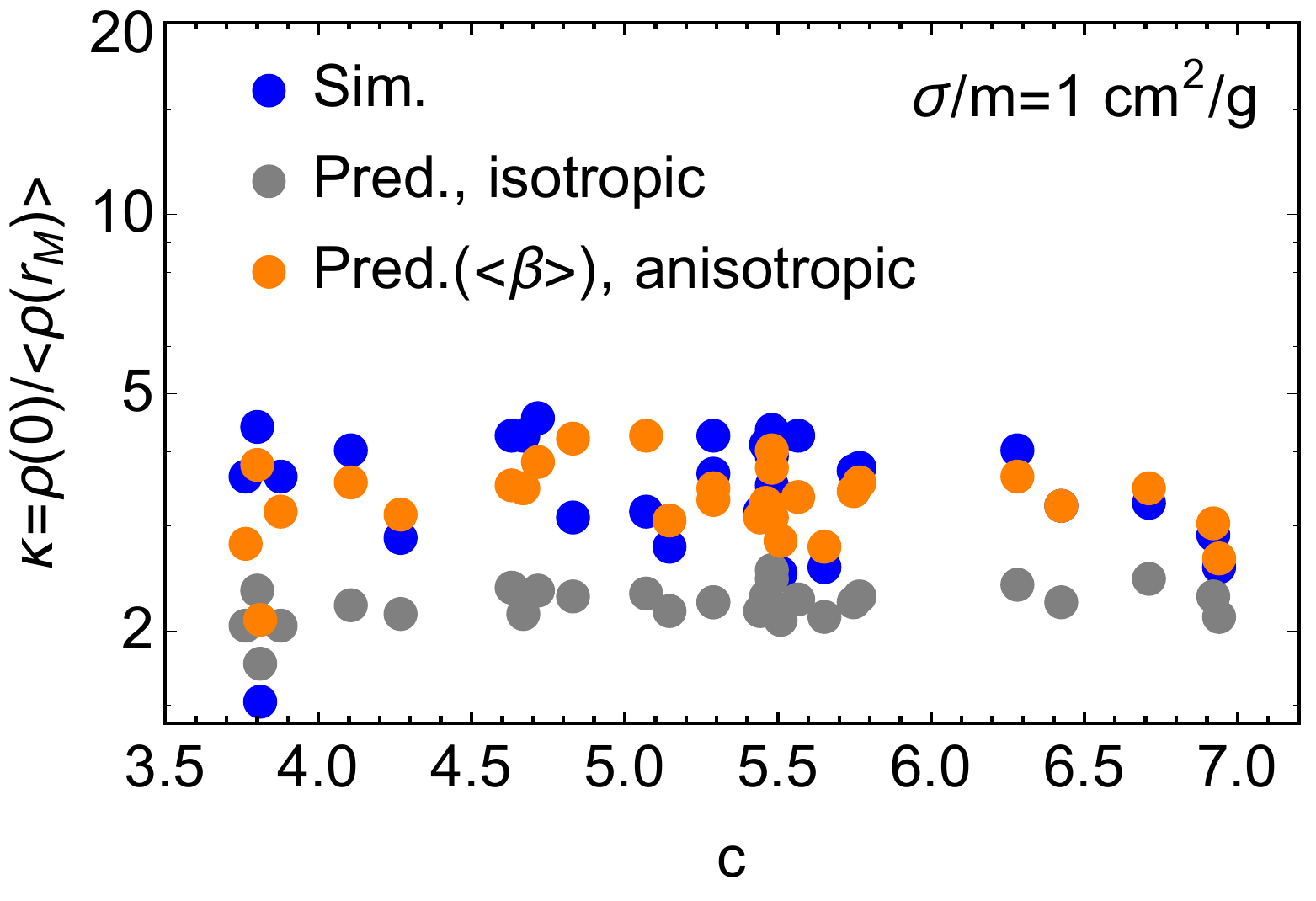}
    \caption{\textit{Left panel:} Density profiles versus radius for a SIDM halo with $\sigma/m=1$ cm$^2/$g from simulations (blue) and the prediction for the SIDM profile from our isotropic (dashed gray line) and anisotropic models (orange line). The red dot indicates the $r_M$ radius. \textit{Right panel:} $\kappa=\rho(0)/\langle\rho(r_M)\rangle$ as a function of halo concentration calculated from the simulation data for $\sigma/m=1$ cm$^2/$g (blue), and our isotropic/anisotropic model (gray/orange).}
    \label{fig:jeans_solution_aniso}
\end{figure}

Although there is no equilibrium inside $r_M$, we can still use the Jeans equation if we take into account the velocity anisotropy $\beta(r)$. The anisotropic Jeans equation for the radial velocity dispersion~$\sigma_r$ is~\cite{1987gady.book.....B}
\begin{equation}
    \frac{d}{d r}\left(\frac{r^2}{\rho}\frac{d}{d r} \left(\rho \sigma_{r}^2\right) + 2 r \beta \sigma_{r}^2 \right) 
    = - 4 \pi G r^2 \rho \, ,
    \label{eq:jeans_aniso}
\end{equation}
where  the radial velocity dispersion $\sigma_r$ is connected to the total velocity dispersion $\sigma_{\bm{v}}$ as
\begin{equation}
    \sigma_{\bm{v}}^2 \equiv \sigma_r^2 + \sigma_{\theta}^2 + \sigma_{\phi}^2 = \sigma_r^2(3-2\beta) \, .
\end{equation}
In Eq.~\ref{eq:jeans_aniso} we can still use the assumption that the total velocity dispersion is a constant as this is consistent with the simulated data.
With the addition of velocity anisotropy into the Jeans equation, we significantly improve the accuracy of our model (see Fig.~\ref{fig:jeans_solution_aniso}). Moreover, the predictions for the SIDM density profiles using our ansatz for $\beta(r)$ now becomes very similar to using the actual velocity anisotropy directly from each simulated halo. 

\begin{figure}[t!]
  \centering
    \hspace{-0.7cm}\includegraphics[width=0.61\textwidth]{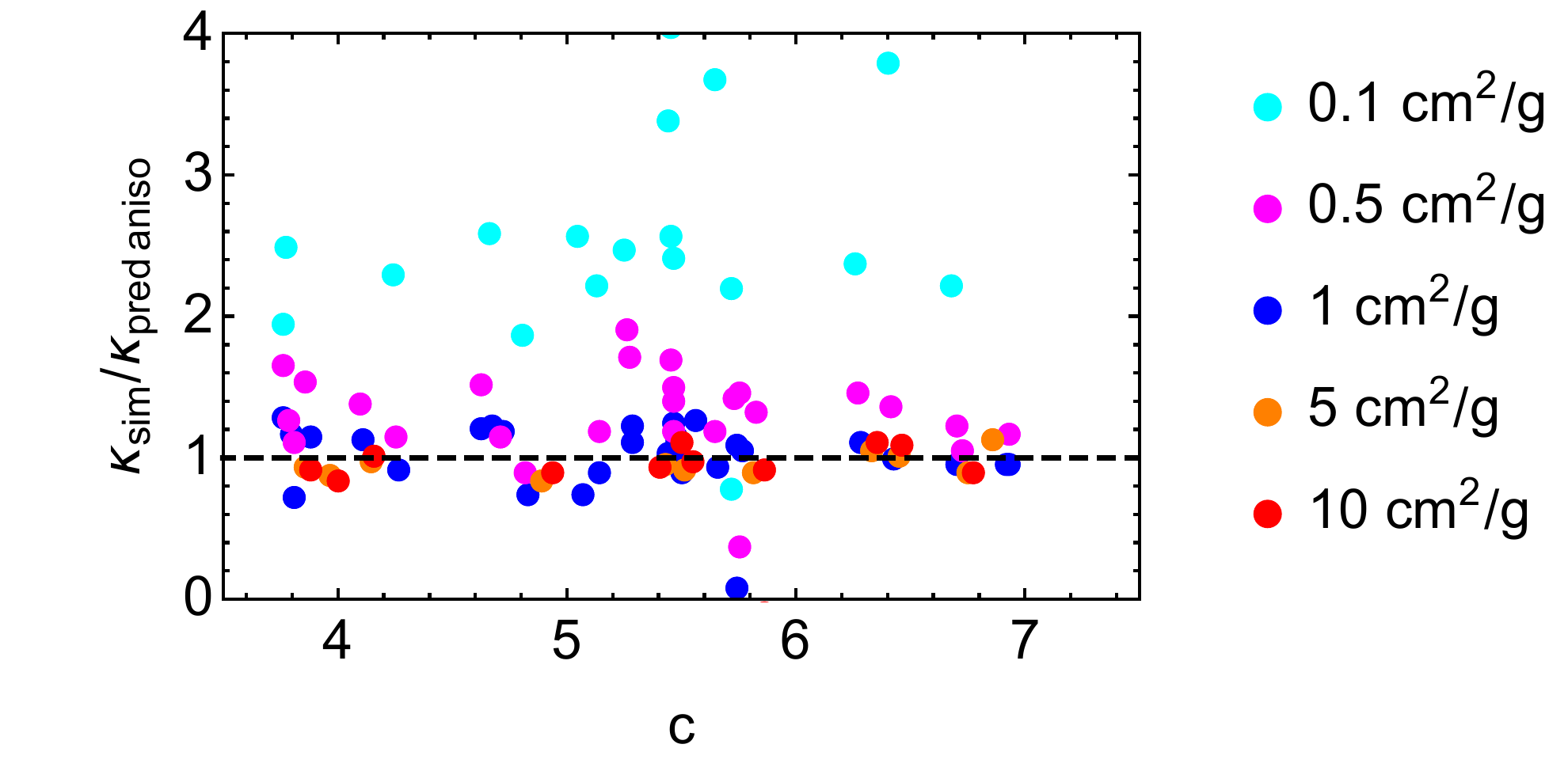}~\includegraphics[trim={0 -0.21cm 0 0},clip,width=0.4\textwidth]{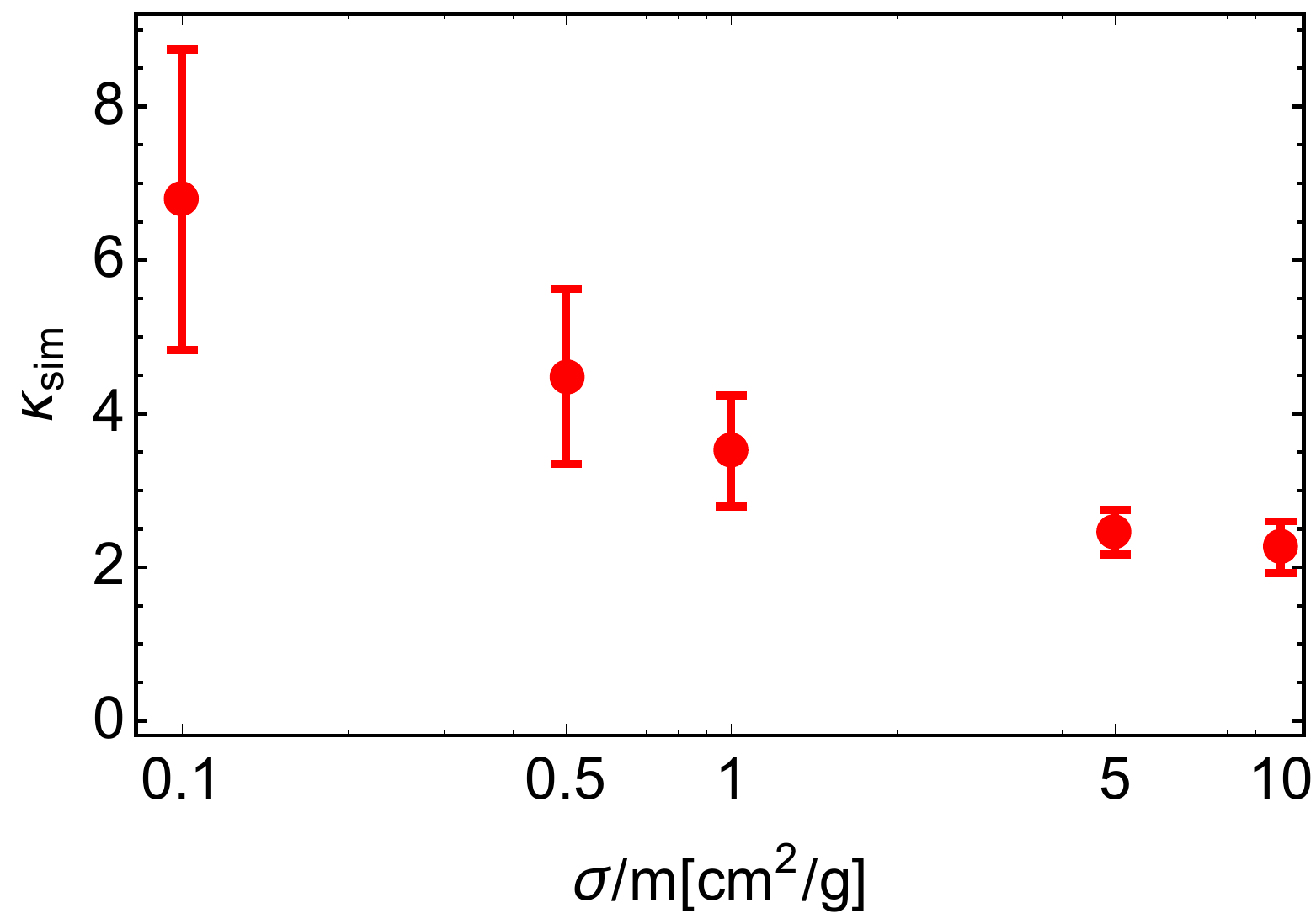}
    \caption{\textit{Left panel}: The ratio of the simulated $\kappa$ values in SIDM haloes to the predicted ones for our anisotropic model for different values of the self-interaction cross-section $\sigma/m$ as a function of halo concentration. \textit{Right panel}: The value $\kappa_{\text{sim}}=\rho(0)/\langle\rho(r_M)\rangle$ in our SIDM haloes for different values of the self-interaction cross-section $\sigma/m$. The central point is the mean $\kappa_{\text{sim}}$ value for a given cross-section in simulations, while error bars correspond to one standard deviation.}
    \label{fig:kappaallaniso}
\end{figure}

As demonstrated in Fig.~\ref{fig:kappaallaniso}, the prediction for the density profile with the anisotropic Jeans equation for the other SIDM cross sections is also consistent with the simulated data. This indicates a good agreement between the anisotropic modelling and the simulations. We note, however, that the modelling is less accurate at smaller cross sections, particularly for $\sigma/m\sim 0.1$ cm$^2/$g, where
the central SIDM halo is farther from equilibrium. This is reflected in the dependence of the parameter $\kappa=\rho(0)/\langle\rho(r_M)\rangle$ with cross section, see Fig.~\ref{fig:kappaallaniso}. 

\subsection{Connecting our model to the CDM NFW parameters}

Now we can try to emulate an actual data analysis with a more realistic approach. At large radii, SIDM profiles are well fitted by NFW profiles and, in this region, we usually have the best observational data. Therefore, we would like to only use the NFW profile to predict the SIDM density profiles for a given cross-section. The only input we will then need from simulations is the radius $r_\text{SIDM}$, which we showed above to be well represented by the radius of equal masses $r_M$. Predicting $r_M$ for every asymptotic NFW profile and each value of the cross-section is a non-trivial task and we leave it for the next section.

\begin{figure}[t!]
  \centering
    \includegraphics[width=0.48\textwidth]{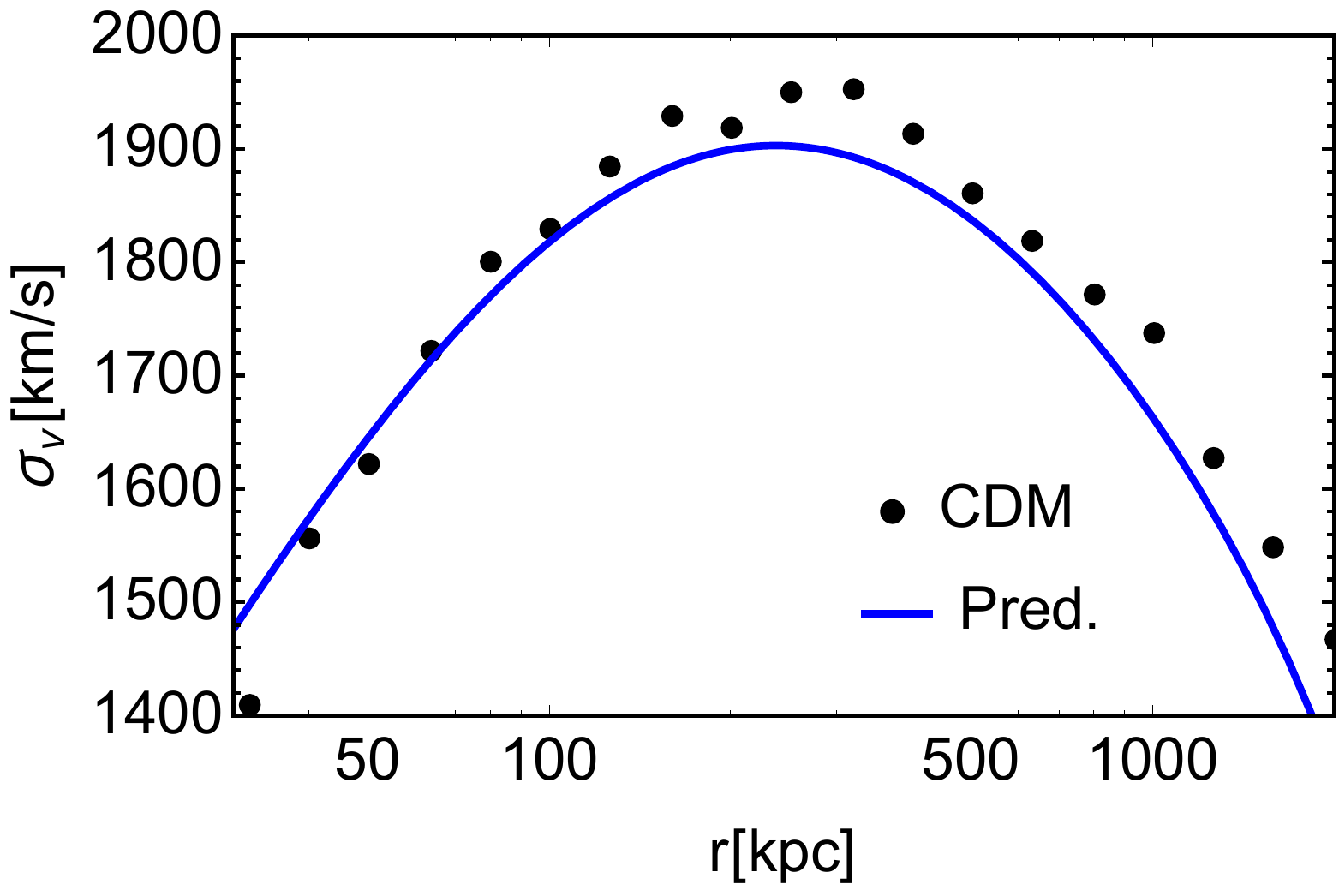}~\includegraphics[width=0.49\textwidth]{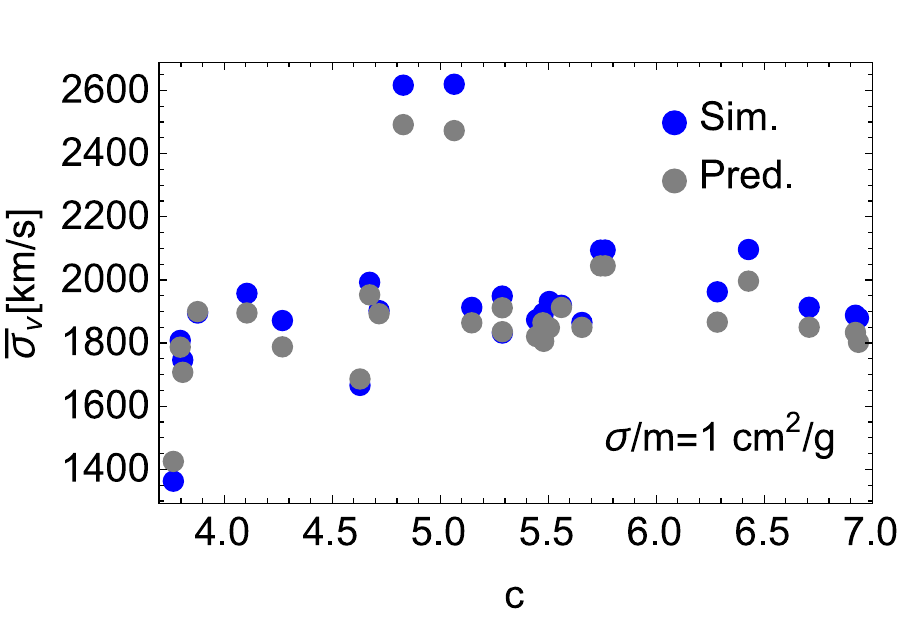}
    \caption{\textit{Left panel:} Total velocity dispersion profile for an example of a simulated CDM halo (black points) and the predictions based on the NFW density profile and the anisotropic Jeans equation (blue line). \textit{Right panel:}
    Central total velocity dispersion $\bar{\sigma}_{\bm{v}}$ inside $r_M$ as a function of concentration in simulated SIDM haloes (blue), and predicted values (gray) for the case $\sigma/m = 1$ cm$^2/$g. To make this prediction, we have used the NFW parameters of the corresponding CDM halo; see text for details.}
    \label{fig:sigmatotfromNFW}
\end{figure}

The method described and tested above also requires the velocity dispersion profile for the corresponding CDM halo as input. We can obtain it by using the NFW profile and solve the anisotropic Jeans equation (see Appendix~\ref{sec:sigmatotNFW}) with the anisotropy $\beta(r)$ described by the ansatz~\eqref{eq:betaansatz} (the same for each CDM halo, similarly to what we did for SIDM), see Table~\ref{tab:beta}. As a result of this procedure, we can obtain the velocity dispersion profile for a given CDM halo (for an example, see the left panel of Fig.~\ref{fig:sigmatotfromNFW}). 
Using the CDM velocity dispersion profile for a given halo, we predict the constant value for the corresponding central SIDM velocity dispersion $\sigma_{\bm{v}}^{\text{SIDM}}$, see right panel of Fig.~\ref{fig:sigmatotfromNFW}.

\begin{figure}[t!]
  \centering
    \includegraphics[width=0.51\textwidth]{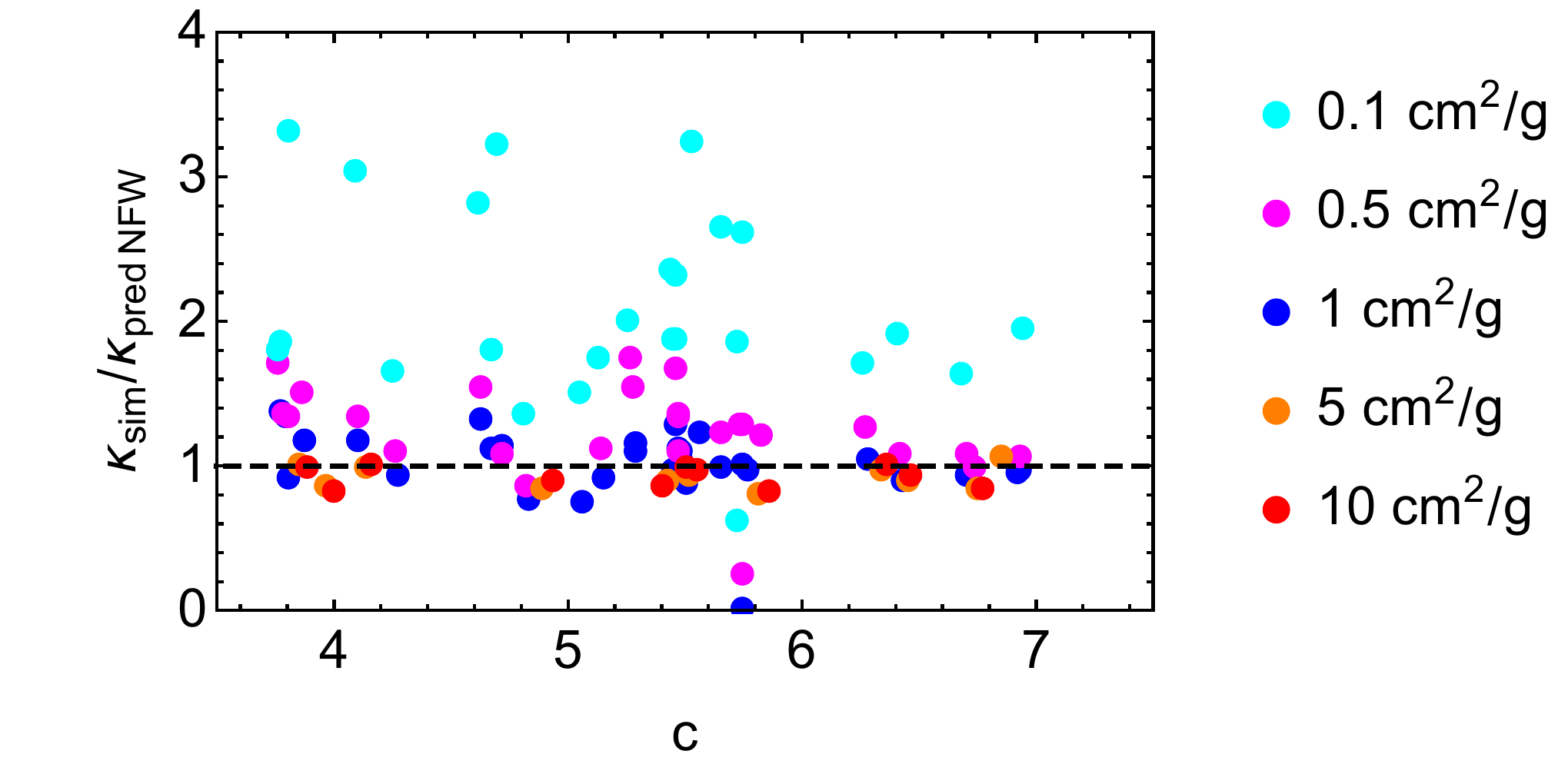}~\includegraphics[width=0.5\textwidth]{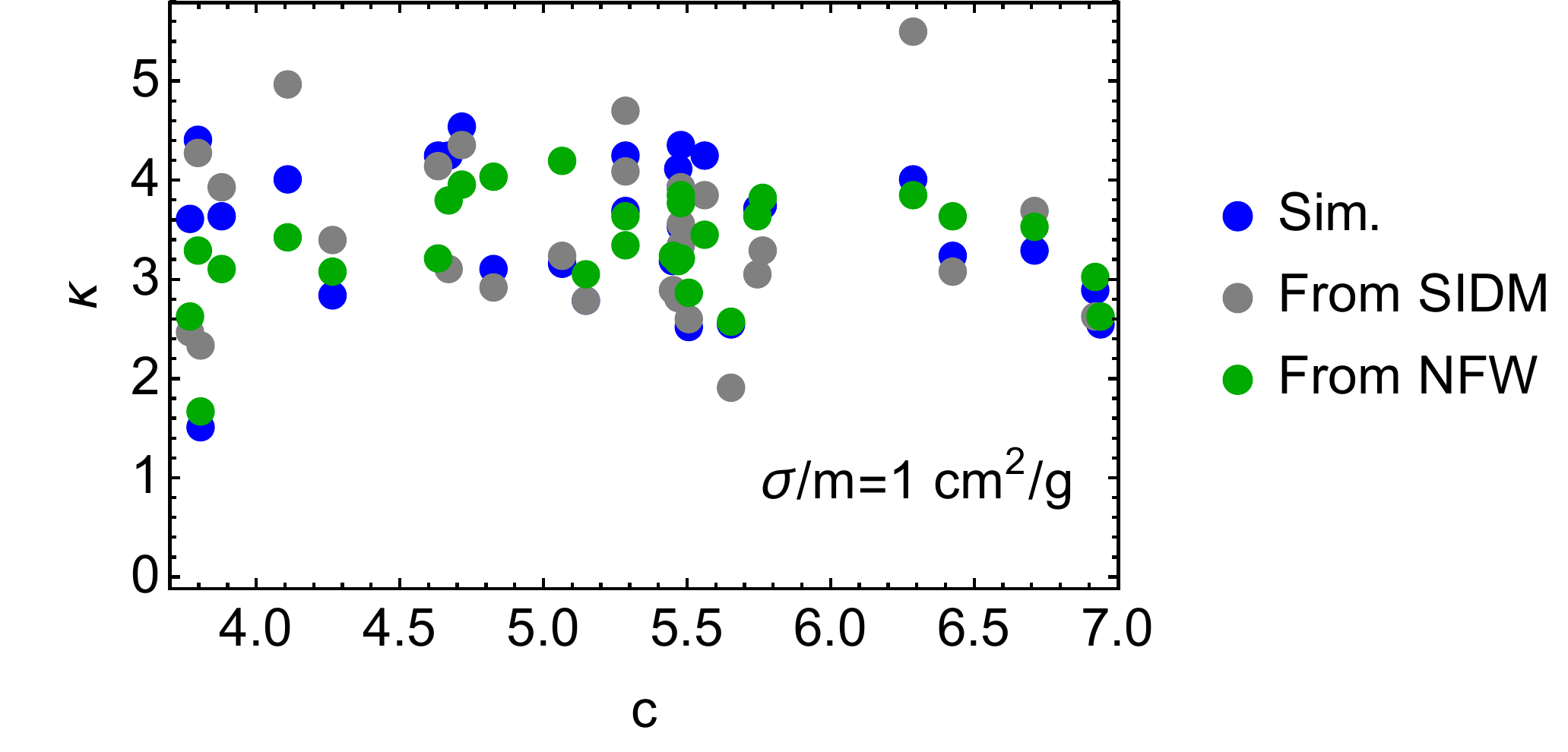}
    \caption{\textit{Left panel}: The ratio between simulated and predicted values of $\kappa$ in SIDM haloes as a function of halo concentration for our anisotropic model using as  input only the NFW parameters of the corresponding CDM haloes. \textit{Right panel}: Density ratios $\kappa$ versus halo concentrations, for $\sigma/m = 1$ cm$^2/$g, taken directly from our simulations (blue), from our anisotropic model using data from the SIDM simulations (gray), and with our full model using only the NFW parameters of the corresponding CDM simulations as input (green).}
    \label{fig:kappafromNFWleft}
\end{figure}

After this, we solve the anisotropic Jeans equation~\eqref{eq:jeans_aniso} with the average velocity anisotropy ansatz for a given cross-section to obtain the prediction of our model. We show the comparison between the $\kappa$ values in SIDM simulations and from this model in Figure~\ref{fig:kappafromNFWleft}. We see that the results are as good as Fig.~\ref{fig:kappaallaniso}, where we used the SIDM simulation data directly. Therefore, for a cross-section of $1~\text{cm}^2/\text{g}$ we are able to predict the density at the center of the SIDM halo with an average accuracy of about 10\%, with a similar distribution for simulated and predicted $\kappa$. 
Like in the previous case, for the cross-sections $\sigma/m = 0.1\text{ cm}^2/\text{g}$ and $0.5\text{ cm}^2/\text{g}$ the 
disagreement with simulations is a bit more significant. The reason is that for these smaller cross-sections equilibrium is not as 
well established  as for larger cross-sections, and a constant $\bar{\sigma}_{\bm{v}}$ is not a good enough approximation 
for the simulated data (see Fig.~\ref{fig:deltasigma}).

\newpage
\section{From the cross section to the radius $r_M$}
\label{sec:rm}

\begin{figure}[t!]
  \centering
  \includegraphics[width=0.48\textwidth]{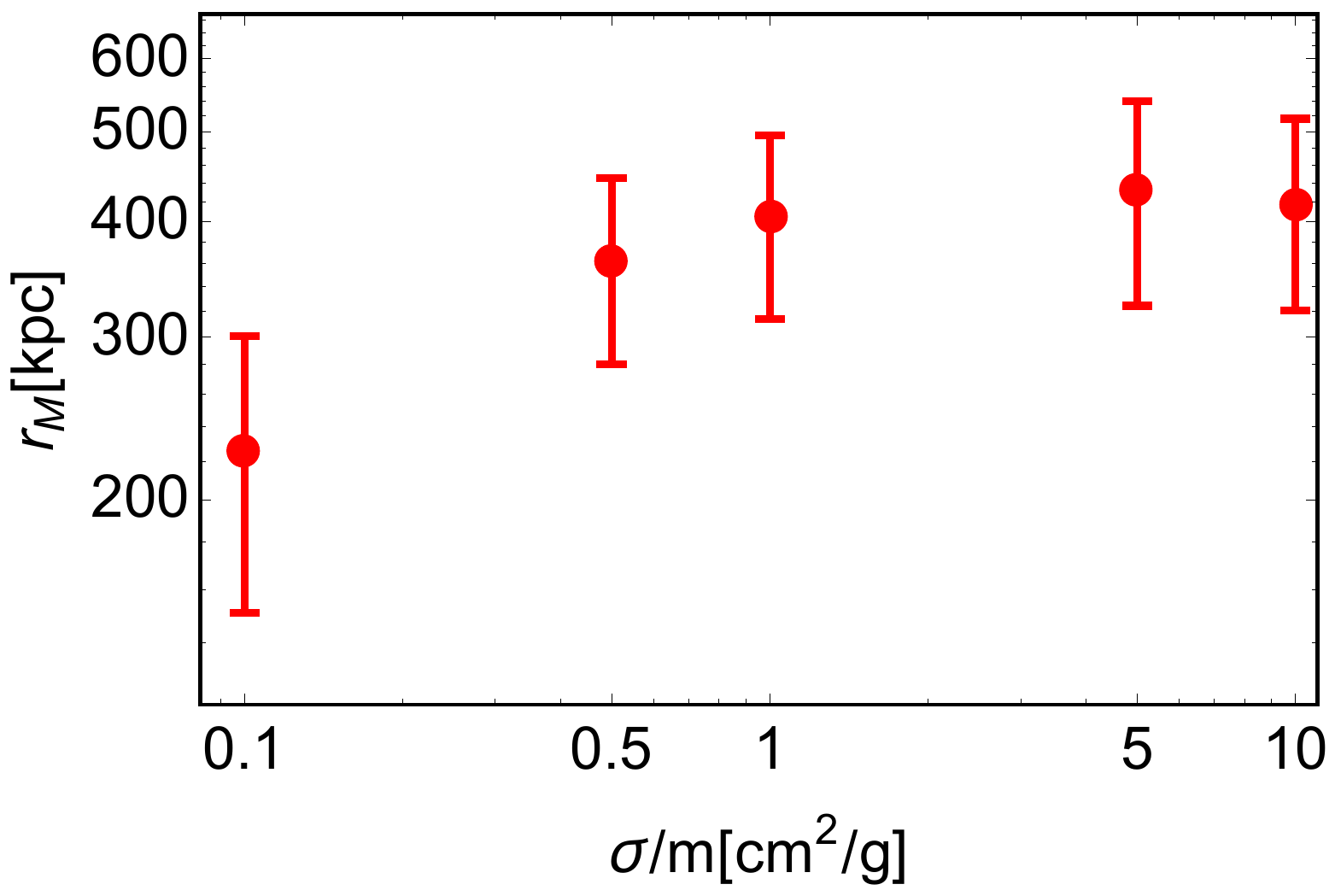}
  \caption{The dependence of $r_M$ on the cross-section. The error bars represent  the standard deviation of the distribution.}
  \label{fig:rMofsigma}
\end{figure}

Ideally, we would like to directly connect the observationally accessible core radius defined in Eq.~\eqref{eq:coredef}, see also 
Fig.~\ref{fig:rcoreofsigma}, to the predictions from our SIDM model.
In the previous sections, we have checked that the radius $r_{\text{SIDM}}$, where equilibrium is assumed to be established, could be chosen as the radius $r_M$, where the enclosed mass and kinetic energy in SIDM haloes are the same as in their CDM counterparts. The dependence of $r_M$ on cross-section is shown in Fig.~\ref{fig:rMofsigma}. Using this radius, one is able to predict, with sufficient accuracy (see the discussion in the previous section), the density profile for a SIDM halo using the data for a CDM halo with the same initial conditions. This means that we can relate the "observed" core radius to $r_{\text{SIDM}}$. To complete the picture, we now need to connect $r_{\text{SIDM}}$ with the self-interaction cross section $\sigma/m$.

\begin{figure}[t!]
  \centering
  \includegraphics[width=0.47\textwidth]{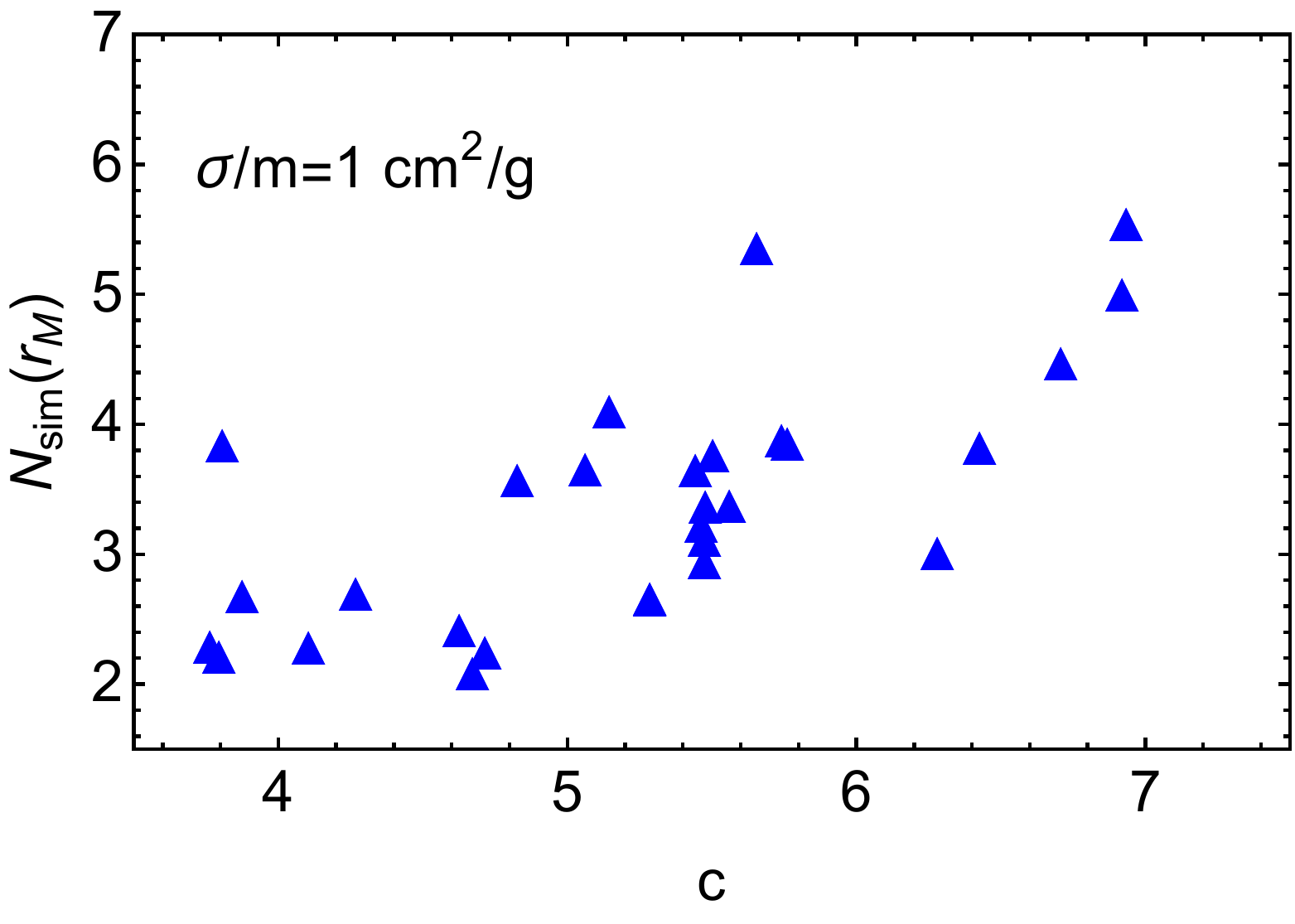}~\includegraphics[width=0.48\textwidth]{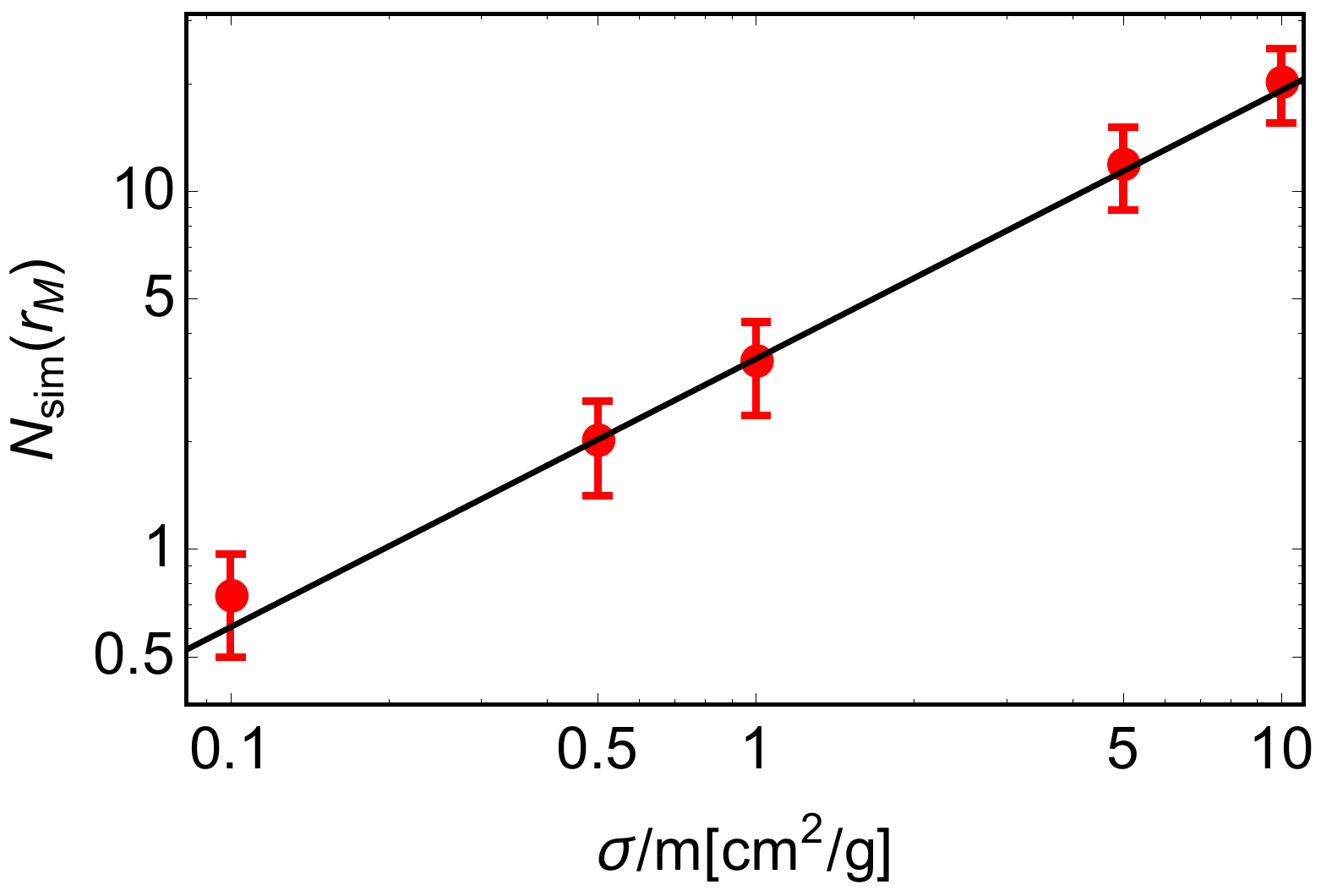}
  \caption{\textit{Left panel:} The average number of collisions at  radius $r_M$ as a function of halo concentration for $\sigma/m = 1$ cm$^2/$g. \textit{Right panel:}  The average number of collisions at  radius $r_M$ for different cross-sections. The black line shows the best fit power law dependence, $N(r_M)\propto(\sigma/m)^{0.75}$.}
  \label{fig:NatrM}
\end{figure}

In the literature (see e.g.~\cite{Kaplinghat:2015aga,Tulin:2017ara}), the relation between these quantities is often defined by the requirement of having at least one collision per particle inside $r_{\text{SIDM}}$, over the halo age $t_{\text{age}}$. We would like to check to what extent this condition is satisfied in simulations,
so we show in Fig.~\ref{fig:NatrM} the average number of collisions per particle at radius $r_M$. 
This clearly demonstrates that $N_{\text{sim.}}(r_M)$ varies significantly in our simulation suite. 
We also find that $N_{\text{sim}}(r_M)$ does not scale exactly linearly with $\sigma/m$, instead we find the best-fit slope of the power law to be  $0.75$ (right panel of Fig.~\ref{fig:NatrM}).

\begin{figure}[t!]
  \centering
  \includegraphics[width=0.48\textwidth]{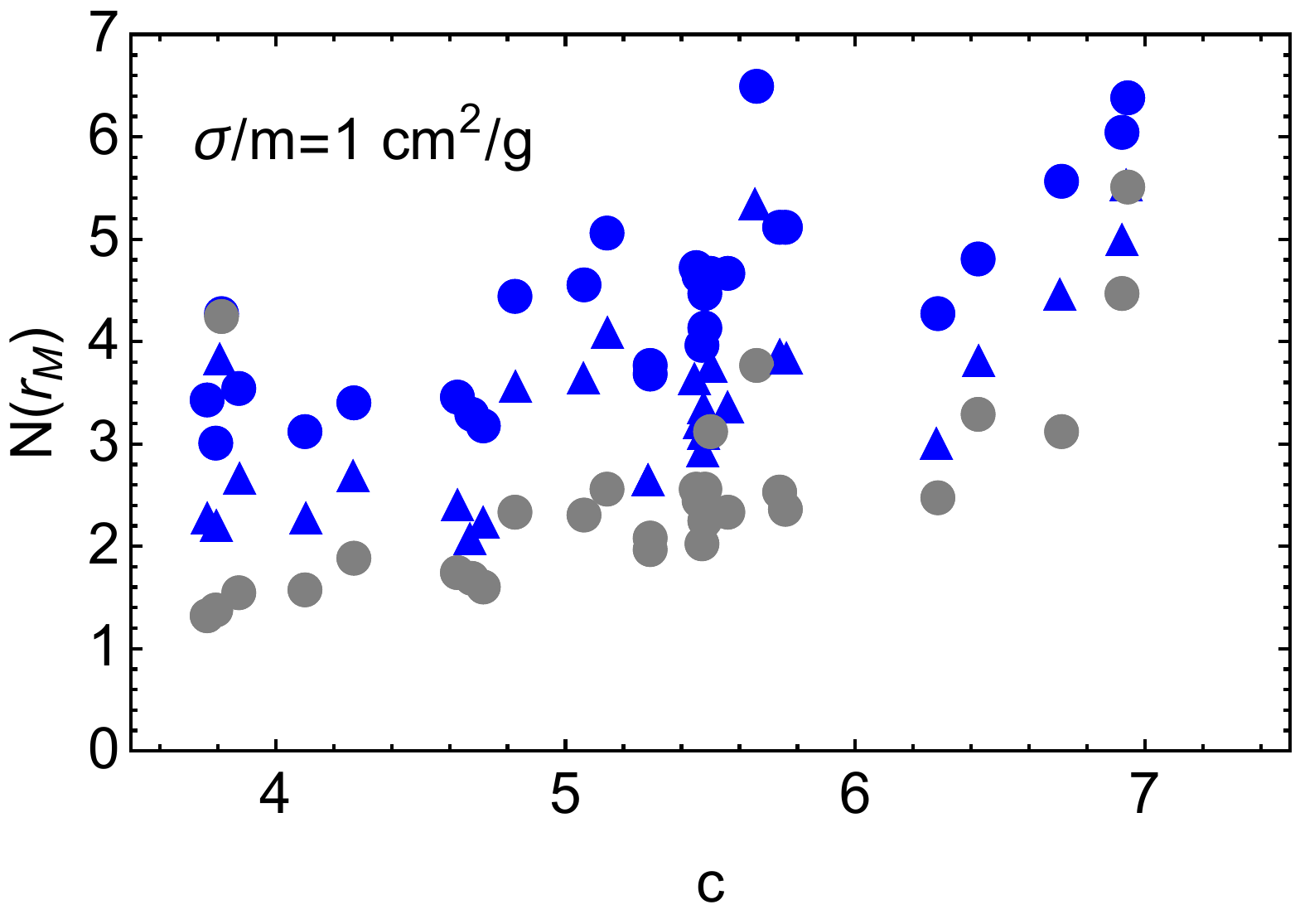}~\includegraphics[width=0.48\textwidth]{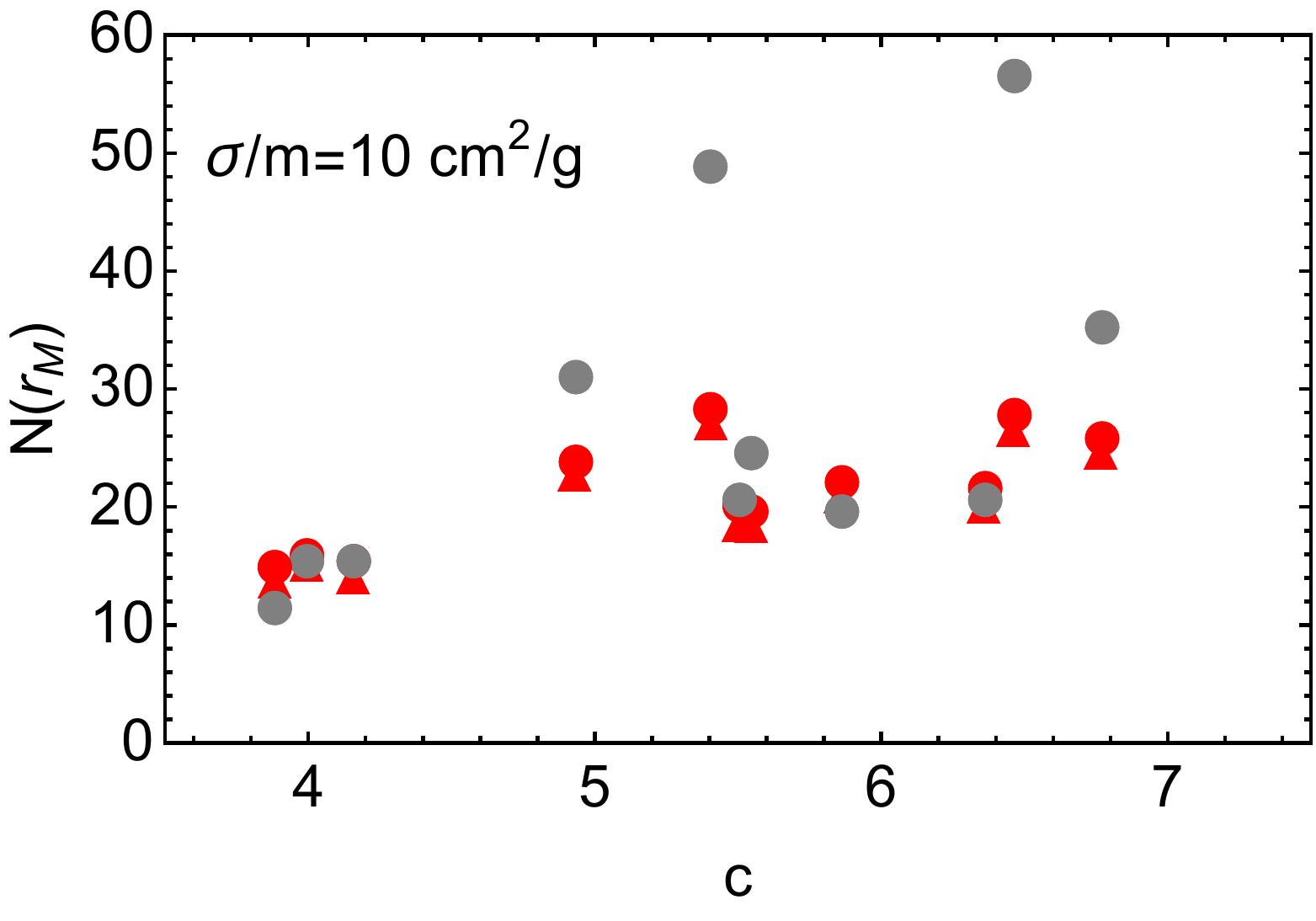}
   \includegraphics[width=0.48\textwidth]{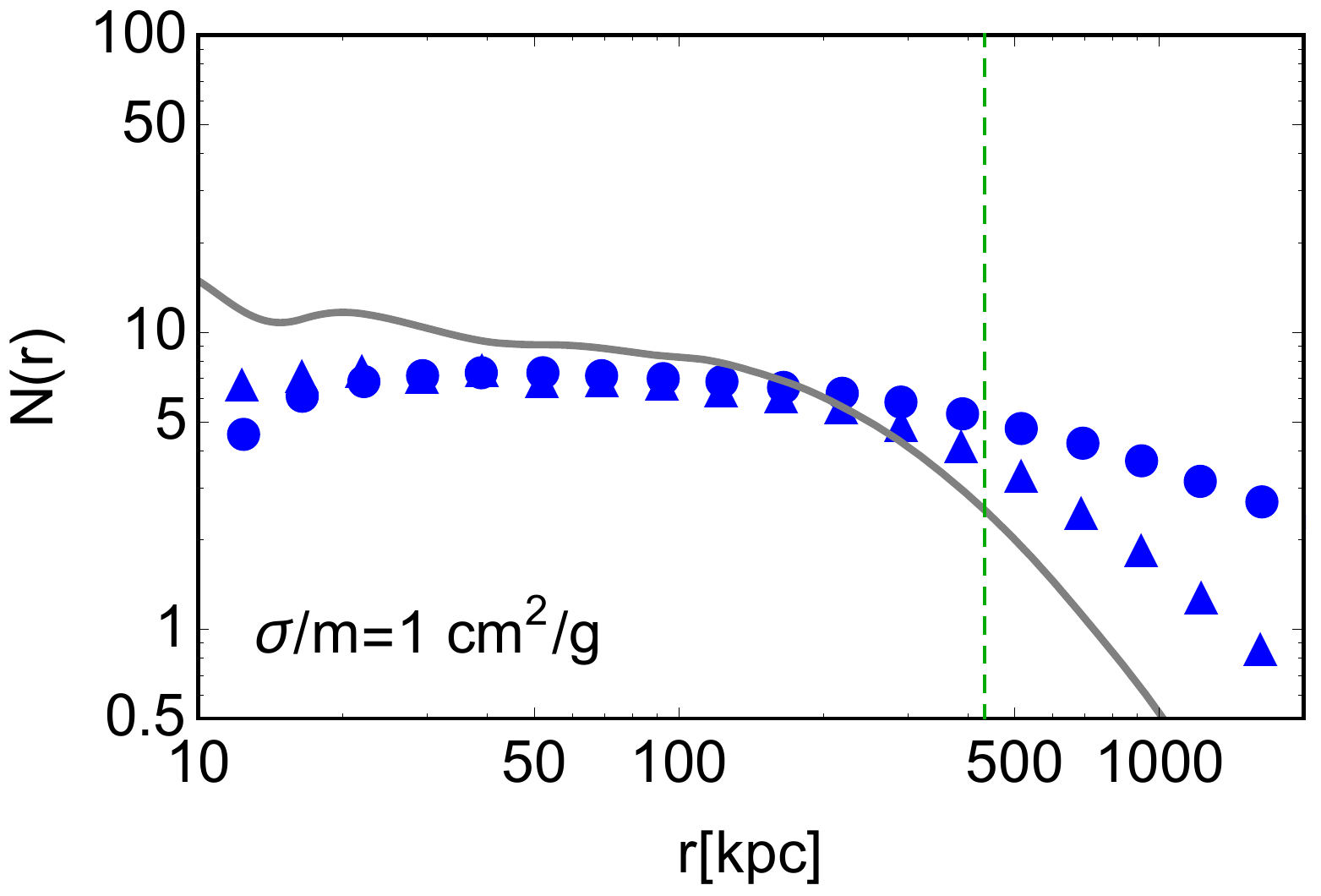}~\includegraphics[width=0.48\textwidth]{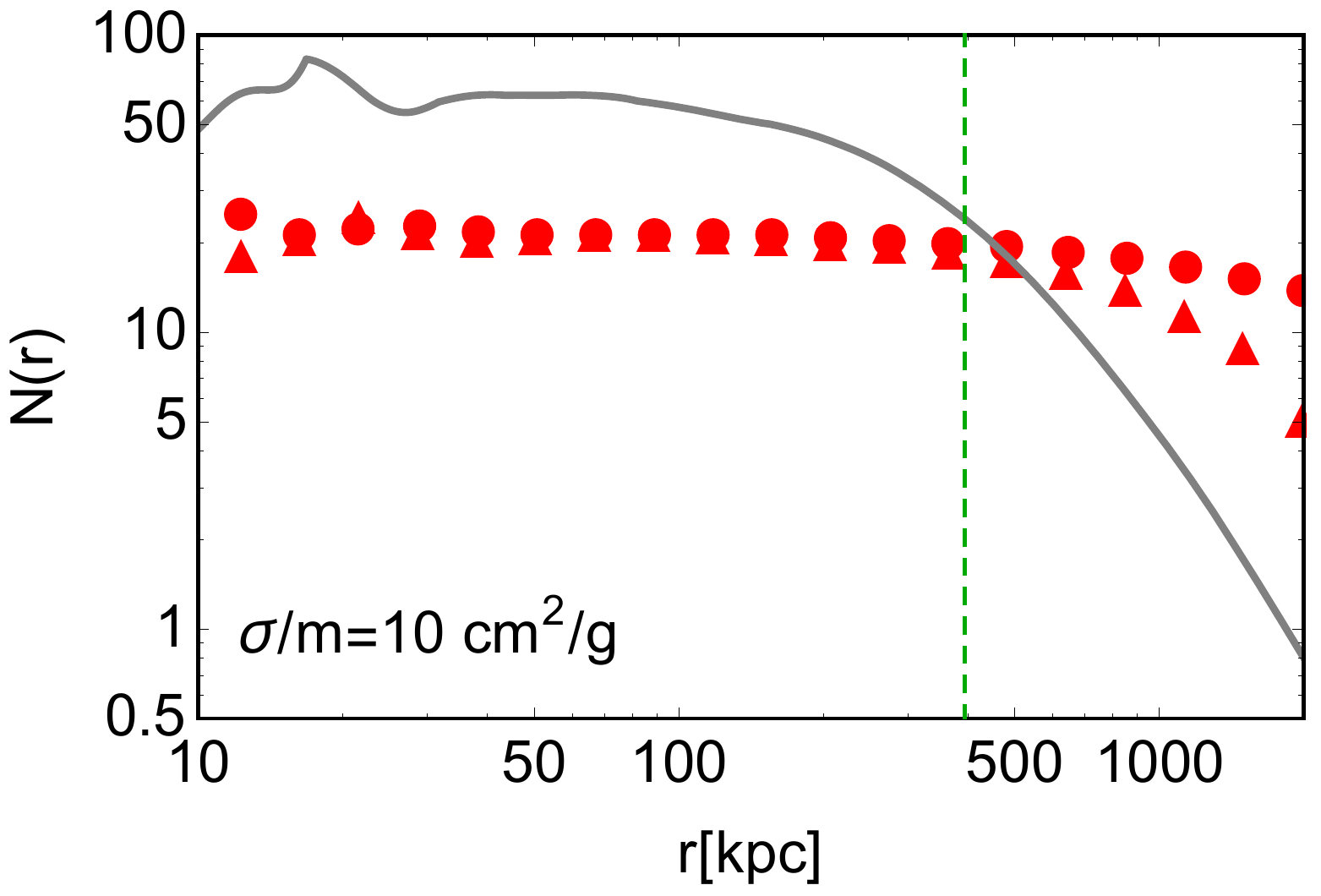}
  \caption{The average number of collisions {\it inside} radius $r_M$ (dots) and {\it at} radius $r_M$ (triangles) versus halo concentration for $\sigma/m = 1$ cm$^2/$g (top left) and $10$ cm$^2/$g (top right). The lower panels show example halos with 
  $\sigma/m = 1$ cm$^2/$g  (lower left) and $10$ cm$^2/$g (lower right) where we plot, as a function of radius $r$, the number of collisions inside (dots) and at $r$ (triangles).
  Blue and red points are the simulation data, while the grey dots (upper panels) and grey lines (lower panels) are the predictions from Eq.~(\ref{eq:Nofrfromeq}). The green dashed line in the lower panels marks the radius $r_M$.}
  \label{fig:NatrM2}
\end{figure}

In Refs.~\cite{Kaplinghat:2015aga,Tulin:2017ara} the number of collision per particles at radius $r$ is estimated as 
\begin{equation}
    N_{\text{l}}(r) \equiv  \frac{\sigma}{m} \rho(r)
    v(r) t_{\text{age}}\,,
    \label{eq:Nofrfromeq}
\end{equation}
where $v(r)$ is the average relative velocity of DM particles at radius $r$ and $t_{\text{age}}$ we take as half-mass formation time, see Ref.~\cite{Bondarenko:2017rfu} for details. The condition of one collision per particle per halo age 
thus translates to the often quoted
\begin{equation}
   \frac{\sigma}{m} \rho(r_{\text{SIDM}})
   v(r_{\text{SIDM}}) t_{\text{age}} = 1\,.
\label{eq:xi}
\end{equation}
Substituting $v(r) = (4/\sqrt{3\pi})\sigma^{\text{SIDM}}_{\bm{v}}(r)$, 
we have compared the values of $N_{\text{l}}(r_M)$ predicted from Eq.~(\ref{eq:Nofrfromeq}) with the  
simulation results.
We find reasonable agreement for cross-sections $5\text{ cm}^2/\text{g}$ and $10\text{ cm}^2/\text{g}$ (the latter is shown in the top right panel of Fig.~\ref{fig:NatrM2}), but the predictions from Eq.~(\ref{eq:Nofrfromeq})  are systematically lower for smaller cross-sections (see top left panel of Fig.~\ref{fig:NatrM2} for the case with $\sigma/m=1\text{ cm}^2/\text{g}$).
In the bottom panel of Fig.~\ref{fig:NatrM2} we furthermore show typical examples of the radial dependence of the number of collisions, $N(r)$,
found in the simulations. Clearly, Eq.~(\ref{eq:Nofrfromeq}) provides an incorrect prediction of this dependence.

\begin{figure}[t!]
  \centering
  \includegraphics[width=0.47\textwidth]{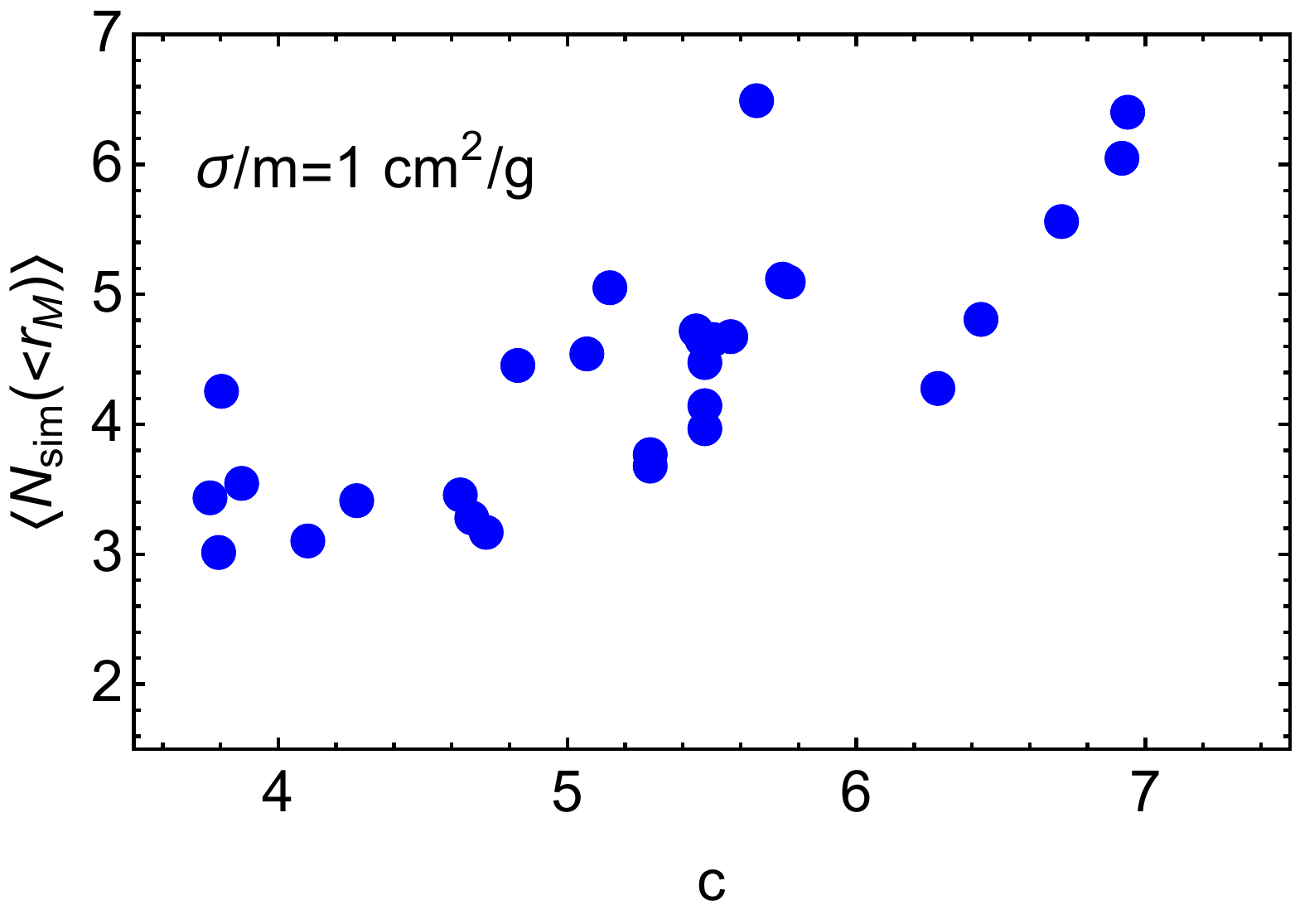}~\includegraphics[width=0.48\textwidth]{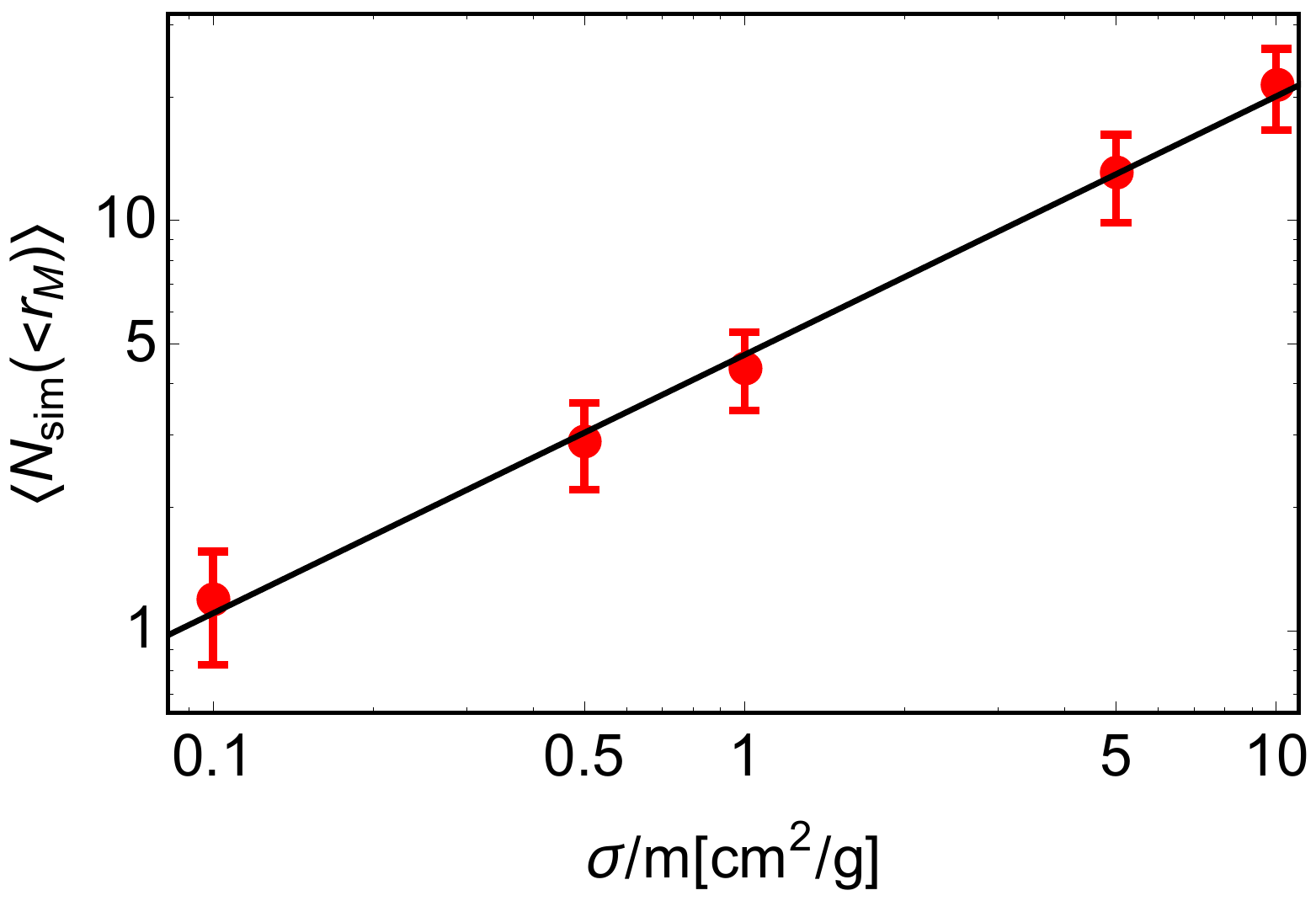}
  \caption{\textit{Left panel:} Average number of collisions inside radius $r_M$ as a function of halo concentration for $\sigma/m = 1$ cm$^2/$g. \textit{Right panel:}  Average number of collisions inside $r_M$ for different cross-sections. The black line shows the best fit power-law dependence, $\langle N(<r_M) \rangle\propto(\sigma/m)^{0.63}$.}
  \label{fig:NinsiderM}
\end{figure}

Following Ref.~\cite{Bondarenko:2017rfu} we have also checked a modified Eq.~\eqref{eq:xi} for the number of collisions per particle \emph{inside} $r_{\text{SIDM}}$,
\begin{equation}
    \frac{\sigma}{m} \langle\rho(r_{\text{SIDM}})\rangle
    v(r_{\text{SIDM}}) t_{\text{age}} = 1\,,
\label{eq:xi2}
\end{equation}
where $\langle\rho(r_{\text{SIDM}})\rangle = 3 M(r_{\text{SIDM}})/(4\pi r_{\text{SIDM}}^3)$ is the average density of the core. In analogy to Eq.~\eqref{eq:Nofrfromeq}, we thus have
\begin{equation}
    N (r) \equiv  \frac{\sigma}{m} \langle\rho(r)\rangle
    v(r) t_{\text{age}}\,,
    \label{eq:Nofrfromeq2}
\end{equation}
We repeated the same analysis for the average number of collision inside the radius $r_M$, $\langle N(<r_M) \rangle$, and found results similar to the previous case, see Fig.~\ref{fig:NinsiderM}. We conclude that neither Eq.~(\ref{eq:Nofrfromeq}) nor Eq.~\eqref{eq:Nofrfromeq2} 
can be reliably used to connect $r_{\text{SIDM}}$ with $\sigma/m$. Below, we discuss a simple toy model that improves upon this situation by qualitatively explaining the behaviour of $N(r)$ at large radii.

\subsection{Radial infall model for how the number of collisions depends on the radius}

Let us for simplicity consider the case of a stationary halo in which the DM particles are only moving on radial orbits. The orbit period $T$, with maximal radius $r_{\max}$, is then determined by the gravitational field as 
\begin{equation}
    T(r_{\max}) = 2 \int_0^{r_{\max}} \frac{dr}{v(r)} \, ,
\end{equation}
where the velocity $v(r)$ follows from energy conservation as
\begin{equation}
    U(r_{\max}) = U(r) + \frac{v^2(r)}{2} \quad \text{ with } \quad U(r) = \int_0^{r} \frac{G M(r)}{r^2} dr \, .
    \label{eq:rmaxenergyconservation}
\end{equation}

During the halo age $t_{\text{age}}$, a particle travels through the center of a halo $t_{\text{age}}/T$ times. The average number of scatterings per center crossing is
\begin{equation}
    N_T(r_{\max}) = 2 \int_0^{r_{\max}} \frac{\sigma}{m} \rho(r) dr \, ,
\end{equation}
where we neglected the change of the particle  trajectory after scattering. Therefore, the total number of collisions for this particle during its lifetime is
\begin{equation}
    N_{\max}(r_{\max}) = 
    \frac{t_{\text{age}}}{T(r_{\max})} N_T(r_{\max})
    \label{eq:Nmax} \, .
\end{equation}

Eq.~\eqref{eq:Nmax} gives the number of collisions for a particle with the maximal radius $r_{\max}$, while from the simulations we can extract the average number of collisions per particle for particles that are found at a given radius~$r$ at $z=0$. To connect these two numbers we determine the maximal radius of a particle with a given velocity $r_{\max}(r,v)$, using~\eqref{eq:rmaxenergyconservation}, and then average over the velocity distribution of the DM particles $f(r,v)$ at  radius~$r$, 
\begin{equation}
    N(r) = \int N_{\max}(r_{\max}(r,v)) f(r,v) dv \, .
    \label{eq:Nrdistibution}
\end{equation}
This formula reduces to Eq.~\eqref{eq:Nofrfromeq} in the case \mbox{$v(r)=\text{const}$}, \mbox{$\rho(r)=\text{const}$}.

\begin{figure}[t!]
  \centering
  \includegraphics[width=0.48\textwidth]{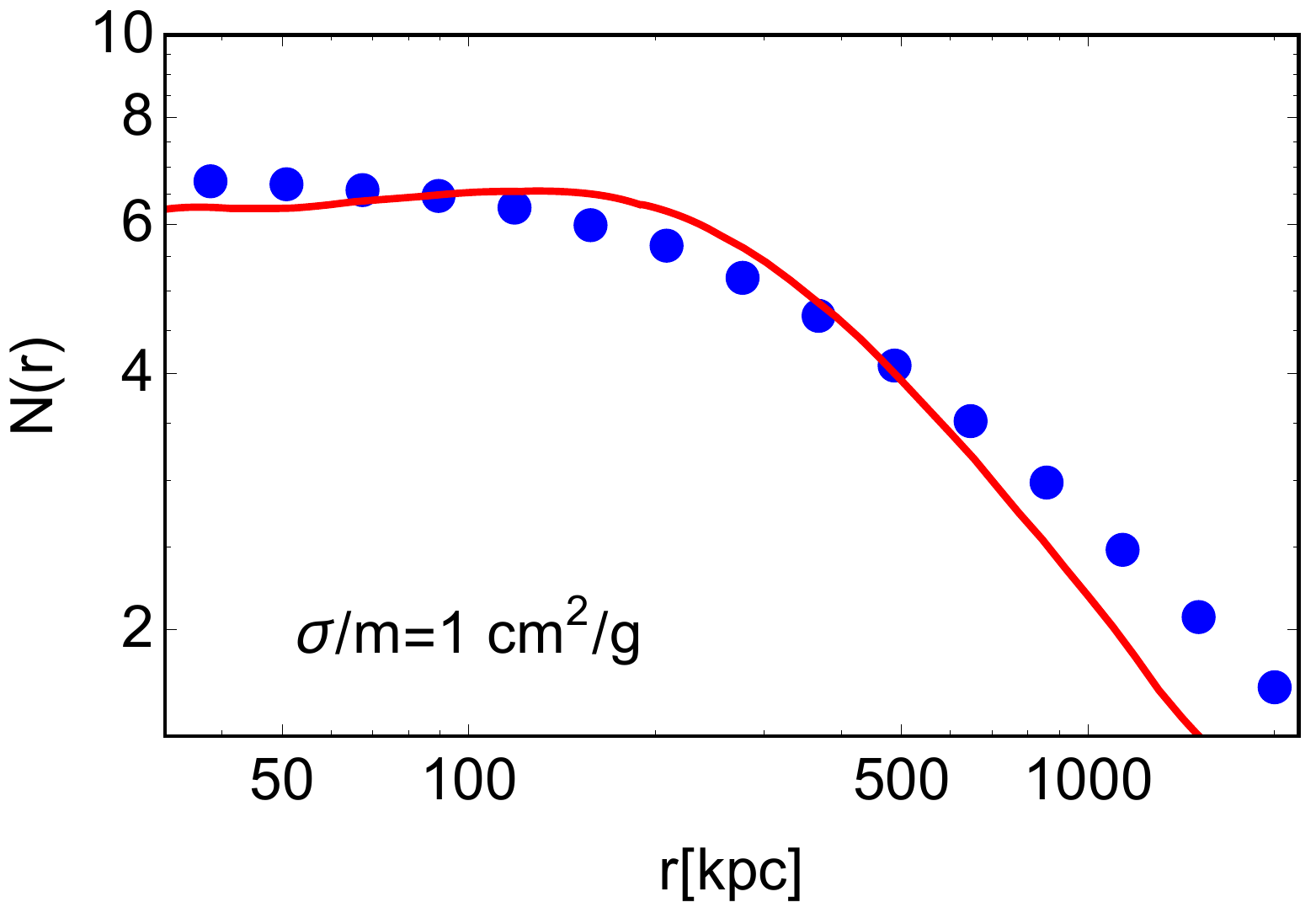}~ \includegraphics[width=0.48\textwidth]{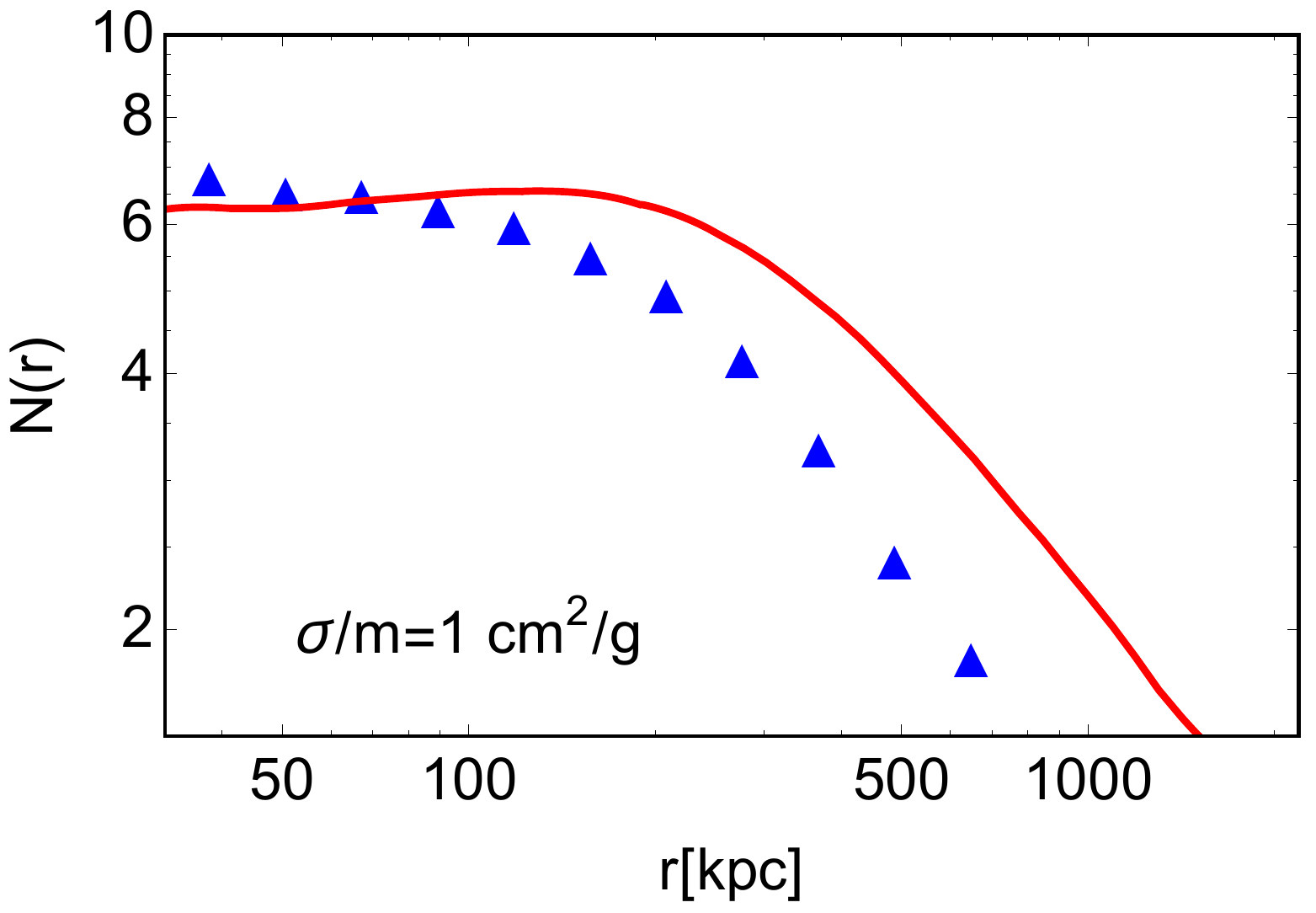}
  \caption{The average number of collisions per particle {\it inside}  (left) and {\it at}  (right) radius $r$  for $\sigma/m = 1$ cm$^2/$g (blue points). The red line is the prediction from a simple radial infall model, see Eq.~\eqref{eq:Nransatz}.}
  \label{fig:N2datapluspred}
\end{figure}

Since the velocity distribution $f(r,v)$ is not known a priory in our modelling, however, we are in general forced to use the radial velocity dispersion $\sigma_r$
instead of averaging as in Eq.~(\ref{eq:Nrdistibution}). This introduces an uncertainty which can be 
parameterized by an unknown factor $C$ of  order one:
\begin{equation}
    N_{\text{our}}(r) = C N_{\max}(r_{\max}(r,\sigma_r)) \,.
    \label{eq:Nransatz}
\end{equation}
In Fig.~\ref{fig:N2datapluspred} we show an example of how well this ansatz works for $C=2.5$ and $\sigma/m=1\text{ cm}^2/\text{g}$. We generally find that the ansatz~\eqref{eq:Nransatz} provides a good description for the average number of collisions {\it inside} a given radius $\langle N_{\text{sim.}}(<r)\rangle$ (left panel), while it works not so well for the number of collisions {\it at} a given radius $N_{\text{sim.}}(r)$ (right panel). Although the qualitative behaviour of the simulations is well described by this simple ansatz, it is clear that it requires improvement for a fully quantitative description.

\subsection{Phenomenological modelling for the radial profile of the number of collisions}

\begin{figure}[t!]
  \centering
  \includegraphics[width=0.48\textwidth]{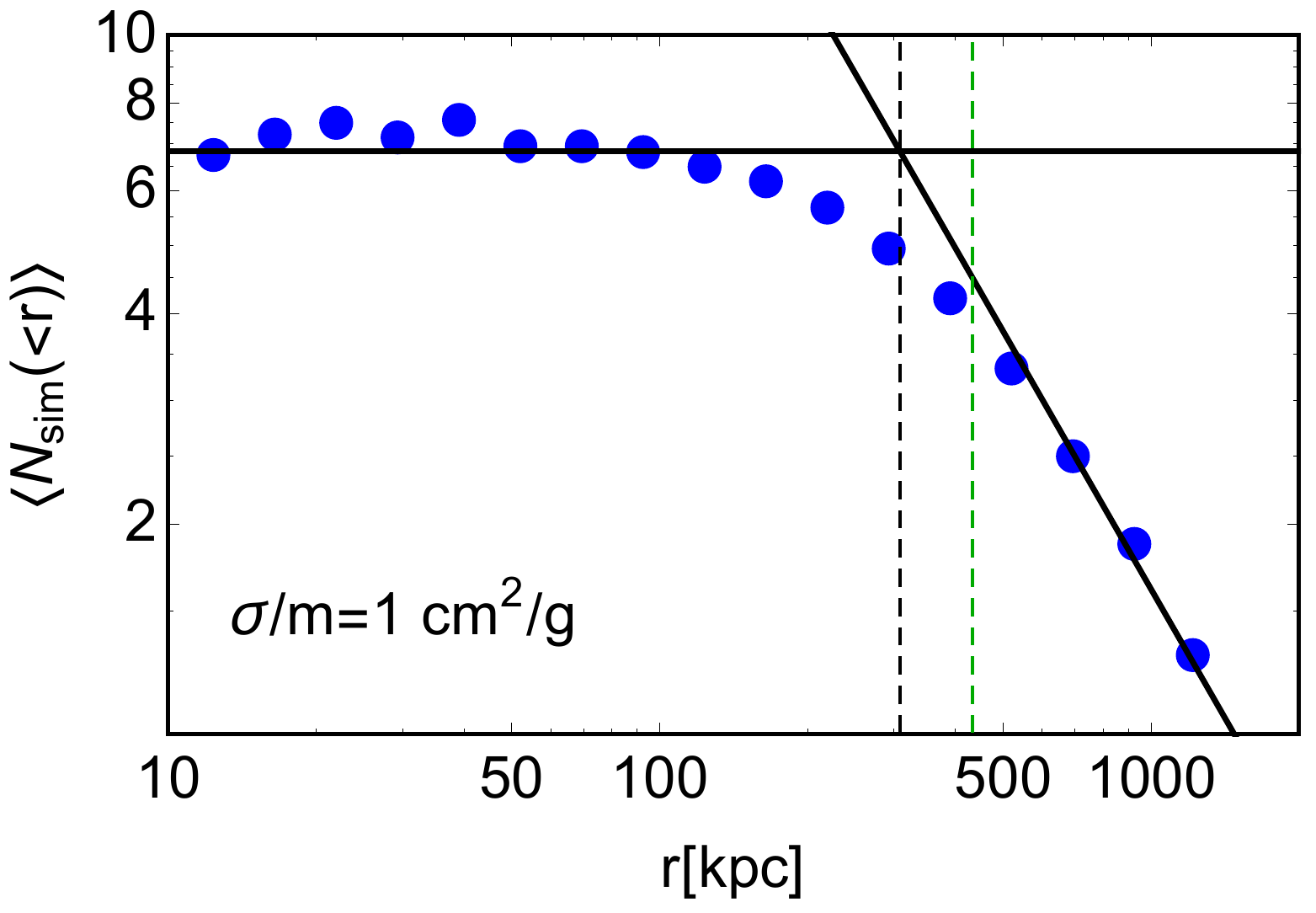}~\includegraphics[width=0.48\textwidth]{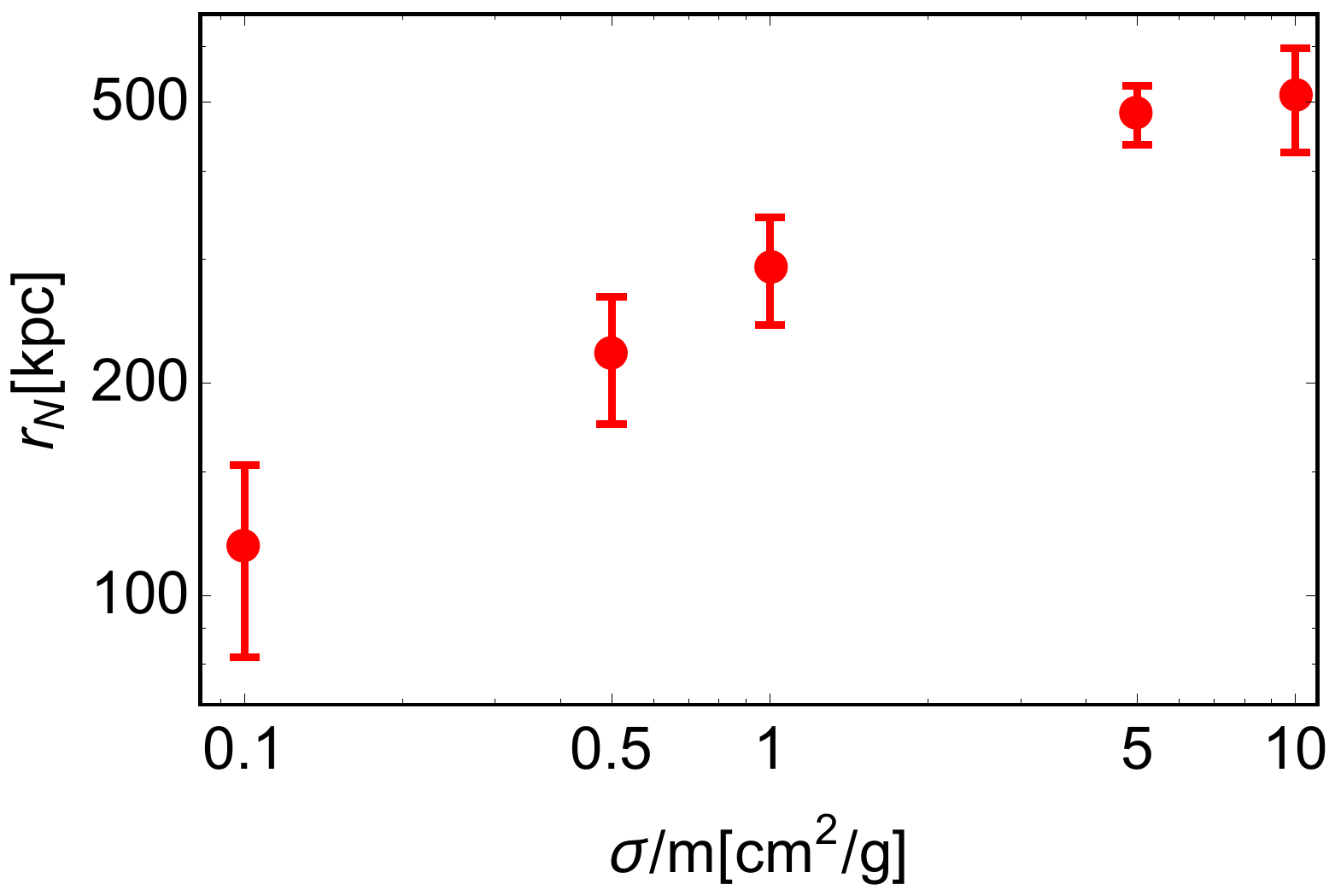}
  \caption{\textit{Left panel:} Illustration of how we have defined the radius $r_N$ (black dashed line) for the number of collisions profile $N(r)$, as the radius where the asymptotic behaviour (modeled as power laws) of the small and large radii cross each other (black solid lines). The blue points are the average number of collisions inside a given radius for $\sigma/m = 1$ cm$^2/$g.
  \textit{Right panel:} The dependence of the radius $r_N$ on the cross-section for our simulated suite. The error bars represent the standard deviation of the distribution.}
  \label{fig:rN}
\end{figure}

\begin{figure}[t!]
  \centering
  \includegraphics[width=0.6\textwidth]{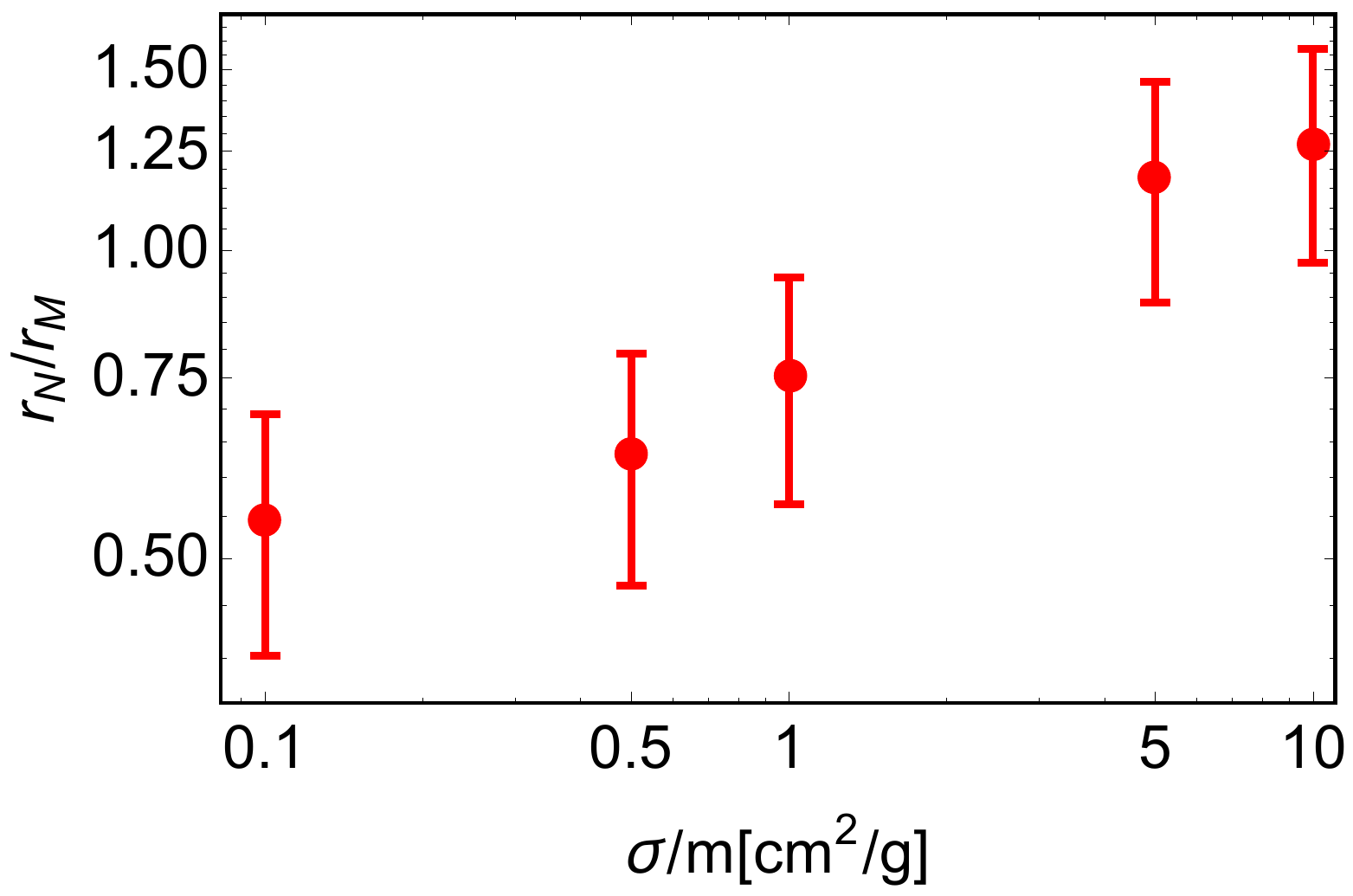}
  \caption{The cross-section dependence of the ratio of the two radii $r_N$ (connected to the radial dependence of the average number of collisions) and $r_M$ (defined as the radius within which the mass and kinetic energy are the same as in for CDM). The error bars represent the standard deviation of the distribution.}
  \label{fig:rNtorMofsigma}
\end{figure}

Since the model described above is not accurate enough to describe $N(r)$, we can try instead to use directly the profile $N(r)$ measured in the simulations to make a connection with $r_{\text{SIDM}}$ (or $r_M$) and thus relate it to $\sigma/m$ phenomenologically.
Looking at the simulation results, we observe that the radial dependence on $N(r)$ is very flat, hence, conditions of the type $N(r_M) = {\rm const}$ would always produce a large uncertainty on the estimated value of $r_M$.
However, $N(r)$ seems to have a characteristic radius $r_N$, where the slope of the profile changes substantially. This scale can be defined, for example, as the radius where the power laws of the asymptotic behaviour at small and large radii cross each other. An illustration of this definition of $r_N$ is shown in the left panel of Fig.~\ref{fig:rN}, where we used the average number of collisions inside a given radius $\langle N_{\text{sim.}}(<r)\rangle$ as a proxy for $N(r)$ in simulations. The scaling of $r_N$ with the cross section is shown in the right panel of the same figure. Although the behaviour is similar to the behaviour of $r_M$ as a function of $\sigma/m$ (see Fig.~\ref{fig:rNtorMofsigma}), $r_M$ and $r_N$ do not coincide. The necessary conditions for a single scaling radius $r_{\text{SIDM}}$ in our SIDM model, which we discussed in Section \ref{sec:verification}, are satisfied more poorly at $r_N$ than they are at $r_M$, hence, we cannot replace $r_M$ by $r_N$. Moreover, the ratio between $r_N$ and $r_M$ depends on the cross-section (see  Fig.~\ref{fig:rNtorMofsigma}). In turn, the dependence of $r_M$ on the 
cross-section is rather weak (see Fig.~\ref{fig:rMofsigma})
and it cannot be used to define $\sigma/m$ with confidence: {\it a small error in estimating $r_M$ results in a large uncertainty for the estimated $\sigma/m$}.

We conclude that, although we are able to construct an improved model that can relate the radius $r_{\text{SIDM}}$ in SIDM haloes to observables (the inner density profile and, in particular, the core radius), we can still not robustly relate this radius (and hence a potential observable) to the self-interaction cross-section $\sigma/m$. This limitation is caused by a complex dependence of the properties of simulated SIDM haloes on the value of the cross-section, which conflicts with the simple estimate used in previous analytical models in the literature, based on a constant average number of collisions per particle and halo time in the SIDM halo core.
 
\section{Comparison with previous approaches}
\label{sec:tulin}

Let us finally compare  predictions from our method with those based on the method commonly adopted in the literature~\cite{Tulin:2017ara},
with a focus on observationally accessible quantities like the core radius.
As a reminder of our discussion in Section~\ref{sec:anal_model}, 
the Jeans equation requires two boundary conditions (and has one free parameter $\bar{\sigma}_{\bm{v}}$), which we choose  
as $\rho'(0)=0$ (cored solution) and $M_{\text{CDM}}(r_{\text{SIDM}}) = M_{\text{SIDM}}(r_{\text{SIDM}})$. This still leaves the determination 
of the transition scale $r_{\text{SIDM}}$. 
In the method used in this work both predicted  $\rho_{\text{SIDM}}(0)$ and $\bar{\sigma}_{\bm{v}}$ agree with simulations (see  e.g. Figs.~\ref{fig:sigmatot} and \ref{fig:kappaallaniso}).

\begin{figure}[t!]
  \centering
  \includegraphics[width=0.6\textwidth]{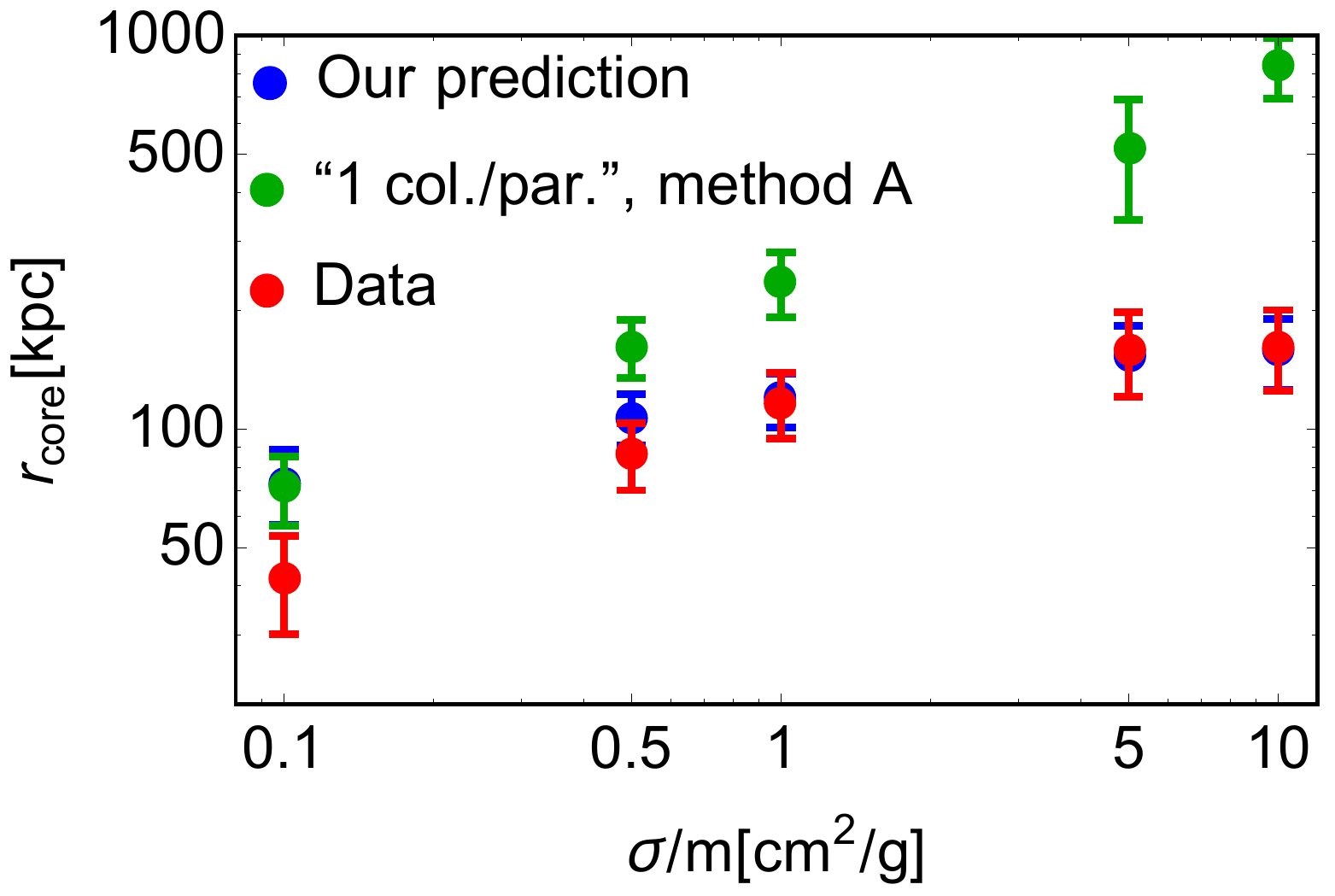}
  \caption{We compare the core radius~\eqref{eq:coredef} from simulation data (red), our predictions from this work (blue) and the predictions obtained by following what we describe as \textit{method A} in the text (green), using an isotropic Jeans equation and imposing the  equal mass boundary condition at the “one collision per particle” radius.}
  \label{fig:r_core}
\end{figure}

An alternative approach is to use the isotropic Jeans equation (as in Ref.~\cite{Tulin:2017ara}) and impose the equal mass 
boundary condition at $r_{\text{SIDM}}$ defined by “one collision per particle” and halo time, cf.~Eq.~\eqref{eq:xi}. 
To fix the density profile we also need to know $\bar{\sigma}_{\bm{v}}$, which we take directly from the SIDM simulations. We call this approach \textit{method A}. 
In Fig.~\ref{fig:r_core}, we compare simulated core radii with those predicted by the two methods. Here, we choose
our standard definition of the core radius, Eq.~\eqref{eq:coredef}, but stress that the qualitative features of this figure would not change with alternative definitions. Clearly, the predictions from \textit{method A} are not consistent with the simulation data. Our method, on the other hand, is in excellent agreement with the data for 
$\sigma/m_\chi\gtrsim 1$\,cm$^2$/g; for smaller cross sections, it slightly overpredicts the expected core size.

\begin{figure}[t!]
  \centering
  \includegraphics[width=0.5\textwidth]{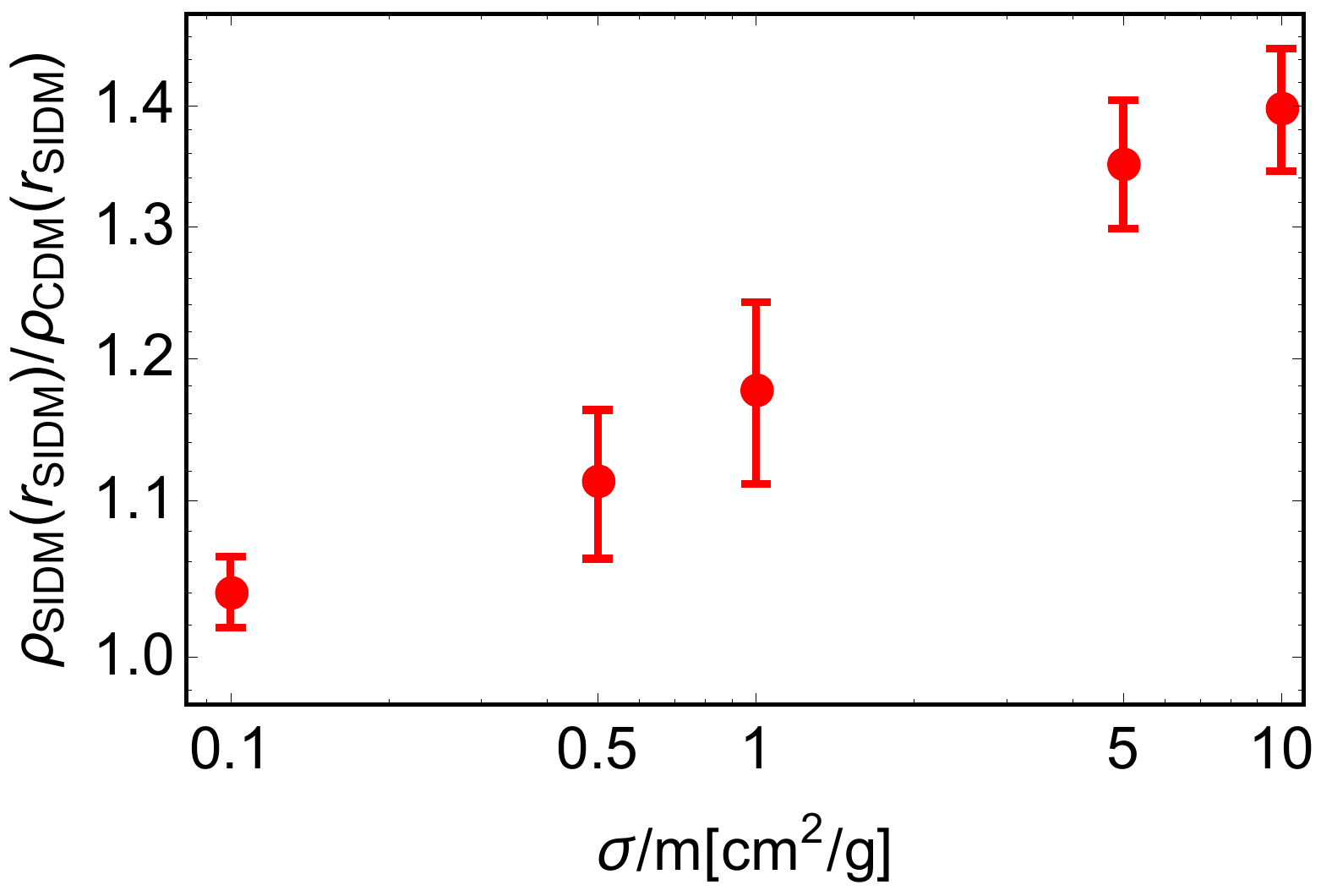}
  \caption{Density ratios of simulated SIDM and CDM halos, at radius $r_{\text{SIDM}}$, versus cross-section.}
  \label{fig:rhoSIDMtorhoCDMatrM}
\end{figure}

\begin{figure}[t!]
  \centering
  \includegraphics[width=0.47\textwidth]{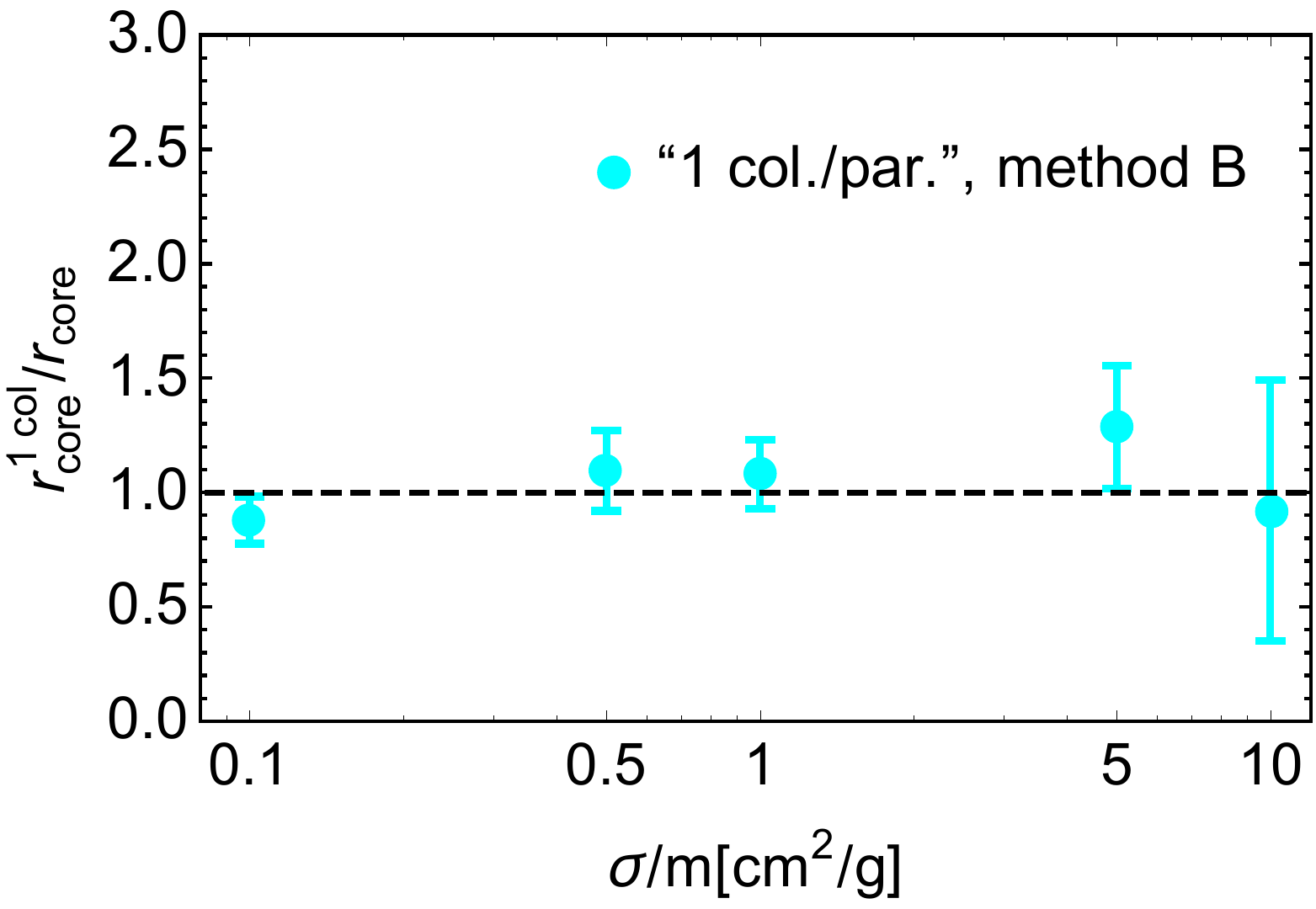}~\includegraphics[width=0.49\textwidth]{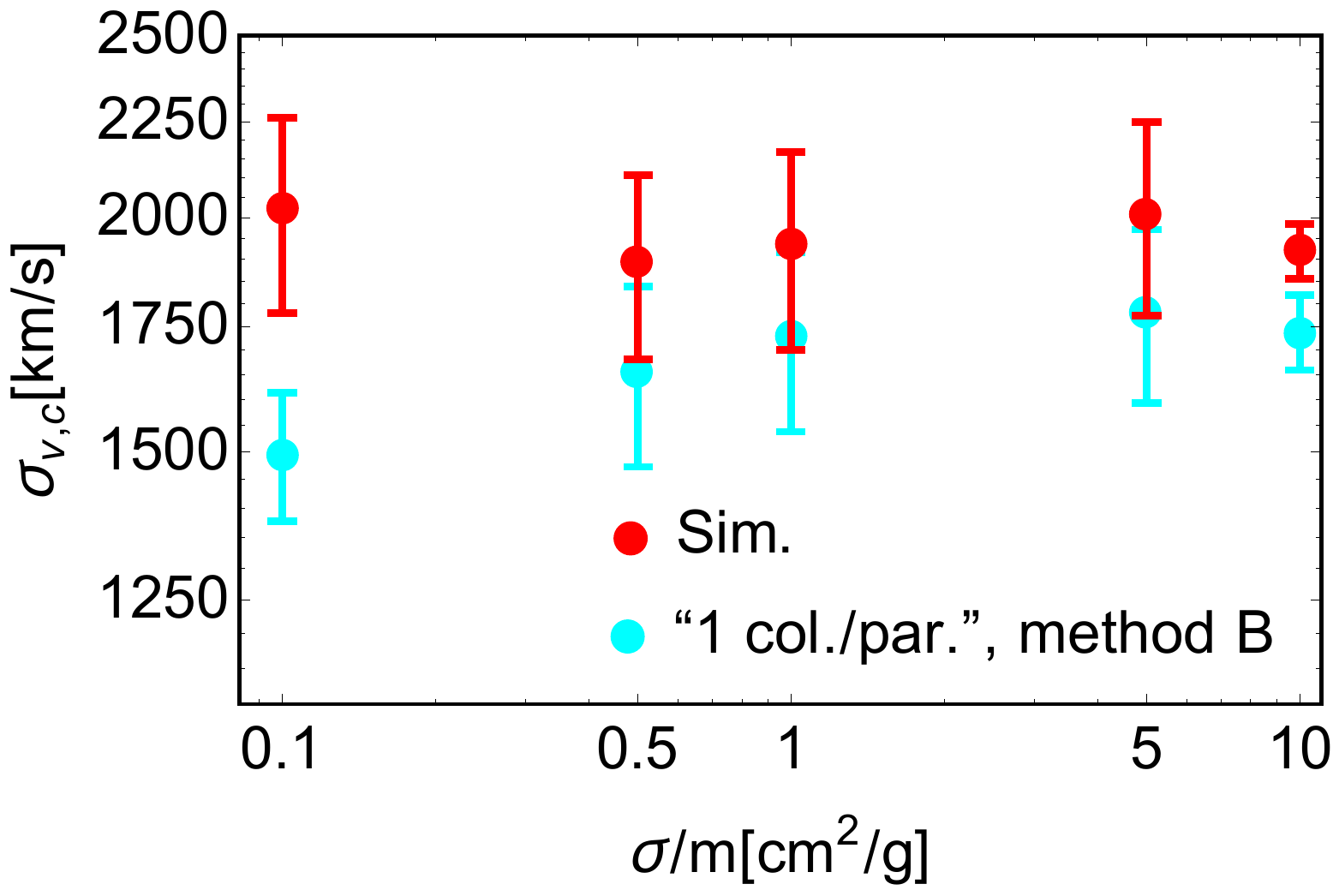}
  \caption{\textit{Left panel:} Ratio between core radii of SIDM halos predicted from CDM halo profiles using the commonly
  adopted method in the literature (\textit{method B}) and core radii in SIDM simulations versus cross-section. 
  \textit{Right panel:} Average velocity dispersion 
  inside $r_{\text{SIDM}}$ for simulation data (red) and  predictions (cyan) for \textit{method B}.}
  \label{fig:rcore_tulin_our}
\end{figure}

The approach {\it actually} implemented in Ref.~\cite{Tulin:2017ara}, and largely followed in the literature, is to implicitly
fix $\bar{\sigma}_{\bm{v}}$ by an additional boundary condition rather than taking it from simulations. 
Concretely, this additional condition is to assume a continuous density profile at $r_{\text{SIDM}}$, i.e.~imposing
$\rho_{\text{CDM}}(r_{\text{SIDM}}) = \rho_{\text{SIDM}}(r_{\text{SIDM}})$ (where  $r_{\text{SIDM}}$ is again defined by 
the “one collision per particle” condition). We call this approach \textit{method B}. Let us remark that the condition of a 
continuous density profile must be satisfied {\it iff} the transition zone between the region of equilibrium (which is
well described by the Jeans equation) and the outer region (described by a standard CDM halo) is infinitely thin. 
In reality, one would expect a more extended transition region where self-interactions neither fully thermalize the halo
nor leave it completely unaffected. Around the point $r_{\text{SIDM}}$ where the boundary is formally placed, both the solution of the Jeans equation and the outer (typically NFW) profile are then only extrapolations that do not describe the actual density profile; hence, it is not obvious why these profiles should match exactly at $r=r_{\text{SIDM}})$.
An explicit comparison with simulations, as shown in Fig.~\ref{fig:rhoSIDMtorhoCDMatrM}, reveals that this is indeed not the case.
In the left panel of Fig.~\ref{fig:rcore_tulin_our}, we plot the ratio of the core radius obtained with method {\it B}
to that found in our simulations. As pointed out previously, e.g.~Refs.~\cite{Rocha:2012jg,Peter:2012jh, Kaplinghat:2015aga}, this leads to very good agreement. However, as demonstrated in the right panel of Fig.~\ref{fig:rcore_tulin_our}, the predictions for the velocity dispersion 
$\bar{\sigma}_{\bm{v}}$ are clearly {\it not} compatible with the simulation results. In view of the failures to correctly reproduce both $\bar{\sigma}_{\bm{v}}$ and the density ratio at the transition point, we are thus lead to conclude that the success of method {\it B} in predicting the core radii is at least partially based on a numerical coincidence.

In other words, the success of the commonly adopted method to correctly describe core sizes as a function
of the self-interaction cross section in DM-only simulations can {\it not} be taken as supporting evidence that the method adequately captures the underlying dynamics of SIDM halos. 
This is particularly relevant when extending this method to more realistic halos that also include baryons, 
simply by replacing the gravitational potential due to DM with the total (DM + baryonic) gravitational potential in the Jeans 
equation~\cite{Kaplinghat:2013xca,Kaplinghat:2015aga,Kamada:2016euw,Valli:2017ktb}. 
While we do not offer an alternative way of modelling the effect of baryons,
our findings suggest that existing conclusions about $\sigma/m$ that are derived from the observation of systems
where the effect of baryons is important should, at this point, also be interpreted with care.

\section{Summary and conclusion}
\label{sec:conc}
In order to constrain the SIDM cross-section from observed dynamical properties of galaxies or clusters of galaxies, one can adopt either of the following two methods:

\begin{itemize}
    \item Use a large number of numerical simulations and a careful mapping between direct observables and simulated quantities. This approach captures the relevant physical processes and gives a full prediction for the structure of SIDM haloes. 
    \item Use an analytical model that accurately captures the effect of dark matter self-scattering on observables. This approach has the advantage of requiring much less computational time, while being able to compare models to data for a wide range of halo masses, but requires a physically motivated model with several assumptions, which are tuned and tested against simulations.
\end{itemize}

In this paper, we have taken a revised look at the second approach, with the goal of improving the analytical modelling of SIDM haloes. A summary of our main findings is as follows:
\begin{itemize}
    \item The models currently used in the literature do not explain the simulation results in a satisfactory manner. In particular the basic underlying formula~\eqref{eq:xi} for one collision per particle and halo age is not supported by simulations, see Figures~\ref{fig:NatrM} and \ref{fig:NatrM2}. Also, we found that the velocity anisotropy is not zero (see Fig.~\ref{fig:beta}), which significantly changes the predictions for the density profiles  (see Figures~\ref{fig:jeans_solution_aniso}, \ref{fig:kappaallaniso}).
    \item We have introduced an improved model, which takes as input the
    large radii behaviour of SIDM haloes that asymptotically reaches the CDM predictions (parameterized with the NFW profile), and matches it to the solution of the anisotropic Jeans equation with a constant velocity dispersion at a radius $r_{\text{SIDM}}$ (see Fig.~\ref{fig:r_core} for the comparison of our analytical results with the data and analytical predictions made by the isotropic Jeans equation with equal mass boundary conditions at “one collision per particle radius”), see Section~\ref{sec:tulin} for details. For a given halo with fixed NFW parameters, our model gives a good prediction of the properties of the corresponding SIDM halo (core size and density) as a function of the cross section {\it if} the radius $r_{\text{SIDM}}$ is taken from the simulated halo.
    The boundary condition imposed to define $r_{\text{SIDM}}$ in our model is to match the mass of the CDM and SIDM haloes within $r_{\text{SIDM}}$. A second boundary condition, imposing the same kinetic energy within $r_{\text{SIDM}}$, fixes the central velocity dispersion, which finally closes the system allowing us to find a unique SIDM profile.
    \item We have checked that the assumptions of our model are to 
    a good approximation satisfied in the simulated SIDM haloes. Inside $r_{\text{SIDM}}$, the \textit{total} velocity dispersion profile is flat enough, $\langle \delta \sigma_{\bm{v}} \rangle/\bar{\sigma}_{\bm{v}}\sim 0.01$ (on average, see Fig.~\ref{fig:deltasigma}), with a scatter that becomes smaller for larger cross-section, but even for $\sigma/m=0.5~\text{cm}^2/\text{g}$ it is small enough ($\langle \delta \sigma_{\bm{v}} \rangle/\bar{\sigma}_{\bm{v}}\sim 0.02-0.03$). Masses and kinetic energies of SIDM and CDM haloes are also the same inside $r_{\text{SIDM}}$, with a precision of about 5$\%$. 
    \item Our model improves upon isotropic models by allowing for a radially-dependent velocity anisotropy $\beta(r)$, taking into account that the SIDM simulations show a non-zero  anisotropy  within $r_{\text{SIDM}}$ for $\sigma/m<5~\text{cm}^2/\text{g}$. Indeed, $\beta$ could be up to 0.2 in the region of interest. This demonstrates that equilibrium is not fully established inside $r_{\text{SIDM}}$, even if the total velocity dispersion is quite close to constant. We have taken this effect into account by using a simple ansatz for the velocity anisotropy, which is the same for all haloes (but with parameters that depend on the cross-section and are fitted to the simulations; see Table 1).
\end{itemize}

Our model would be complete if, for each halo and for a given cross-section, we could predict $r_{\text{SIDM}}$. This has been done by fixing the (radially-dependent) number of collisions per particle $N(r)$, as $N(r_{\text{SIDM}}) = 1$. We have found, however, that we could not complete our model, in this sense, for the following reasons: 
\begin{itemize}
    \item  In the simulations, $N(r_{\text{SIDM}})$ is not equal to 1, but instead depends on the concentration of the halo and on the cross-section. $N(r)$ in simulated haloes is a slowly-varying monotonic function of radius, flattening at the center. Because of this, a condition like $N(r_{\text{SIDM}})\sim3-5$  would fix $r_{\text{SIDM}}$ with an uncertainty of up to an order of magnitude. Therefore, even if we were able to measure $r_{\text{SIDM}}$ from observations, this would not help us to fix the cross-section, $\sigma/m$, as the uncertainty in the relation between $r_{\text{SIDM}}$ and $\sigma/m$ via $N(r)={\rm const.}$ is too large. 
    \item We have tried to model the radial dependence of the number of collisions by defining a scale radius $r_N$ where the slope of $N(r)$ changes most abruptly, effectively separating a central flat behaviour from an outer power law. Although this radius correlates strongly with the cross-section, the ratio of $r_N$ and $r_{\text{SIDM}}$ also changes with the cross-section. Because of this, although we built a simple model that roughly explains the radial dependence of $N(r)$, we ultimately cannot relate $\sigma/m$ to observables (e.g the core size) in an accurate way.
    \item This limitation seems to be fundamentally driven {\it i)} on the low-end of the cross-sections studied here by the lack of full thermalization of the core (up to the maximum size it can take), and {\it ii)} at the high-end of the cross sections due to the saturation of the core and the triggering of an on-setting gravothermal collapse. Therefore, the range of cross sections where an equilibrium model can in principle be used, even with our suggested improvements, is indeed quite narrow.
\end{itemize}

We conclude that despite the improvements we have made to the analytical modelling of SIDM haloes, the limitations are relevant enough that
the most reliable method to put constraints on the self-scattering cross-section remains a direct comparison between
the results from full simulations and quantities that are in principle observable, such as the core radius and the surface density (see Fig.~\ref{fig:rcoreofsigma}).
In the absence of sufficient simulation data to do so, in particular for systems including baryons, current limits and 'measurements' of
the self-scattering cross section should be interpreted with some care.

\section*{Acknowledgements}

JZ acknowledge support by a Grant of Excellence from the
Icelandic Research Fund (grant number 173929$-$051). Research of KB and AB was supported by the Netherlands Science Foundation (NWO/OCW) and the  European  Research  Council  (grant number 694896). Th.B. was supported by the Deutsche Forschungsgemeinschaft through the graduate school ``Particle and Astroparticle Physics in the Light of the LHC'' and through the individual grant ``Cosmological probes of dark matter properties''. TB wishes to thank McGill university for hospitality,
where part of this manuscript was completed.
This research was supported in part by Perimeter Institute for Theoretical Physics. Research at Perimeter Institute is supported by the Government of Canada through the Department of Innovation, Science and Economic Development and by the Province of Ontario through the Ministry of Research, Innovation and Science.
The new simulations for this paper have been done using the facilities provided by the RWTH High Performance Computing cluster under the project \texttt{rwth0113}.

\newpage
\appendix

\begin{figure}[t!]
  \centering
    \includegraphics[width=0.48\textwidth]{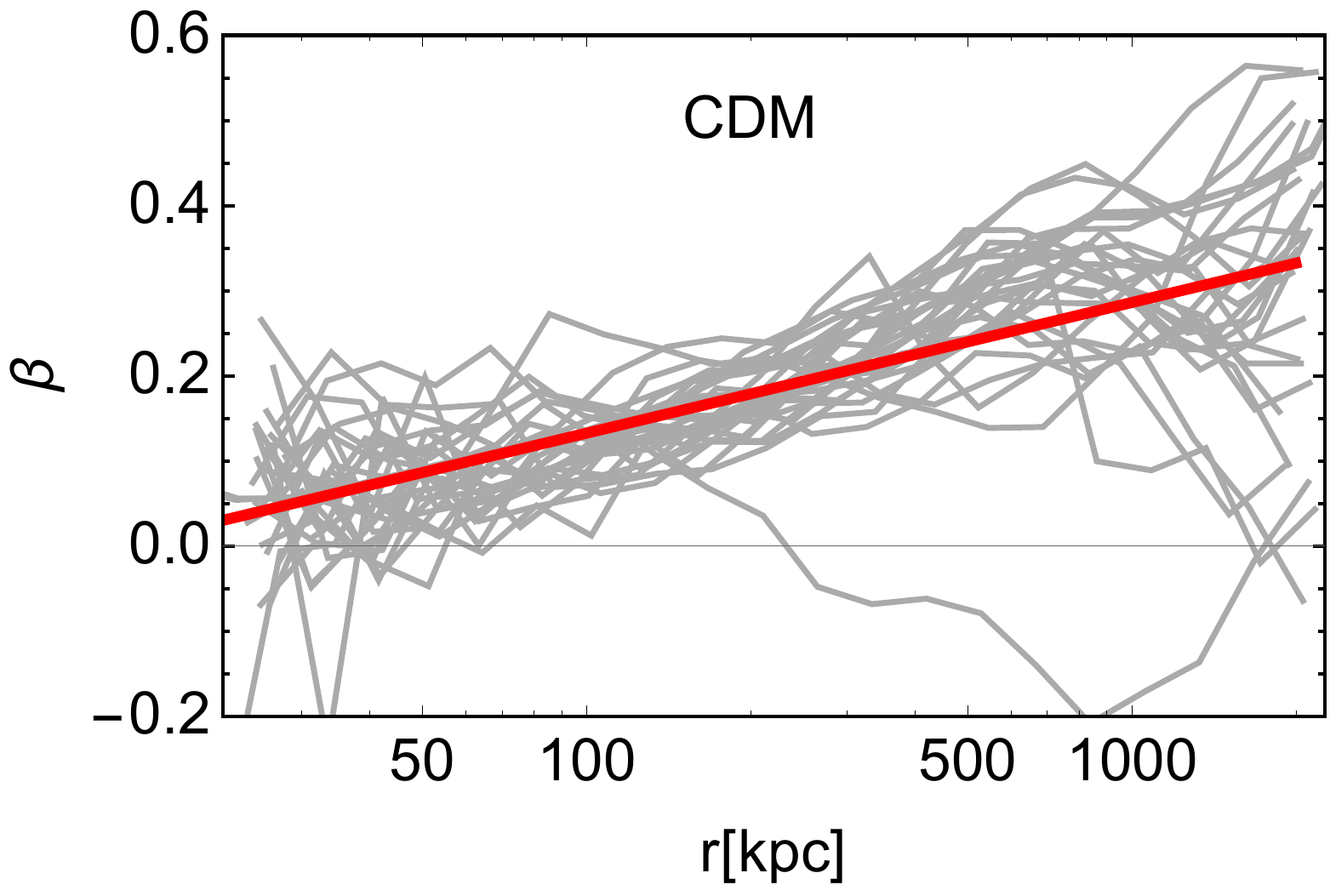}~\includegraphics[width=0.48\textwidth]{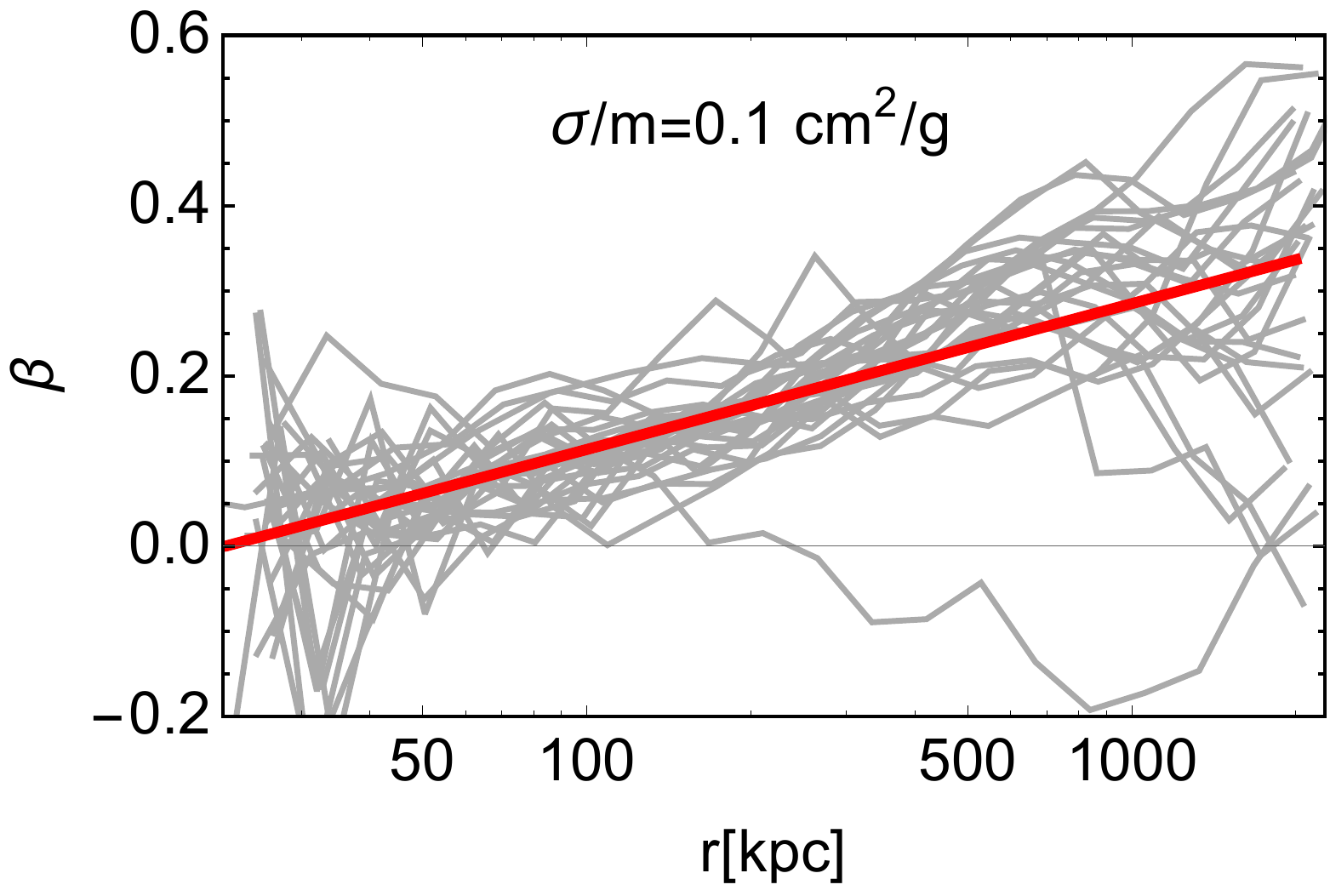} \\ \includegraphics[width=0.48\textwidth]{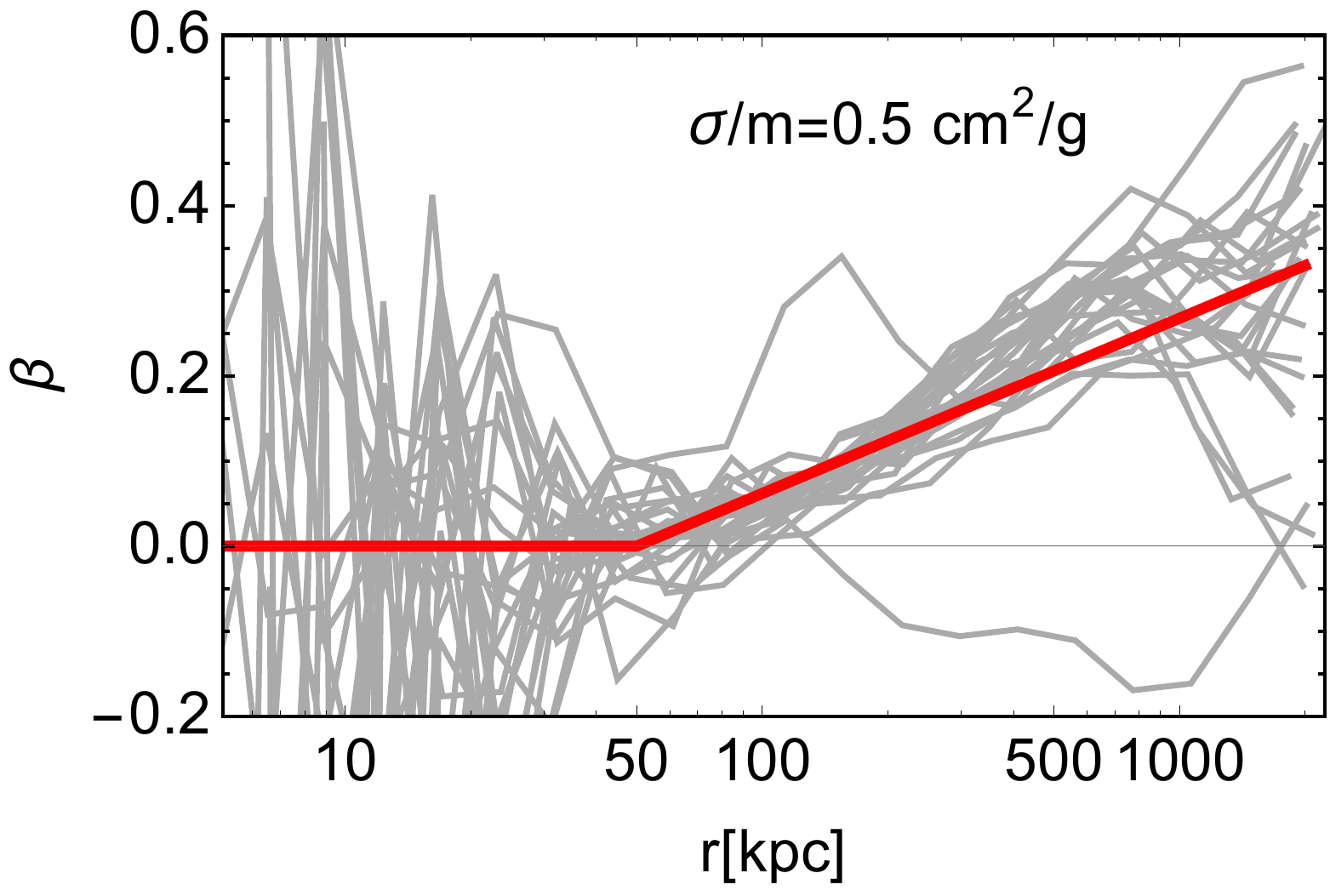}~\includegraphics[width=0.48\textwidth]{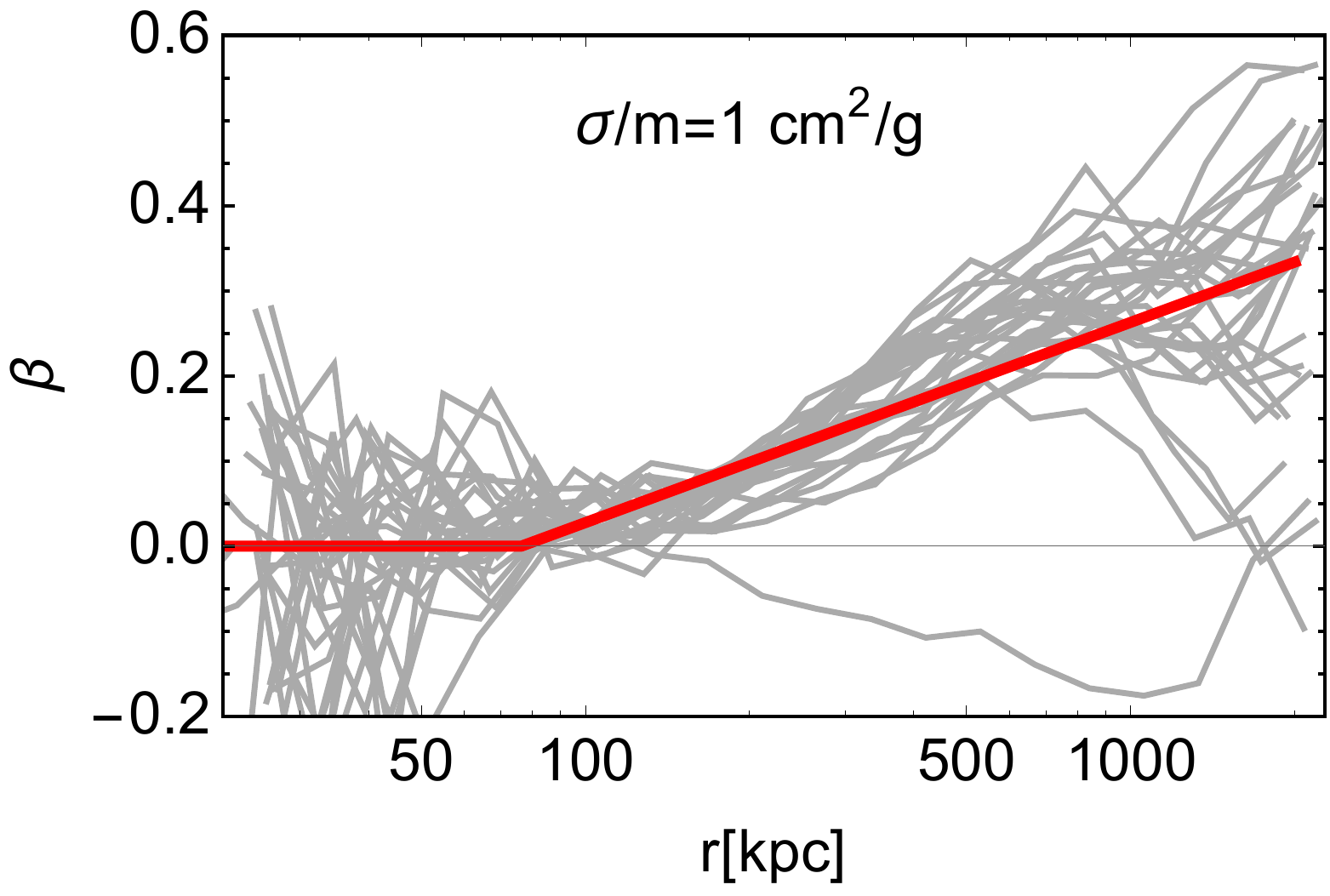} \\
    \includegraphics[width=0.48\textwidth]{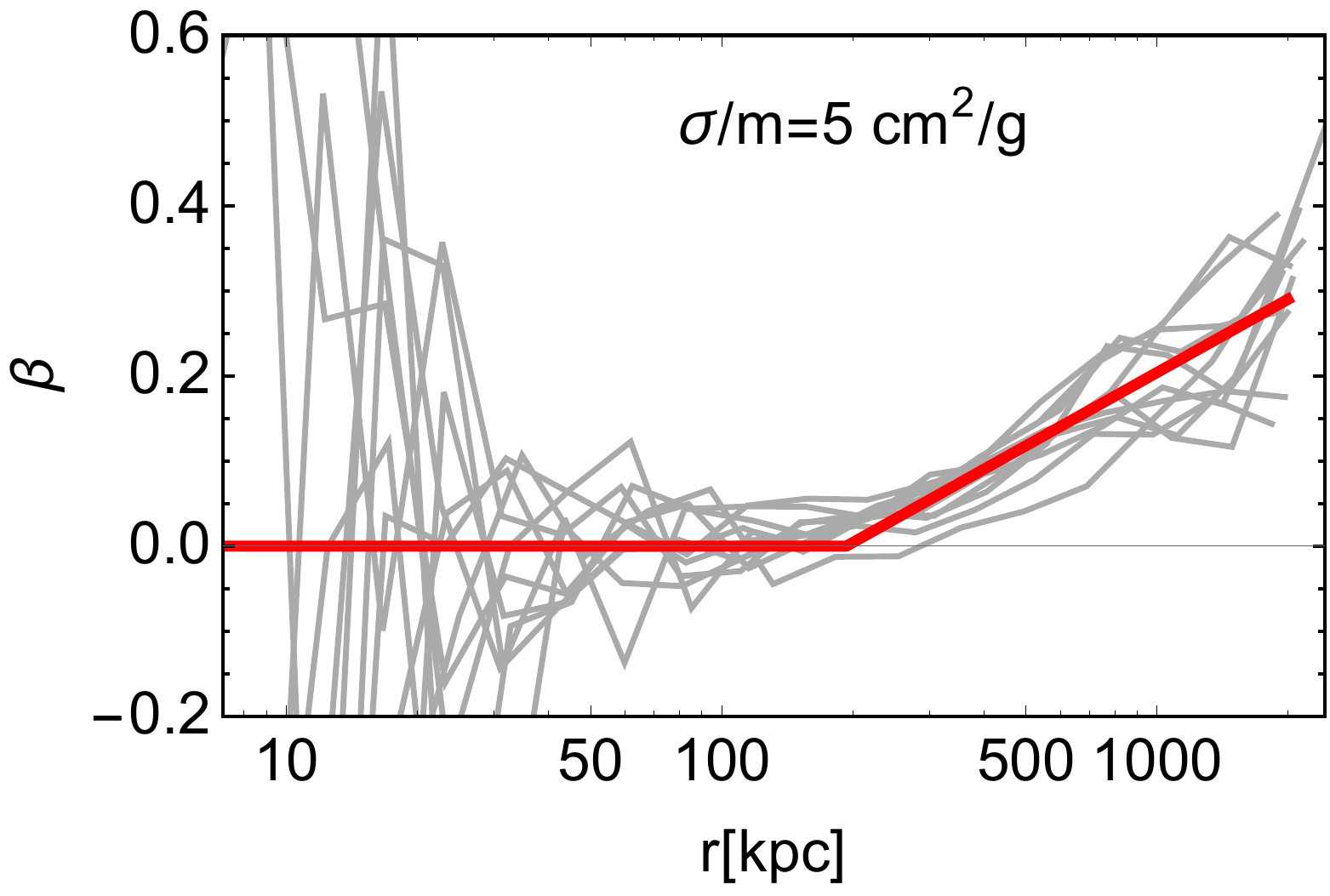}~\includegraphics[width=0.48\textwidth]{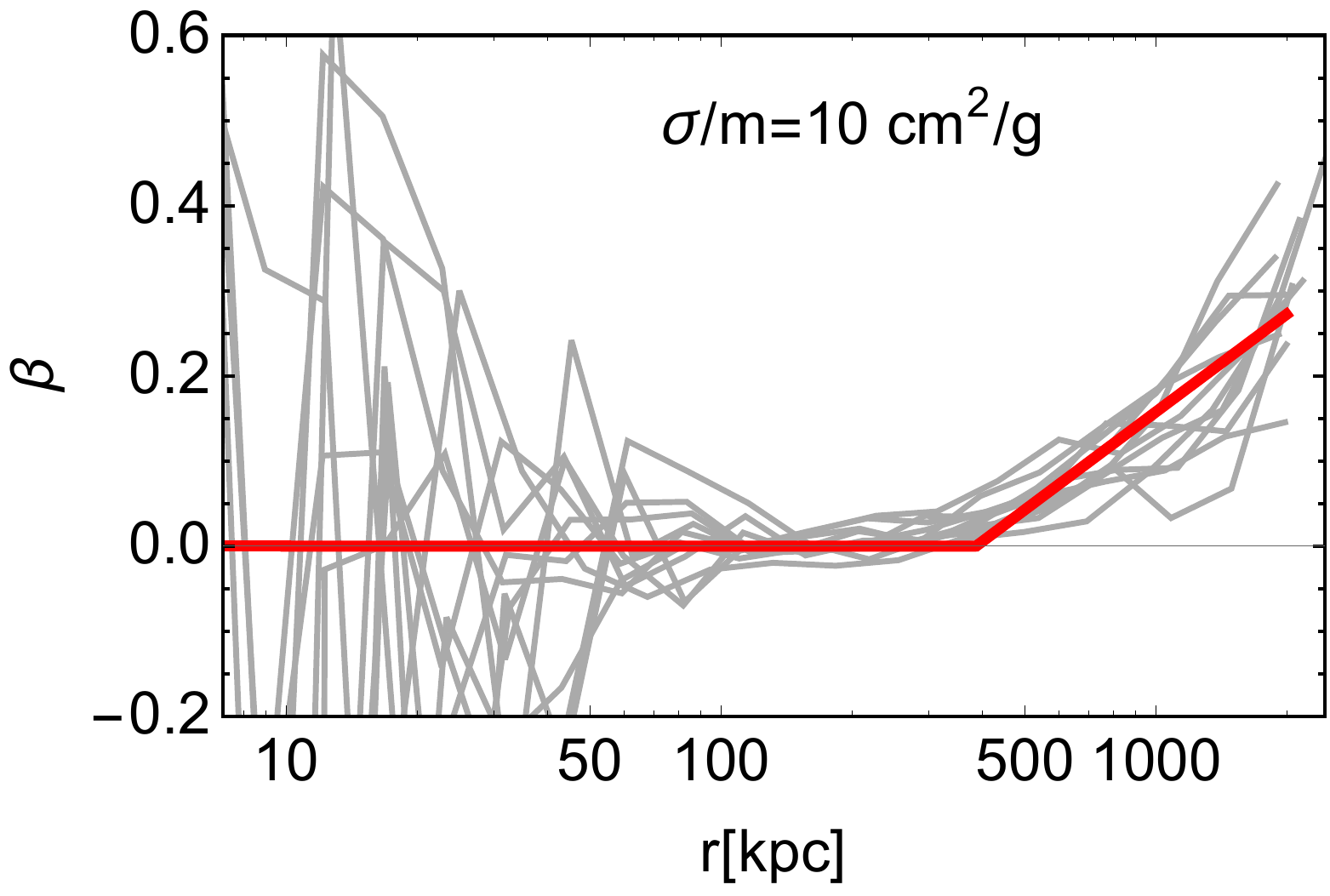}
    \caption{Velocity anisotropy profiles for all haloes in our simulations (gray lines), for CDM and SIDM with cross-sections $\sigma/m = 0.1$ cm$^2/$g, $0.5$ cm$^2/$g, $1$ cm$^2/$g, $5$ cm$^2/$g and $10$ cm$^2/$g. The red line is given by the ansatz~\eqref{eq:betaansatz} with the best-fit parameters stated in Table~\ref{tab:beta}.}
    \label{fig:allbetafits}
\end{figure}

\section{Radial velocity anisotropy profile $\beta(r)$}
\label{sec:betafit}

We have adopted a simple two-parameter ansatz to fit the mean behaviour of the velocity anisotropy profile in the ensemble of our simulated haloes:\\[-1.5ex]
\begin{equation}
    \beta(r) = \left\{ \begin{matrix}
    A \ln (r/r_{\beta}),&\text{ for }r\ge r_{\beta} \\
    0,&\text{ for }r< r_{\beta}
    \end{matrix}
    \right.
\end{equation}\\[-1.5ex]
We fitted this formula to all haloes for a given cross section, and for the CDM model. 
In Fig.~\ref{fig:allbetafits} we show the 
result of this procedure along with the data
 (the best fit values of the parameters $A$ and $r_{\beta}$ are presented in Table~\ref{tab:beta}).

\section{From the NFW parameters of the CDM halo to the model of the SIDM halo}
\label{sec:sigmatotNFW}

The density, radial velocity dispersion, and velocity anisotropy profiles of collisionless CDM haloes are connected through the Jeans equation:
\begin{equation}
        \frac{d}{dr}\left(\sigma_{r}^2(r) \rho(r) \right)  
        +  \frac{2}{r} \beta(r) \sigma_r^2(r) \rho(r)
        = - \rho(r) \frac{G M(r)}{r^2} \
\label{eq:ApB1}
\end{equation} 
Therefore, one can find the velocity dispersion profile in CDM haloes using as input the profiles for density and velocity anisotropy. To show this, we introduce the function $f(r)=\sigma_{r}^2(r) \rho(r)$ to write Eq.~\ref{eq:ApB1}:
\begin{equation}
        \frac{df}{dr}  
        +  \frac{2}{r} \beta(r) f(r)
        = - \rho_{\text{DM}}(r) \frac{G M_{\text{DM}}(r)}{r^2} \ ,
\label{eq:ApB2}
\end{equation}
and we use the method of variation of constants to solve this equation. The solution of the homogeneous equation
\begin{equation}
        \frac{df}{dr}  
        +  \frac{2}{r} \beta(r) f(r)
        = 0
\end{equation}
is
\begin{equation}
        f(r) = C_1 e^{-2\int_{r_0}^r \frac{\beta(y)}{y} dy} \ .
\end{equation}
We substitute this solution in Eq.~\ref{eq:ApB2} with $C_1\to C_1(r)$ and get
\begin{equation}
        \frac{dC_1}{dr} e^{-2\int_{r_0}^r \frac{\beta(y)}{y} dy}
        = - \rho(r) \frac{G M(r)}{r^2} \ .
\end{equation}
The general solution for $C_1(r)$ in this equation is
\begin{equation}
        C_1(r) 
        = C - \int_{r_0}^r e^{2\int_{r_0}^x \frac{\beta(y)}{y} dy} \rho(x) \frac{G M(x)}{x^2} dx \ .
\end{equation}
Thus, the velocity dispersion profile is given by
\begin{equation}
        \sigma_r^2(r) \rho(r) = 
        C e^{-2\int_{r_0}^r \frac{\beta(y)}{y} dy}
        -
        \int_{r_0}^{r} e^{2 \int_r^x \frac{\beta(y)}{y} dy} 
        \frac{G M (x)}{x^2} \rho(x) dx \ . 
\label{eq:ApB3}
\end{equation}
The constant $C$ can be fixed with the values of the density and velocity dispersion at a radius $r=r_0$ in Eq.~\ref{eq:ApB3}:
$C = \sigma_r^2(r_0) \rho_{\text{CDM}}(r_0)$. Thus, we finally have: 
\begin{equation}
        \sigma_r^2(r) \rho(r) = 
        \sigma_r^2(r_0) \rho(r_0) e^{-2\int_{r_0}^r \frac{\beta(y)}{y} dy}
        -
        \int_{r_0}^{r} e^{2 \int_r^x \frac{\beta(y)}{y} dy} 
        \frac{G M (x)}{x^2} \rho(x) dx \ .
        \label{eq:sigmarho}
\end{equation}

In principle, we have the problem that we do not know the value of $\sigma_r(r)$ at any finite radius $r_0$, but if we assume that $\sigma_r^2(r) \rho(r) \to 0$ as $r\to\infty$, which is reasonable for CDM haloes, then we can choose the boundary condition $\sigma_r^2(10r_s) \rho(10r_s) = 0$. Therefore, we can use the NFW profile for $\rho(r)$ and the ansatz for $\beta(r)$ described in Appendix~\ref{sec:betafit} to estimate $\sigma_{\bm{v}}$. An example of the resulting velocity dispersion profiles is shown in the left panel of Fig.~\ref{fig:sigmatotfromNFW}. The quality of the match to the simulated data is comparable in all the cases to this examples and we can see that the fit is reasonable.

\begin{figure}[t!]
  \centering
    \includegraphics[width=0.48\textwidth]{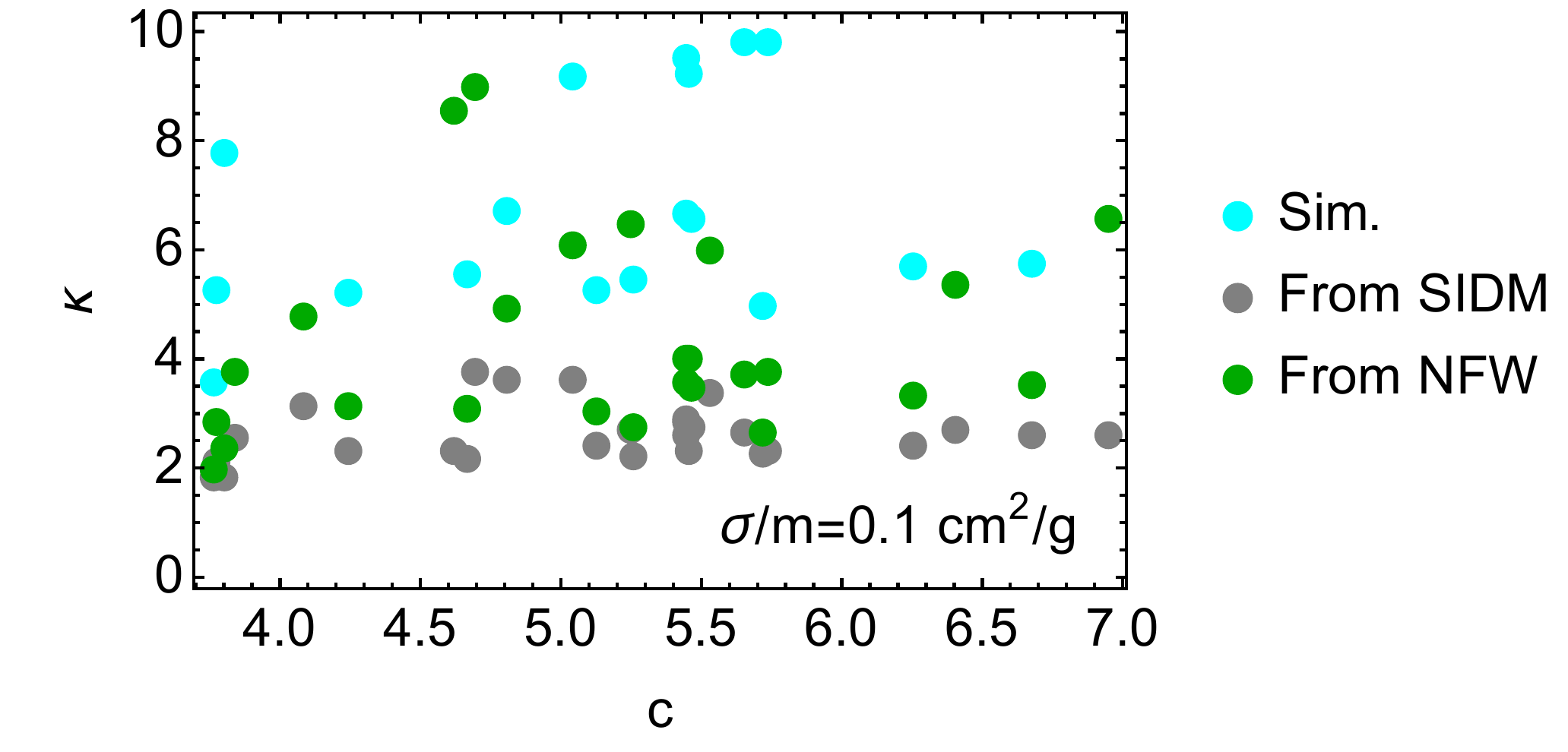}~\includegraphics[width=0.48\textwidth]{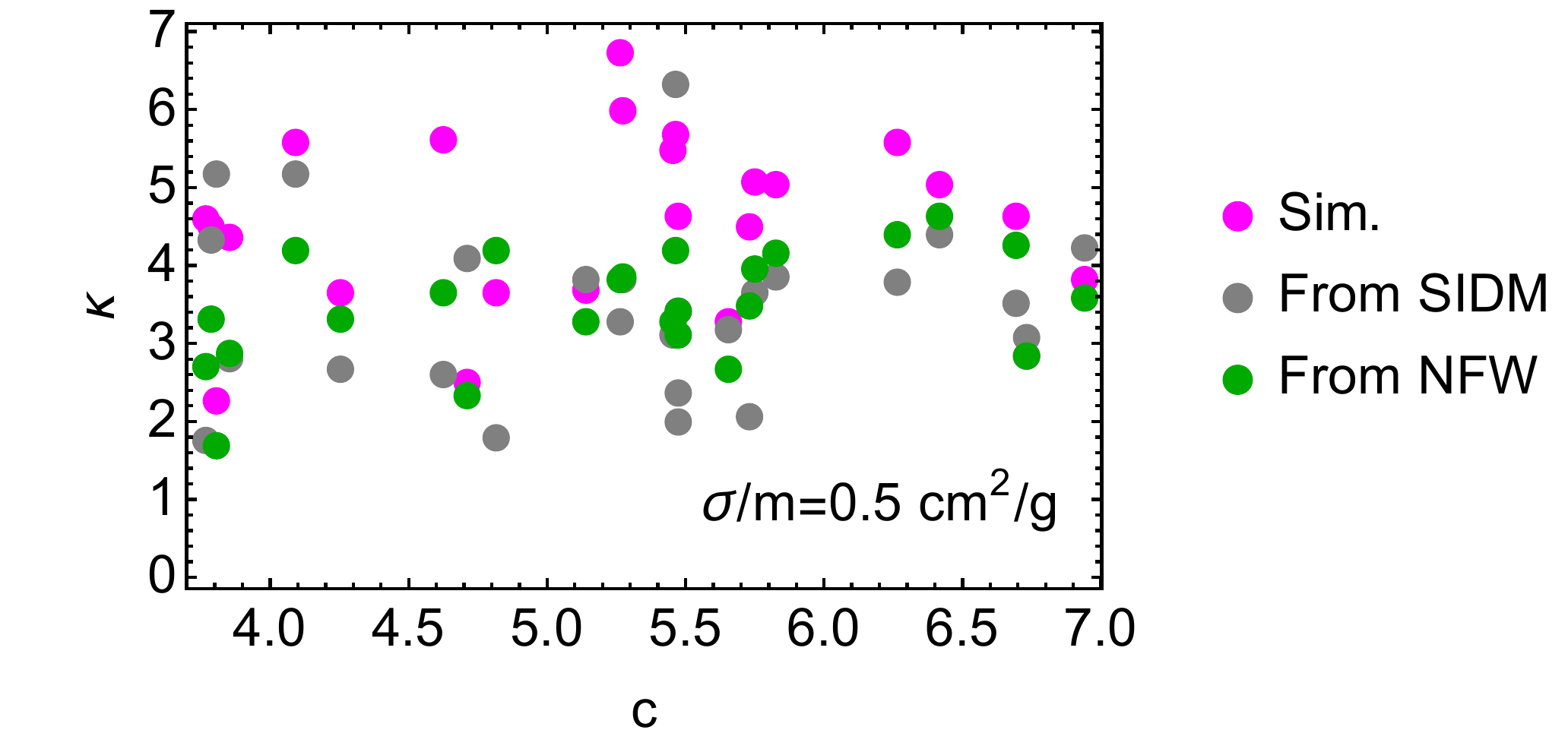} \\
    \includegraphics[width=0.48\textwidth]{plots/kappafromNFW.pdf}~\includegraphics[width=0.48\textwidth]{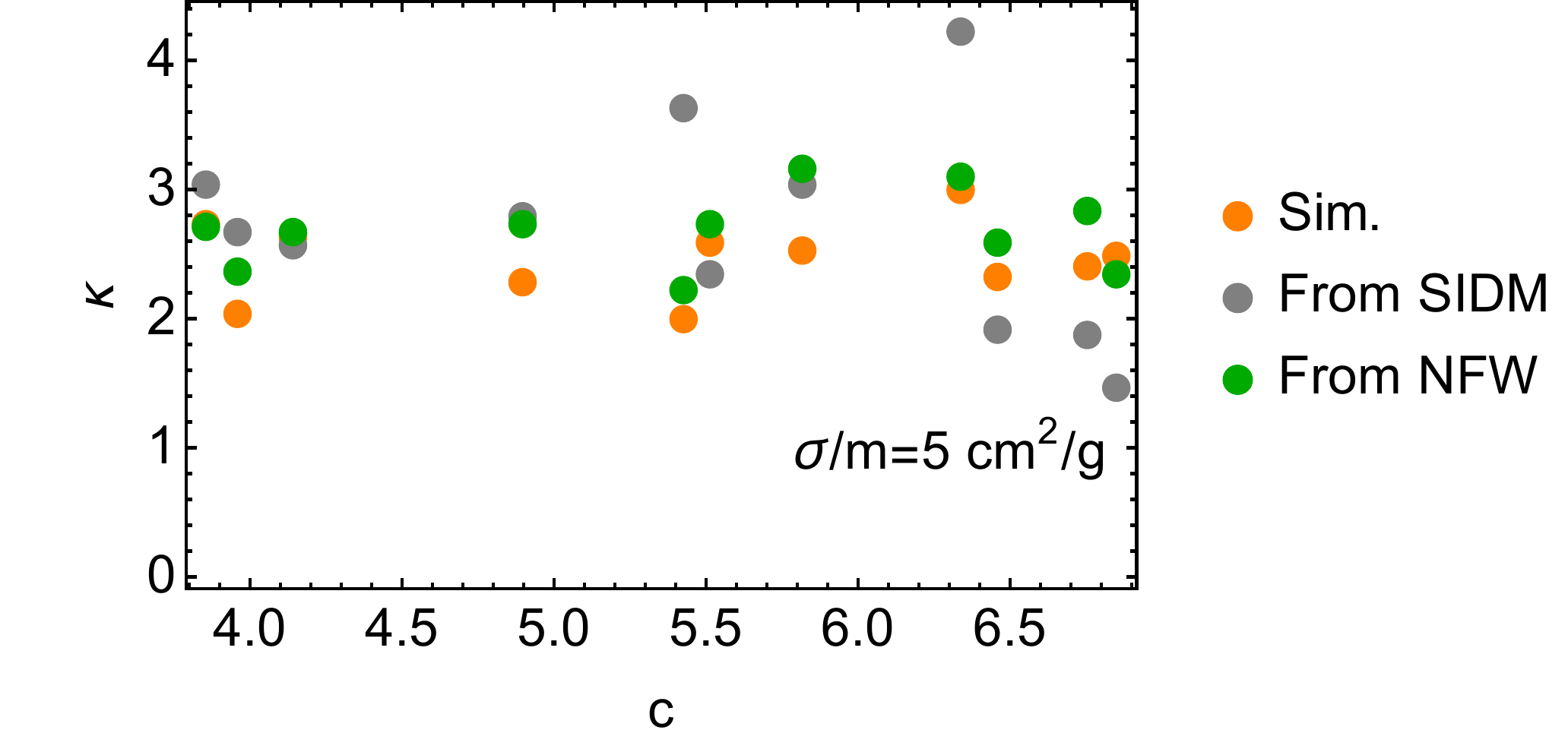} \\
    \includegraphics[width=0.48\textwidth]{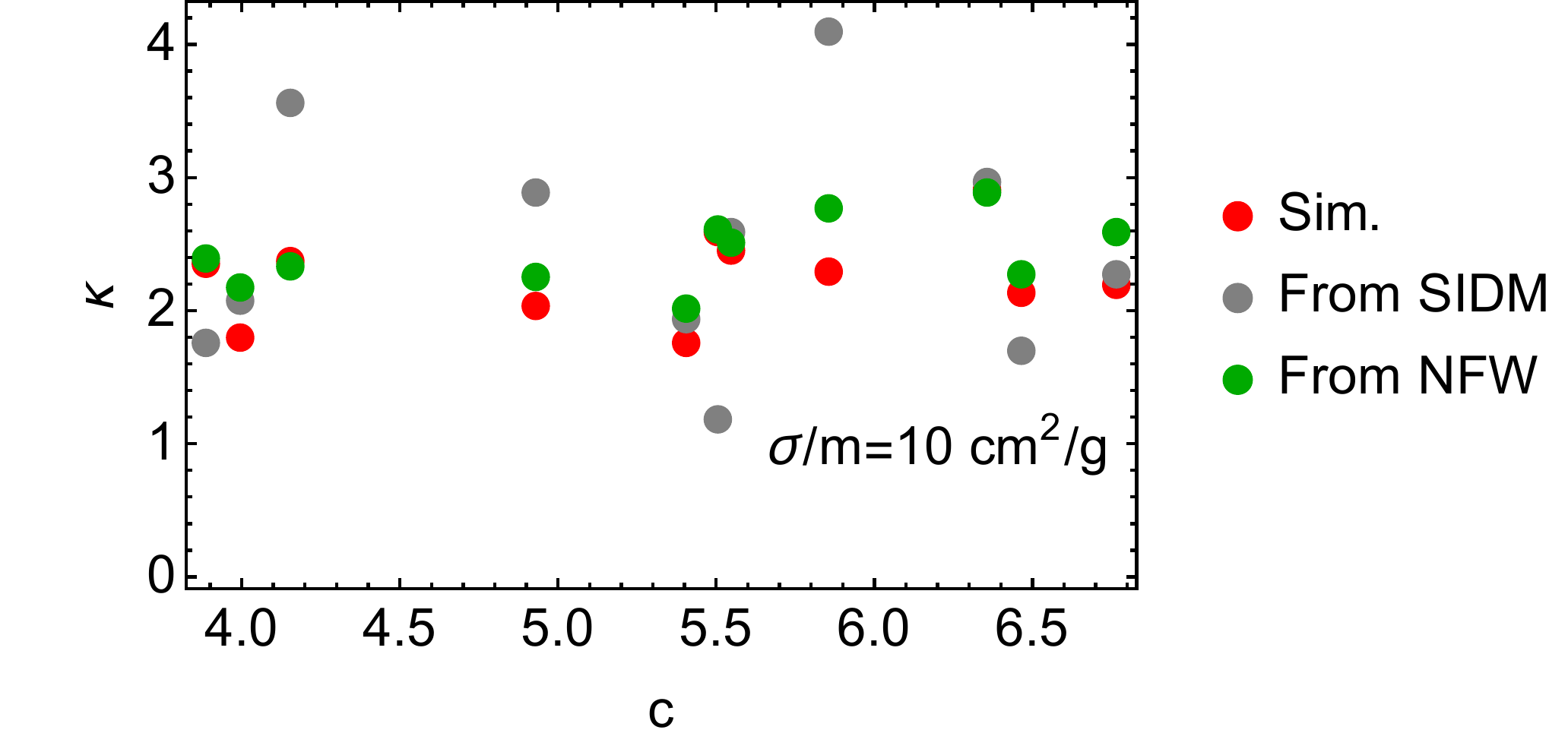}
    \caption{The ratio $\kappa=\rho(0)/\langle\rho(r_M)\rangle$ in simulated SIDM haloes as a function of halo concentration for the cross-sections $\sigma/m = 0.1$ cm$^2/$g (cyan), $0.5$ cm$^2/$g (magenta), $1$ cm$^2/$g (blue), $5$ cm$^2/$g (orange) and $10$ cm$^2/$g (red). The predicted values from our anisotropic model using input parameters from the SIDM haloes are shown in gray, while the results of the same model using input parameters from the corresponding CDM halo (modeled with a NFW density profile and velocity anisotropy as described in Appendix~\ref{sec:betafit}) is shown in green.}
    \label{fig:kappafromNFW2}
\end{figure}

\bigskip
Having established that we can obtain the velocity dispersion profile of CDM haloes from the (NFW) density and velocity anisotropy profiles, we show in Fig.~\ref{fig:kappafromNFW2} how well this connection allows us to model the properties of the SIDM halo directly from the properties of the corresponding CDM halo.

\section{The precision of the radius $r_M$ in the halo mass profiles}
\label{sec:rM}

\begin{figure}[t!]
  \centering
    \includegraphics[width=0.5\textwidth]{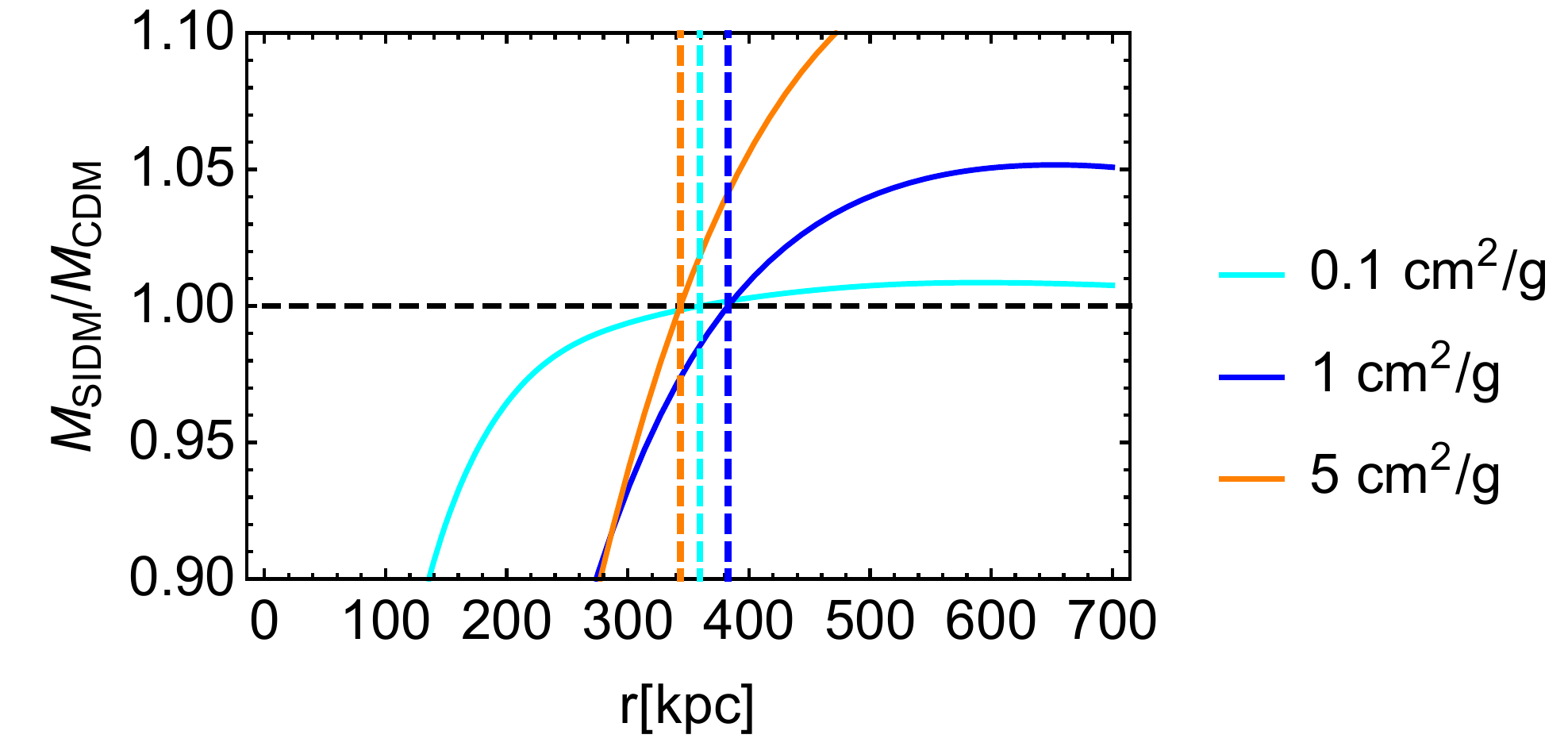}~\includegraphics[width=0.5\textwidth]{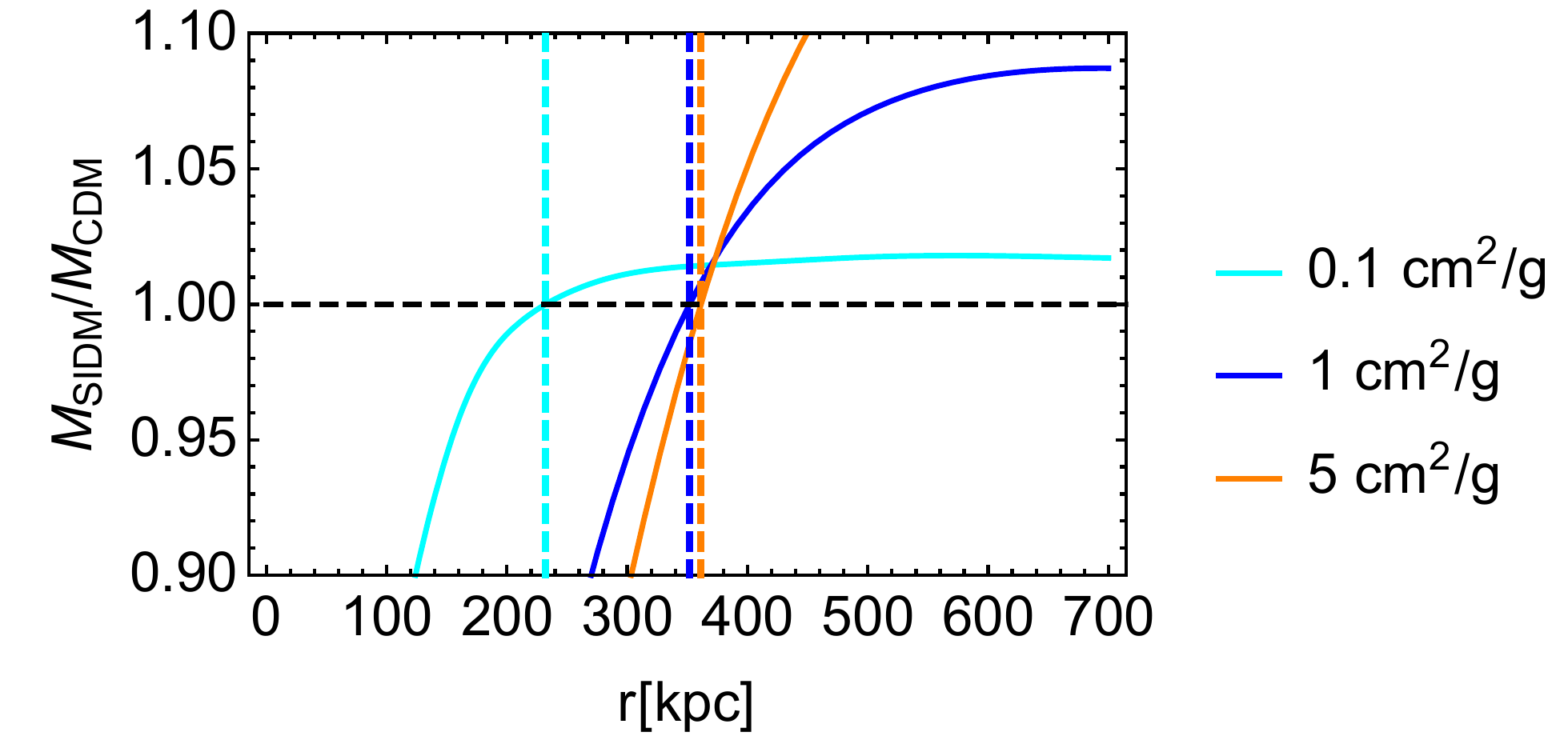} \\
    \includegraphics[width=0.5\textwidth]{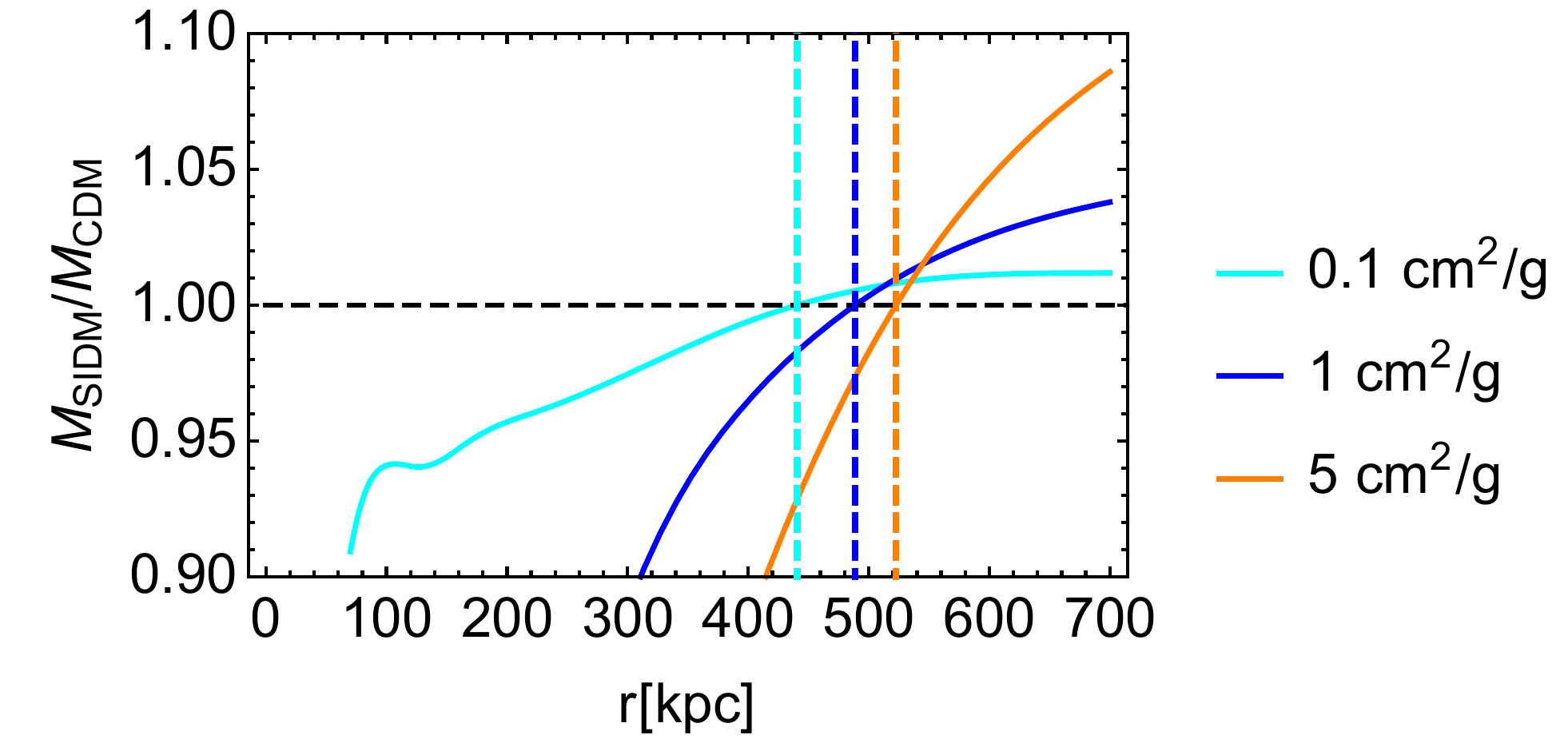}~\includegraphics[width=0.5\textwidth]{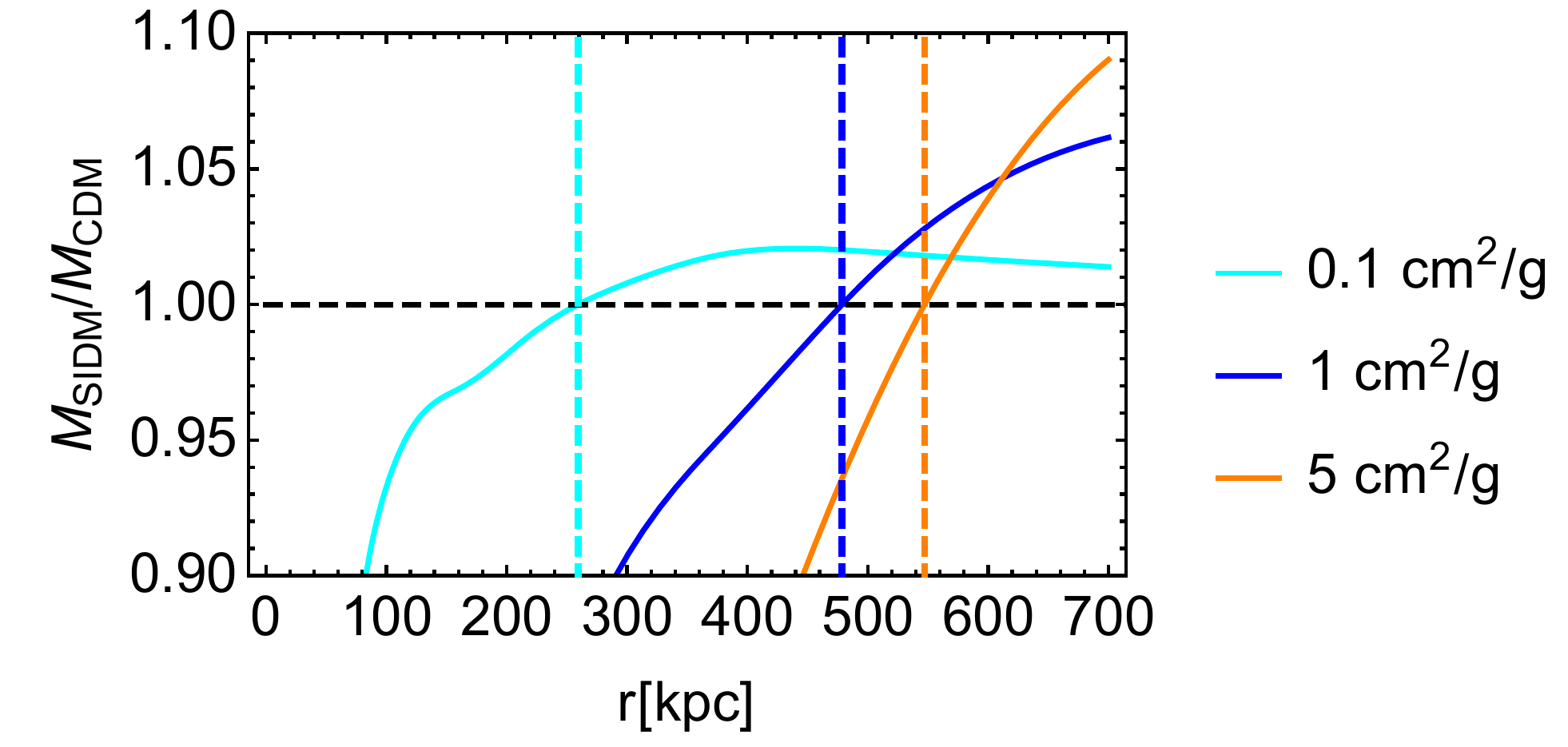}
    \caption{The ratio of SIDM to CDM profile versus radius for different cross-sections. The dashed vertical lines represent the corresponding radii $r_M$.}
    \label{fig:massprofile}
\end{figure}

In Fig.~\ref{fig:massprofile} we present the ratio of SIDM to CDM profiles for the cross-sections $\sigma/m = 0.1,$ $\sigma/m = 1$ and $\sigma/m = 5$ cm$^2/$g.
\begin{figure}[t!]
  \centering
    \includegraphics[width=0.5\textwidth]{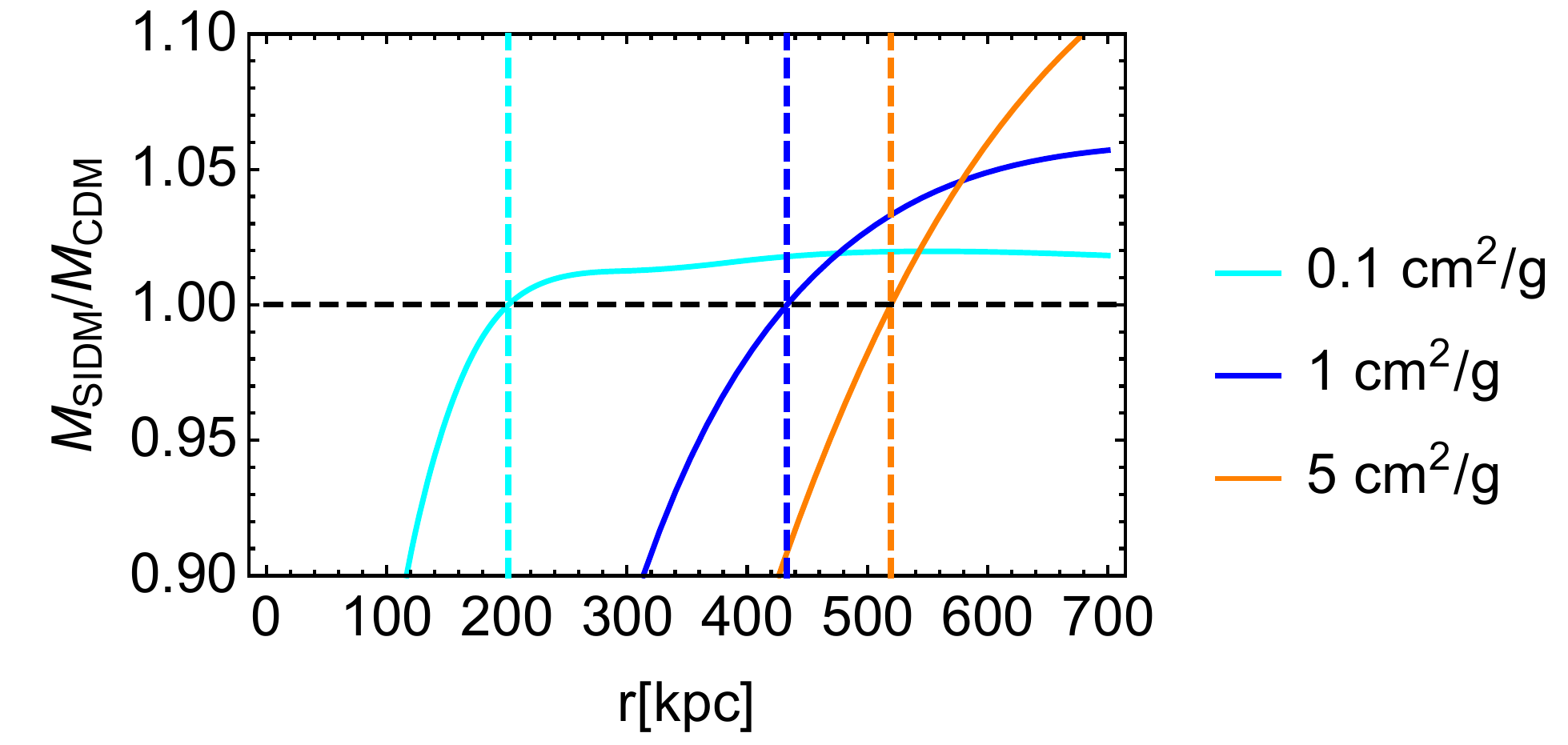}~\includegraphics[width=0.5\textwidth]{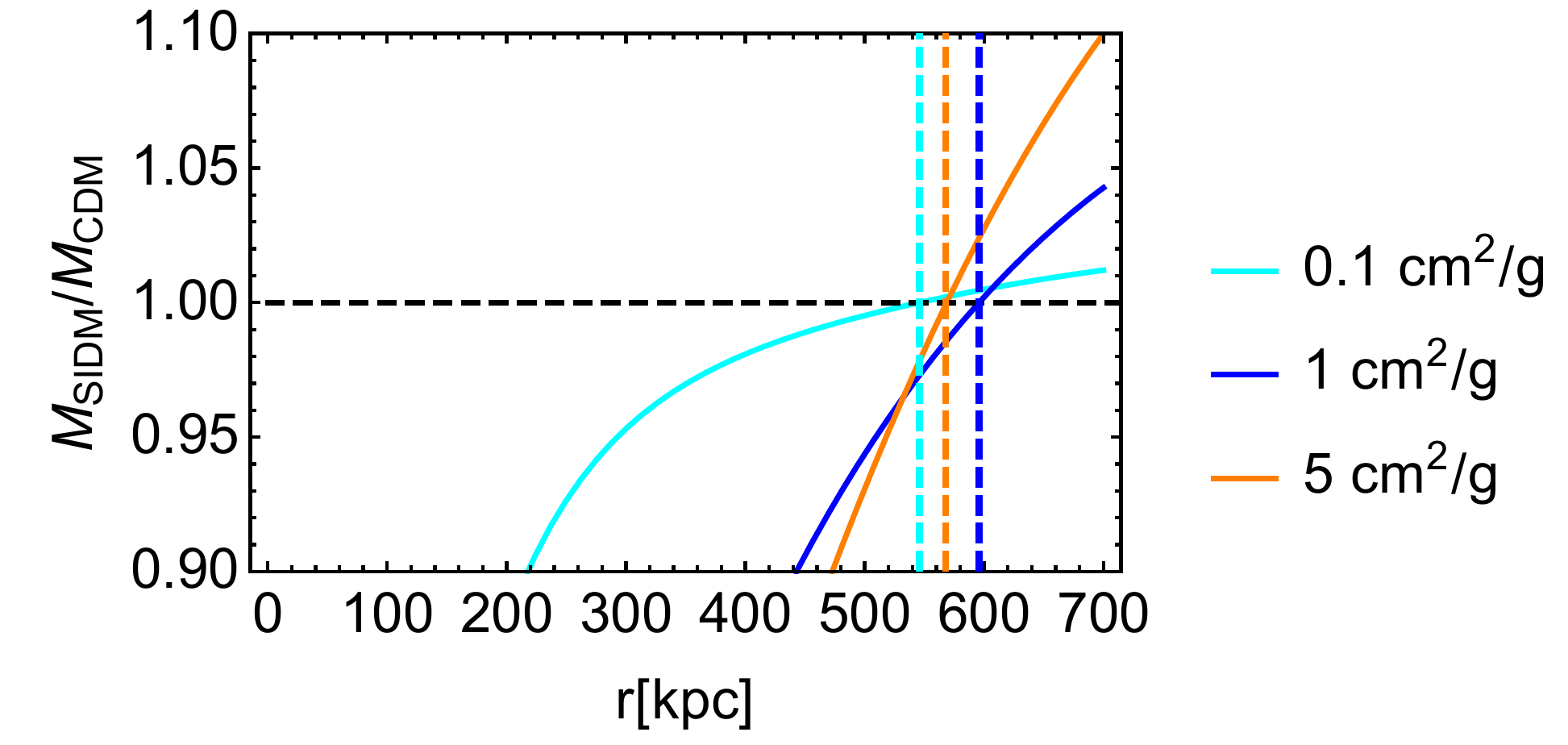} \\
    \includegraphics[width=0.5\textwidth]{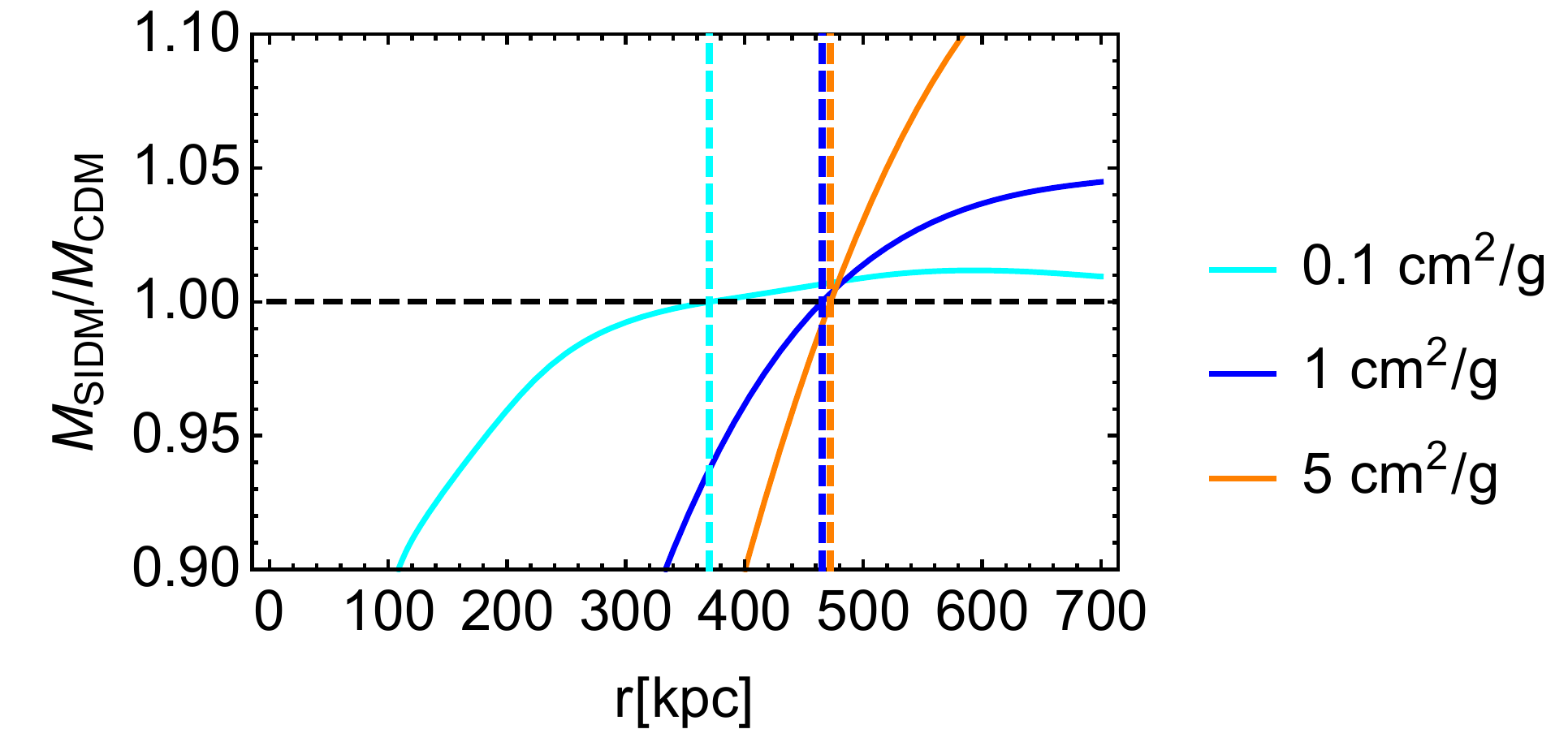}~\includegraphics[width=0.5\textwidth]{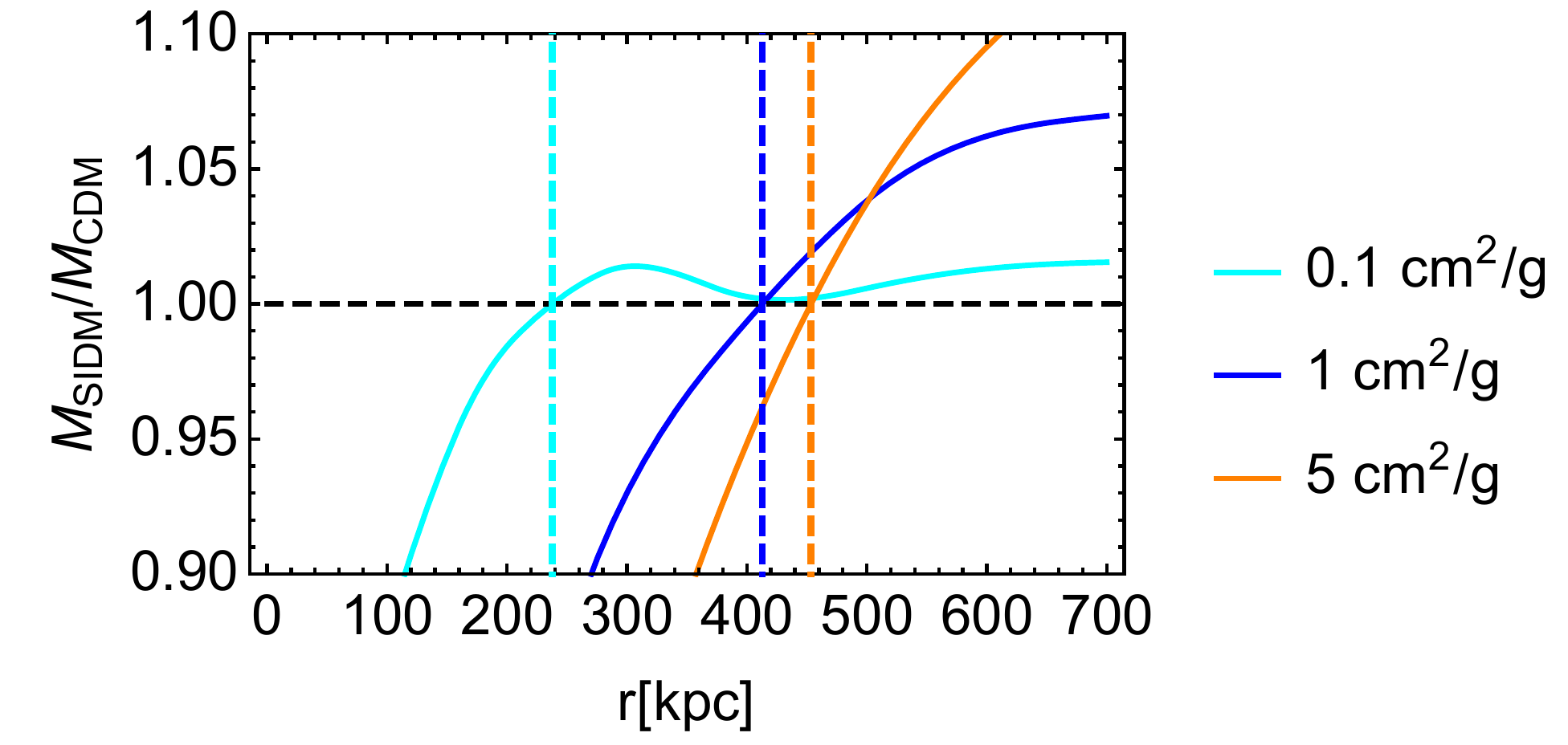}
\end{figure}

\bibliographystyle{JHEP} %
\bibliography{SIDM.bib}

\end{document}